\documentclass[12pt,onecolumn]{article}
\usepackage{color}
\usepackage{amsmath}
\usepackage{amssymb}
\usepackage{amsfonts}
\usepackage{geometry}
\usepackage{setspace}
\usepackage{amsmath,bm}
\usepackage{graphicx}
\usepackage[section]{placeins}% avoid figures cross the section
\usepackage{subfigure}
\usepackage[sort&compress]{natbib}
\usepackage{booktabs}  
\usepackage{makecell}
\usepackage{caption}% caption
\usepackage[colorlinks, citecolor=blue]{hyperref}% link to the superlink
\usepackage{lineno} %line number

\author{Shuwei Zhou$^{1,2}$, Xiaoying Zhuang$^{1,2,*}$, Timon Rabczuk$^{3,4}$}
\title {Phase field modeling of brittle compressive-shear fractures in rock-like materials: a new driving force and a hybrid formulation}
\begin{document}
	%\linenumbers %line number
%\captionsetup[figure]{labelfont={bf},name={Fig.},labelsep=space,justification=raggedright, singlelinecheck = false}
\captionsetup[figure]{labelfont={bf},name={Fig.},labelsep=space}
% referecne style
%\bibliographystyle{unsrtnat}

\bibliographystyle{elsarticle-num-names}
\setcitestyle{numbers,square,aysep={},yysep={,},citesep={,}}
%\bibliographystyle{plainnat}
%%\bibliographystyle{apa}
%%\setcitestyle{authoryear,round,aysep={},yysep={,}}
% referecne style
%\bibliographystyle{unsrtnat}
%\setcitestyle{numbers,square,aysep={},yysep={,}}
% delte the date
% Increase the section level (declaration}
%\setcounter{tocdepth}{4} 
%\setcounter{secnumdepth}{4}

\date{}
\maketitle

\spacing {2}
\noindent
1 Department of Geotechnical Engineering, College of Civil Engineering, Tongji University, Shanghai 200092, P.R. China\\
2 Institute of Continuum Mechanics, Leibniz University Hannover, Hannover 30167, Germany\\
3 Division of Computational Mechanics, Ton Duc Thang University, Ho Chi Minh City, Viet Nam\\
4 Faculty of Civil Engineering, Ton Duc Thang University, Ho Chi Minh City, Viet Nam\\
* Corresponding author: Xiaoying Zhuang (zhuang@ikm.uni-hannover.de)

%\begin{spacing}{2.0}
\begin{abstract}
\noindent Compressive-shear fracture is commonly observed in rock-like materials. However, this fracture type cannot be captured by current phase field models (PFMs), which have been proven an effective tool for modeling fracture initiation, propagation, coalescence, and branching in solids. The existing PFMs also cannot describe the influence of cohesion and internal friction angle on load-displacement curve during compression tests. Therefore, to develop a new phase field model that can simulate well compressive-shear fractures in rock-like materials, we construct a new driving force in the evolution equation of phase field. Strain spectral decomposition is applied and only the compressive part of the strain is used in the new driving force with consideration of the influence of cohesion and internal friction angle. For ease of implementation, a hybrid formulation is established for the phase field modeling. Then, we test the brittle compressive-shear fractures in uniaxial compression tests on intact rock-like specimens as well as those with a single or two parallel inclined flaws. All numerical results are in good agreement with the experimental observation, validating the feasibility and practicability of the proposed PFM for simulating brittle compressive-shear fractures.
\end{abstract}
%\end{spacing}

\noindent Keywords: Phase field model, Rock-like material, Compressive-shear fracture, Driving force, Hybrid formulation, Strain decomposition
% Add a content
%\tableofcontents

%\twocolumn
\section {Introduction}\label{Introduction}

Geomaterial failure dominated by shear or compressive fractures is commonly observed in underground engineering. However, the prediction of fractures in rock-like solids remains a challenging topic especially for complex fracture patterns. One major reason of this is that geological media generally have many intrinsic flaws such as micro cracks, voids and soft minerals \citep{zhou2015analytical, zhou2017numerical, zhou2018phase3}, which drastically increase the complexity of fracture propagation in the geomaterials. One direct way to explore the fracture mechanisms is to establish experimental tests and there have been many studies carried out regarding the fracture propagation in rock-like materials; see for instance the contributions in \citet{lajtai1974brittle, sagong2002coalescence, wong2009systematic, yang2011strength, basu2013rock, xia2015strength, zhou2015damage, zhou2017statistical}. Although experimental tests can observe basic phenomenon and provide guidance for empirical studies, they are generally limited by the loading conditions and observations of the fracture pattern evolution. Furthermore, many physical quantities cannot be measured directly, e.g. stress/strain near crack tips. Therefore, numerical approaches have become important and complementary alternative and are attracting increasing interests because they can help to understand the mechanism and are less expensive than experimental tests.
 
In the past decade, a variety of numerical methods have been developed for capturing fracture phenomena and providing distinctive physical insights that cannot be obtained through experimental tests alone \citep{zhou2018phase3}. These numerical methods can be categorized into two categories, namely the discrete approach and smeared approach according to the different treatments of strong discontinuities in the displacement field across the fracture interfaces. For the discrete approaches, a sharp fracture model is introduced in modeling either explicitly \citep{ingraffea1985numerical} or implicitly \citep{moes2002extended, chen2012extended}. Among these methods, the extended finite element method (XFEM) \citep{moes2002extended, chen2012extended, fu2013boundary, fu2019robust, xi2019efficient}, generalized finite element method (GFEM) \citep{fries2010extended}, and the phantom-node method \citep{chau2012phantom, rabczuk2008new} have been extensively applied and developed in recent years. In contrast to the discrete approaches, the smeared approaches diffuse the fracture by using smooth transitions between the intact material and fully damaged material \citep{zhang2017modification}; therefore the displacement field is in a sense still continuous. Popular smeared approaches include gradient damage models \citep{peerlings1996some}, screened-Poisson models \citep{areias2016damage, areias2016novel} and phase field methods (PFMs) \citep{miehe2010thermodynamically, miehe2010phase, borden2012phase}. 

Among those smeared methods, PFMs have shown great potentials for complex fracture failure especially for multi-field problem \citep{bourdin2008variational, borden2012phase, hesch2014thermodynamically, zhou2018phase, zhou2018phase2, zhou2018phase3, zhou2018propagation, zhou2018adaptive}. The origin of PFMs for brittle fractures can be traced back to \citet{bourdin2008variational} and later on the first thermodynamic consistent formulation of PFM was not proposed until \citet{miehe2010phase}. The original PFM and its successors all introduce an additional scalar field (the so-called phase field) to represent the evolution of fracture. Therefore, the sharp fracture is represented by a diffusive 'damage-like' zone with the zone width is controlled by a length scale parameter. By using the variational approach the phase field modeling of fractures becomes a multi-field problem, namely phase field and mechanical field, and fracture propagation is easily obtained by solving partial differential equations. This results in a natural detection and tracking of fracture paths without any additional crack growth criteria \citep{borden2012phase}, and for this reason PFM is believed to be a promising way to model complex fracture growth such as fracture arrest, deflection, branching, percolation and coalescence.

To date, PFMs have been extensively used to simulate a wide variety of fracture problems including dynamic fracture \citep{borden2012phase, nguyen2018modeling, zhou2018phase}, ductile fracture \citep{miehe2014variational, ambati2016phase, borden2016phase, shanthraj2016phase}, cohesive fracture \citep{verhoosel2013phase, vignollet2014phase, nguyen2018modeling}, fractures in thin shells and plates \citep{ambati2016phase}, fractures in functionally graded materials and composites \citep{natarajan2019modeling, natarajan2019phase}, and fluid-driven fractures \citep{miehe2015phase, lee2016pressure, miehe2016phase, ehlers2017phase, li2019numerical, zhou2019phase}, while some implementations in FEM software such as COMSOL \citep{zhou2018phase} and FEniCS \citep{HIRSHIKESH380} accelerate the application of PFMs. However, the applications and improvements of the PFM for fractures in rock-like materials are relatively limited and far from practical application. For example, only mode I tensile dominated fractures can be simulated such as the Brazilian tests \citep{zhou2018adaptive, zhou2018fracture, zhou2018propagation}, notched semi-circular bend (NSCB) tests \citep{zhou2018phase3}, direct tension tests \citep{zhou2018phase3}, and the hydraulic fractures \citep{zhou2018phase2, miehe2016phase}. Furthermore, in the geological environment, geomaterials are often subjected to compressive principal stresses and compressive-shear fractures are commonly observed. This means that shear fractures will initiate and propagate even when all principal strains of the material are compressive. In contrast to this fact, the compressive-shear fractures cannot be predicted by the current PFMs for rock-like solids \citep{zhang2017modification, choo2018coupled, bryant2018mixed}, where the negative strains and the compressive part of elastic energy are assumed not to contribute to the evolution of phase field. More specifically, the first attempt of the PFM for mixed-mode crack propagation in rock-like materials was proposed by \citet{zhang2017modification}. This method was established by introducing the critical energy release rate of mode II and two historic energy references in the model of \citet{miehe2010phase}. \citet{choo2018coupled} coupled a pressure-sensitive plasticity model with a phase field approach to capture brittle fracture, ductile flow, and their transition in rocks. \citet{bryant2018mixed} proposed a kinematic-consistent phase field approach to model mixed-mode fractures in anisotropic rocks. On the other hand, the contributions in \citet{zhang2017modification, choo2018coupled, bryant2018mixed} also show that the PFMs developed so far do not account for the influence of cohesion and internal friction angle on fracture propagation and the load-displacement curve. This is a missing aspect since it is generally accepted that a rock-like material will have a higher compressive strength if its cohesion or internal friction angle increases. Therefore, it is an imperative task to further develop PFM in order to be able to better predict the phenomena in rock failure.

In this study, we propose a new phase field model for simulating brittle compressive-shear fractures in rock-like materials by introducing a new driving force term, called $H_p$ in the evolution equation of the phase field. By using the spectral decomposition, this new driving force is established using the negative parts of the strain. In addition, the influence of cohesion and internal friction angle is considered in the driving force, and hence a hybrid formulation is developed in the present PFM. The present PFM is validated by reproducing the compressive-shear fractures observed in uniaxial compression tests on intact rock-like specimens as well as the tests on specimens with a single inclined flaw and two parallel inclined flaws, respectively.

The content of this paper is outlined as follows. The anisotropic phase field model for brittle fractures is described in Section \ref{Anisotropic phase field model of fracture}. Subsequently, the new phase field model for brittle compressive-shear fractures is proposed in Section \ref{Modified phase field model for compressive-shear fractures} followed by Section \ref{Numerical implementation} describing the numerical implementation of the proposed phase field model. Numerical examples validating the proposed phase field method are then presented in Section \ref{Numerical examples} before Section \ref{Conclusions} concludes the present work and outlook for future development.

\section{Anisotropic phase field model of fracture}\label{Anisotropic phase field model of fracture}

The basic theoretical aspects of the PFMs for fractures are described in this section. In fact, the PFMs were developed differently in the physics and mechanics communities. In the physics community, a majority of the phase field models are originated from the so-called Landau-Ginzburg phase transition \citep{ambati2015review} and commonly used for simulations of dynamic fractures \citep{aranson2000continuum, karma2001phase, henry2004dynamic}. On the other hand, the PFMs in the mechanics community are commonly regarded as the extension of Griffith's fracture theory and have gained more popularity in recent years \citep{miehe2010phase, miehe2010thermodynamically}. Therefore, as a well-accepted method, the anisotropic phase field model of \citet{miehe2010phase} in mechanics community is used to show the construction of PFM.

\subsection {Variational approach and regularization}\label{Variational approach and regularization}

\begin{figure}[htbp]
	%\centering
	\hspace{4cm}\subfigure[Strong discontinuity]{\includegraphics[height = 5cm]{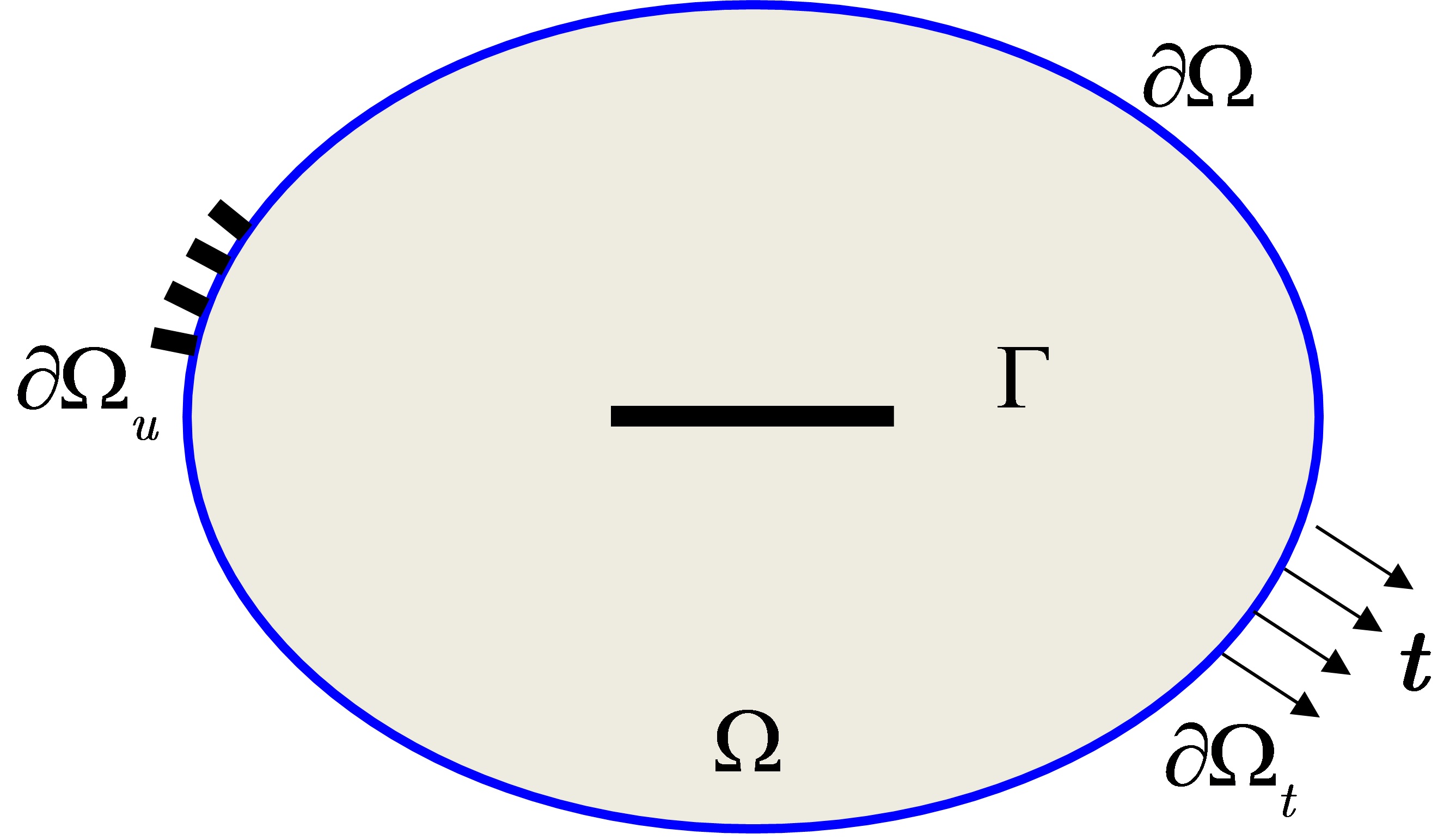}} \\

	\hspace{4cm}\subfigure[Phase field representation]{\includegraphics[height = 5cm]{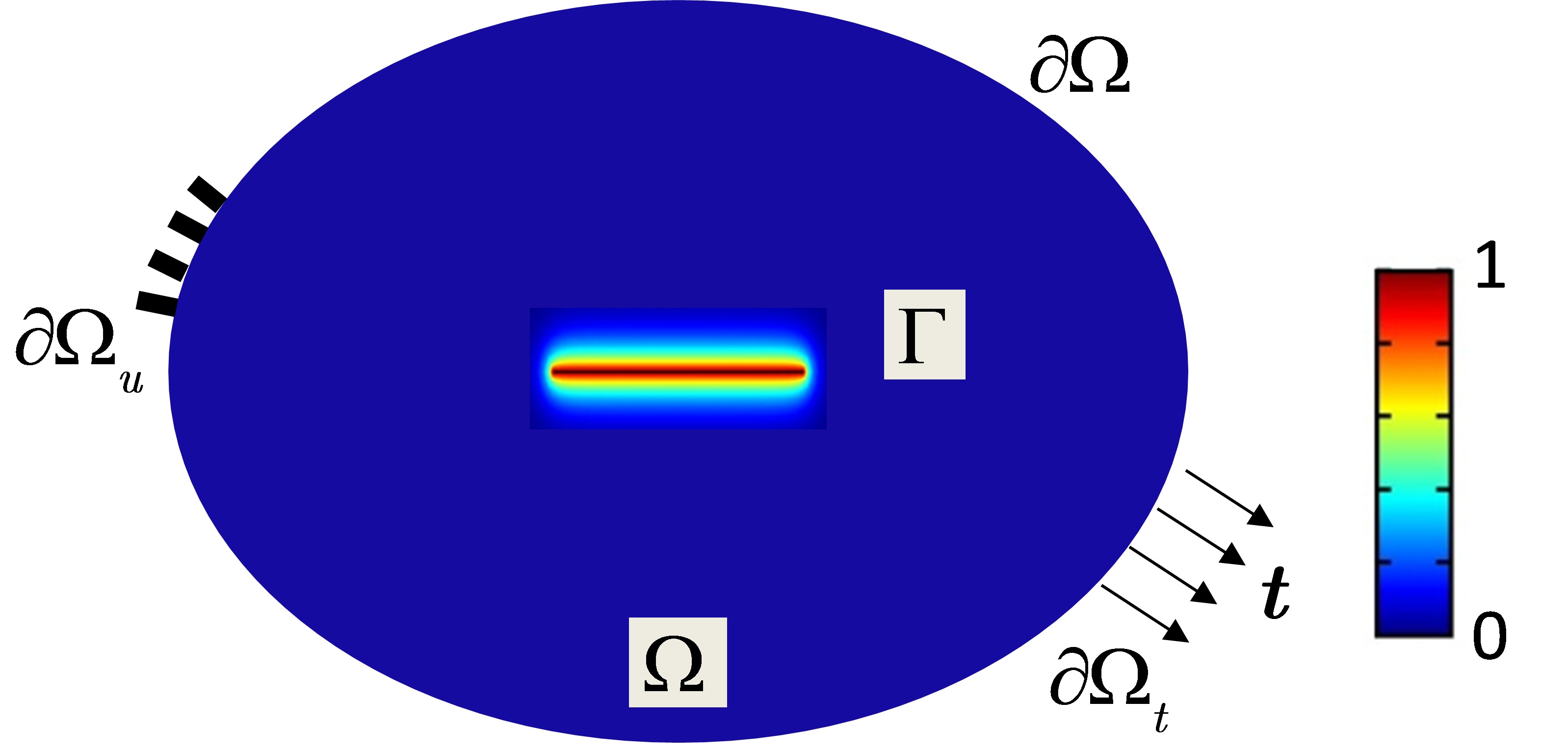}} \\
	\caption{Phase field representation of discrete fractures}
	\label{Phase field representation of discrete fractures}
	\end{figure}

Consider an elastic body $\Omega\subset \mathbb R^d$ ($d\in \{1,2,3\} $) shown in Fig. \ref {Phase field representation of discrete fractures}a; the external boundary and internal discontinuity set of the elastic body are denoted as $\partial \Omega$ and $\Gamma $, respectively. A phase field method is normally facilitated in implementation by coupling it to the variational approach, which requires that the fracture energy $\Psi_f$ can be transformed from the stored elastic energy $\Psi_{\varepsilon}$, and the fracture propagation is regarded as a process to minimize the energy functional $L$:

\begin{equation}
	L = \underbrace{\int_{\Omega}\psi_{\varepsilon}(\bm \varepsilon) \mathrm{d}{\Omega}}_{\Psi_{\varepsilon}}+\underbrace{\int_{\Gamma}G_c \mathrm{d}\Gamma}_{\Psi_f}\underbrace{-\int_{\Omega} \bm b\cdot{\bm u}\mathrm{d}{\Omega} - \int_{\partial\Omega_{t}} \bm t\cdot{\bm u}\mathrm{d}S}_{W_{ext}}
	\label{energy functional L1}
	\end{equation}

\noindent where $\bm u$ is the displacement field, $\psi_{\varepsilon}$ is the elastic energy density, $G_c$ is the critical energy release rate, and $W_{ext}$ denotes the external work with $\bm b$ being the body force and $\bm t$ the surface traction . In addition, the linear strain tensor $\bm\varepsilon = \bm\varepsilon(\bm u)$ is given by

	\begin{equation}
	\bm\varepsilon=\frac 1 2 \left[\nabla \bm u+(\nabla \bm u)^\mathrm{ T }\right]
	\end{equation}

It should be noted that a direct application of the variational approach to fracture is difficult in numerical simulations. A leading difficulty is how to deal with the discontinuous displacement filed at fracture sets and determine the optimal fracture set. Therefore, to overcome these numerical difficulties, \citet{bourdin2000numerical, miehe2010phase} regularized the variational model by using the following energy functional $L$:
\begin{equation}
	L = \int_{\Omega}\psi_{\varepsilon}(\bm \varepsilon, \phi) \mathrm{d}{\Omega}+\int_{\Omega}G_c\left(\frac{\phi^2}{2l_0}+\frac{l_0}2|\nabla\phi|^2 \right) \mathrm{d}\Omega -\int_{\Omega} \bm b\cdot{\bm u}\mathrm{d}{\Omega} - \int_{\partial\Omega_{t}} \bm t\cdot{\bm u}\mathrm{d}S
	\label{energy functional L2}
	\end{equation}

\noindent where $\phi(\bm x,t)\in[0,1]$ ($\bm x$ being a position vector) is an auxiliary field, namely the phase field that is used to smear out the crack surface over the domain $\Omega$ as shown in Fig. \ref{Phase field representation of discrete fractures}b. Correspondingly, in the regularized model, the phase field $\phi(\bm x,t)\in[0,1]$ satisfies the following conditions:

	\begin{equation}
	\phi = 
		\begin{cases}
		0,\hspace{1cm}\text{if material is intact}\\1,\hspace{1cm}\text{if material is fully cracked}
		\end{cases}
	\end{equation}

In addition, in Eq. \eqref{energy functional L2} $l_0$ denotes the length scale parameter which controls the transition region of the phase field. The length scale parameter thereby indirectly reflects the width of the crack. This feature has been proved by a number of numerical simulations \citep{zhou2018phase, zhou2018phase2, zhou2018phase3, zhou2018adaptive}, which show that the crack region has a larger width with an increasing $l_0$ while the phase field will represent a sharp crack when $l_0$ tends to zero. 

\subsection{Governing equations}

In a phase field method, the elastic energy can be split into different parts to model fractures of different mechanisms. For example, compressive fractures are inhibited in recent PFMs \citep{zhou2018phase, zhou2018phase3} and to capture fractures only under tension, \citet{miehe2010thermodynamically} split the elastic energy into tensile and compressive parts based on the spectral decomposition of strain:
	\begin{equation}
	\bm\varepsilon_{\pm}=\sum_{a=1}^d \langle\varepsilon_a\rangle_{\pm}\bm n_a\otimes\bm n_a
	\label{strain decomposition}
	\end{equation}

\noindent where $\bm\varepsilon_+$  and $\bm\varepsilon_-$  are the tensile and compressive strain tensors,  $\varepsilon_a$ is the principal strain, and  $\bm n_a$  is the direction of the principal strain. In addition, the operators $\langle\cdot\rangle_{\pm}$  are defined as: $\langle\cdot\rangle_{\pm}=(\cdot \pm |\cdot|)/ 2$. 

By using the strain decomposition, \citet{miehe2010thermodynamically} defined the elastic energy density in Eq. \eqref{energy functional L2} as
	\begin{equation}
	\psi_{\varepsilon}(\bm \varepsilon, \phi)=\left[(1-k)(1-\phi)^2+k\right]\psi_{\varepsilon}^+(\bm \varepsilon)+\psi_{\varepsilon}^-(\bm \varepsilon)
	\label{psi}
	\end{equation}

\noindent with the positive and negative elastic energy densities defined by
	\begin{equation}
	\psi_{\varepsilon}^{\pm}(\bm \varepsilon) = \frac{\lambda}{2}\langle \mathrm{tr}(\bm\varepsilon)\rangle_{\pm}^2+\mu \mathrm{tr} \left(\bm\varepsilon_{\pm}^2\right) 
	\end{equation}

\noindent where $\lambda>0$ and $\mu>0$ are the Lam\'e constants. In this study, the Lam\'e constants are obtained from the Young's modulus $E$ and Poisson's ratio $\nu$ of the material through the well-known relation:
	\begin{equation}
		  \left\{
	   \begin{aligned}
	\lambda&=\frac{E\nu}{(1+\nu)(1-2\nu)}
	\\ 	\mu&=\frac{E}{2(1+\nu)}
	   \end{aligned}\right.
	\label{relationship}
	\end{equation}

In addition, in Eq. \eqref{psi}, $0<k\ll1$ is a stability parameter for avoiding numerical singularities when the phase field $\phi$ tends to 1. Thus, by combining Eqs. \eqref{strain decomposition} to \eqref{relationship}, the regularized variational model of fracture in Eq. \ref{energy functional L2} can give rise to the governing equations with respect to the displacement $\bm u$ and the phase field $\phi$. Provided that the first variation of the functional $\delta L=0$ holds for all admissible displacement and phase fields, the governing equations of strong form read 
	\begin{equation}
	\left\{
	\begin{aligned}
	\nabla\cdot\bm\sigma+\bm b=\bm 0
	\\ \left[\frac{2l_0(1-k)\psi_{\varepsilon}^+}{G_c}+1\right]\phi-l_0^2\nabla^2\phi=\frac{2l_0(1-k)\psi_{\varepsilon}^+}{G_c}
	\end{aligned}\right.
	\label{governing equation1}
	\end{equation}

\noindent where $\bm\sigma$ is the Cauchy stress tensor calculated by
	\begin{equation}
		\begin{aligned}
		\bm\sigma&=\partial_{\bm\varepsilon} \psi_{\varepsilon}\\
		&=\underbrace{\left [(1-k)(1-\phi)^2+k \right]\left[\lambda \langle tr(\bm\varepsilon)\rangle_+ \bm I+ 2\mu \bm\varepsilon_+ \right]}_{\bm\sigma^+}+\underbrace{\lambda \langle tr(\bm\varepsilon)\rangle_- \bm I+ 2\mu \bm\varepsilon_-}_{{\bm \sigma}^{-}}\\
		\end{aligned}
	\end{equation}

\noindent with a unit tensor $\bm I$ $\in \mathbb R^{d\times d}$.

To avoid fracture healing, \citet{miehe2010phase,miehe2010thermodynamically} replaced the positive energy in Eq. \eqref{governing equation1} by a new driving force $H$. Note that $H=\mathrm{max}(\psi_\varepsilon^+)$ is the maximum positive reference energy and it ensures the crack irreversibility \citep{miehe2010phase,miehe2010thermodynamically}. Therefore, a monotonically increasing phase field can be ensured with the increase of $H$, and the governing equations of strong form are rewritten as 
	\begin{equation}
	\left\{
	\begin{aligned}
	\nabla\cdot\bm\sigma+\bm b= \bm 0
	\\ \left[\frac{2l_0(1-k)H}{G_c}+1\right]\phi-l_0^2\nabla^2\phi=\frac{2l_0(1-k)H}{G_c}
	\end{aligned}\right.
	\label{governing equation2}
	\end{equation}

\noindent subjected to the Dirichlet and Neumann boundary conditions
	\begin{equation}
	\left\{
	\begin{aligned}
	&\bm u = \bar {\bm u} \hspace{2cm} &\mathrm{on}\hspace{0.5cm} \partial\Omega_{u}\\
	&\bm\sigma\cdot\bm n=\bm t &\mathrm{on}\hspace{0.5cm} \partial\Omega_{t}\\
	&\nabla\phi\cdot\bm n = 0  &\mathrm{on}\hspace{0.5cm} \partial\Omega\\
	\end{aligned}
	\right.
	\label{boundary condition}
	\end{equation}

As can be seen from the equation sets in \eqref{governing equation2} and \eqref{boundary condition}, the original variational problem that involves strong discontinuities is now successfully transformed to a standard multi-field problem. For solving this type of problem, conventional finite element methods can be applied, which facilities the implementation of the regularized phase field method. In addition, because the elasticity tensor is calculated by $\mathbb{D} = {\partial\bm \sigma} / {\partial \bm \varepsilon} = {\partial \bm \sigma}^+ / {\partial \bm \varepsilon} +{\partial \bm \sigma}^{-} / {\partial \bm \varepsilon}$, an anisotropic constitutive relationship exists between the stress and strain and the approach in \citet{miehe2010phase} is hence called as an anisotropic phase field method.

\section{Modified phase field model for compressive-shear fractures}\label{Modified phase field model for compressive-shear fractures}

\subsection{Comparison on the current PFMs and their limitations}

In the former section, we use the anisotropic model of \citet{miehe2010phase, miehe2010thermodynamically} as a classical example to show the fundamental theoretical aspects in a phase field model on the basis of the variational principle. However, more generally, the PFMs for quasi-static fractures can be classified into three basic types under the framework of a regularized formulation; expect the anisotropic methods such as the model of \citet{miehe2010phase}, the isotropic and hybrid (isotropic-anisotropic) phase field methods \citep{ambati2015review, zhang2017modification, wu2017unified} are also developed according to the energy references and constitutive models used. Some representative PFMs of the isotropic, anisotropic, and hybrid forms are listed in Table \ref{Different forms of phase field methods}. Note that the isotropic and anisotropic formulations are consistent with the variational principle while the hybrid ones lose fairly some thermodynamic consistency.

\begin{figure}[htbp]
	\centering
	\subfigure[Compressive-shear fracture]{\includegraphics[height = 3.5cm]{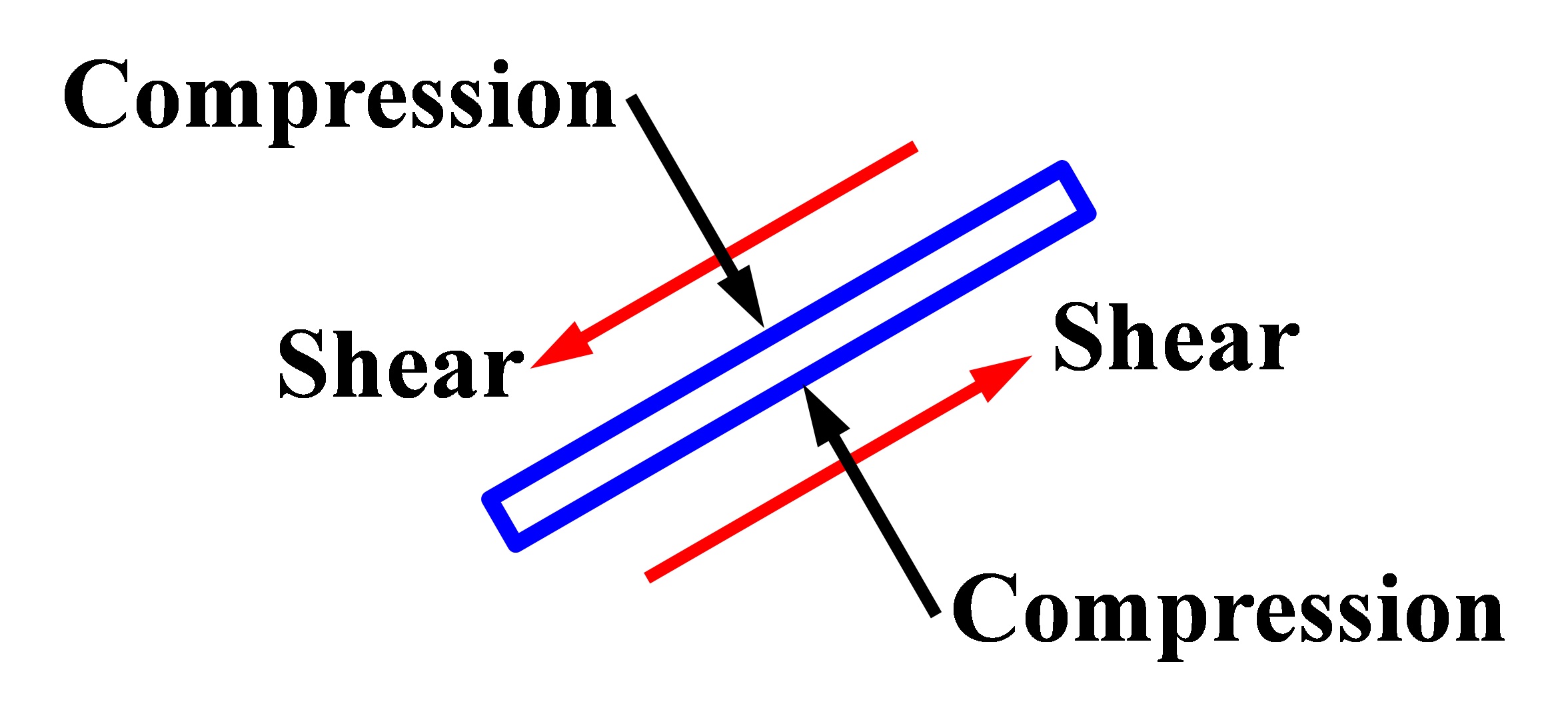}}
	\subfigure[Tension-shear mode]{\includegraphics[height = 3.5cm]{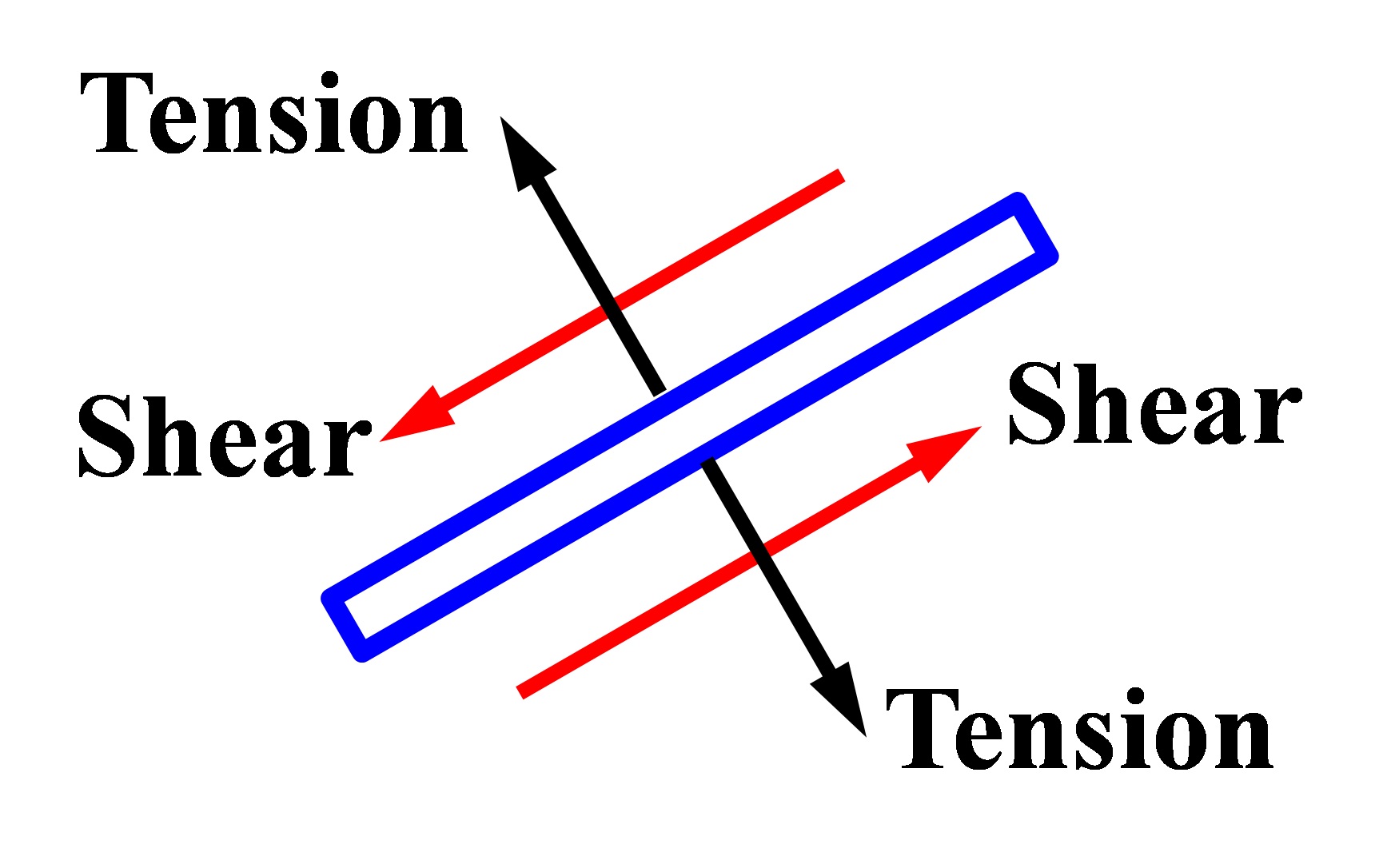}}\\
	\subfigure[Shear mode in \citet{zhou2018phase}]{\includegraphics[height = 6cm]{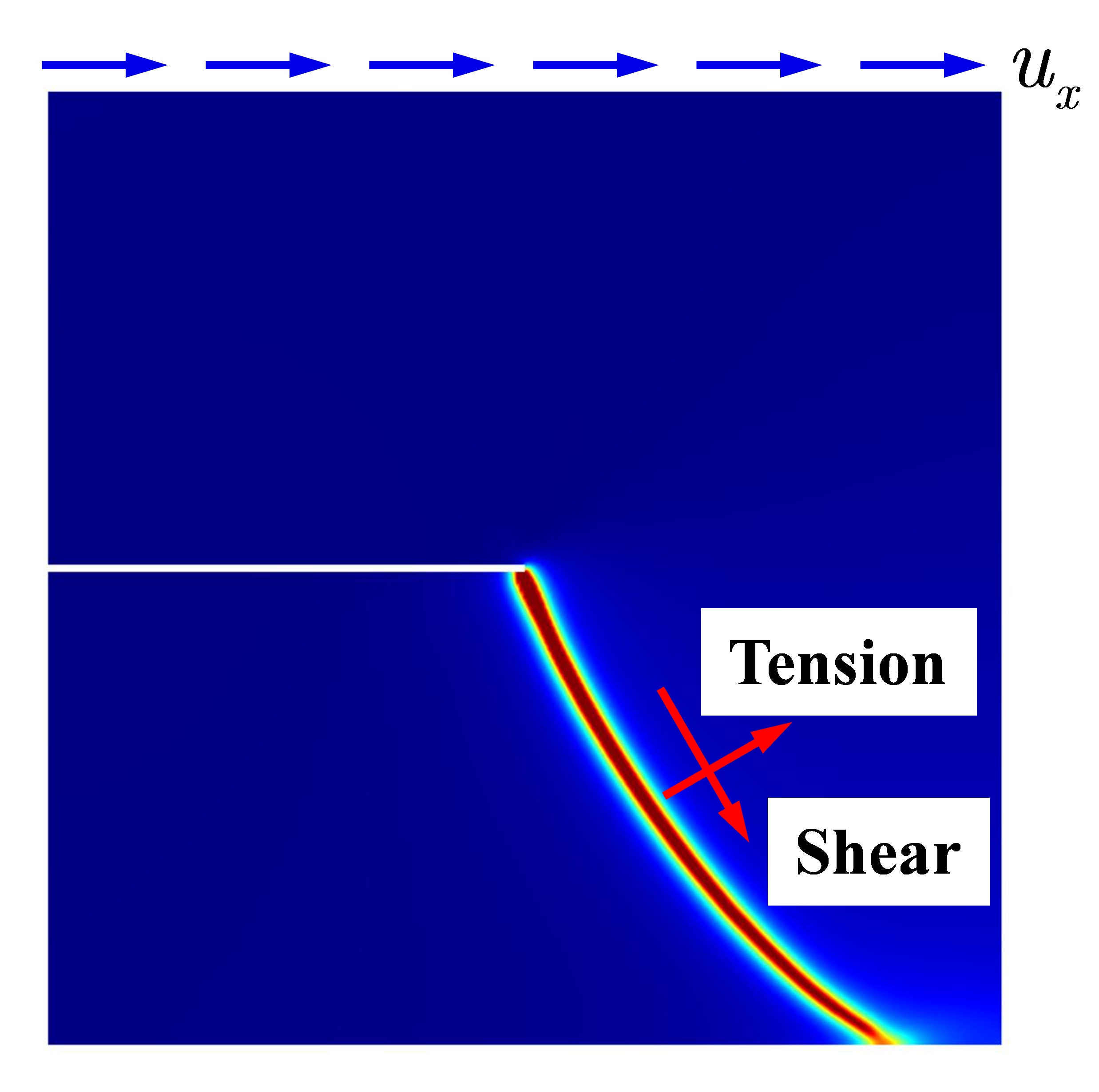}}\\
	\caption{Different shear modes}
	\label{Different shear modes}
	\end{figure}

The three types of PFMs can be applied to capture tensile fractures in rocks, which is indicated by the formulations in Table \ref{Different forms of phase field methods}. However, these PFMs have difficulties in modeling shear fractures, especially those compressive-shear fractures that are commonly observed in rock-like solids. This can be also verified through Table \ref{Different forms of phase field methods} and Fig. \ref{Different shear modes}, which shows different forms of shear fractures for better comparison. In the first place, Table \ref{Different forms of phase field methods} indicates that the typical anisotropic and hybrid phase field methods cannot model the compressive-shear fractures because the compressive part of the elastic energy is not included in the driving force for the evolution equation of phase field. Therefore, only some fractures under tension-shear mode (Fig. \ref{Different shear modes}b and c) can be simulated by using the anisotropic and hybrid PFMs. In the second place, although the isotropic formulations can model fractures under compression, these fractures are reported to be unrealistic \citep{miehe2010phase, miehe2010thermodynamically}. Furthermore, cohesion and internal friction angle are not included in the formerly developed PFMs; this is inconsistent with the experimental observations because it is generally accepted that for geomaterials the internal friction angle and cohesion have a significant effect on the compressive-shear fracture propagation and load-displacement curves \citep{zhou2018theoretical}. Hence, a phase field model for simulating fractures in geomaterials should be able to account for the compressive-shear fractures and also to include the effects of cohesion and internal friction angle. Hence, in the following sections we will describe a new phase field model to simulate compressive-shear fractures in rock-like materials by introducing a new driving force $H_p$ in the evolution equation of phase field.

	\begin{table}[htbp]
	\caption{Different forms of phase field methods}
	\footnotesize
	\label{Different forms of phase field methods}
	\centering
	%\begin{tabular}{p{20pt}p{20pt}p{200pt}p{200pt}}%|p{2cm}<{\centering}|
	\begin{tabular}{m{3cm}<{\centering} m{8cm}<{\centering} m{5cm}<{\centering}}
	%\hline
	\toprule
	Type &  Governing equations &  Definition\\
	%\hline
	\midrule
	Isotropic &\begin{equation*}
\left\{\begin{aligned}
&\nabla\cdot\bm\sigma+\bm b =0\\
&-l_0^2\Delta \phi+\phi=\frac {2l_0}{G_c}(1-\phi)H\\
&\bm\sigma(\bm u,\phi)=(1-\phi)^2\frac {\partial\psi_0}{\partial\bm\varepsilon}\\
&\mathbb{D}=\frac{\partial\bm\sigma}{\partial\bm\varepsilon}\\
\end{aligned}\right.
\end{equation*}&  $H = \max \limits_{\tau\in[0,t]}\psi_0\left(\bm\varepsilon(\bm x,\tau)\right)$ with $\psi_0=\frac \lambda 2 (\mathrm{tr}(\bm\varepsilon))^2+\mu \mathrm{tr}(\bm\varepsilon^2)$\\
	Anisotropic \citep{miehe2010phase} &\begin{equation*}
\left\{\begin{aligned}
&\nabla\cdot\bm\sigma+\bm b =0\\
&\left(\frac{2l_0 H}{G_c}+1\right)\phi-l_0^2\nabla^2\phi=\frac{2l_0 H}{G_c}\\
&\bm\sigma(\bm u,\phi)=(1-\phi)^2\frac {\partial\psi_\varepsilon^+}{\partial\bm\varepsilon}+\frac {\partial\psi_\varepsilon^-}{\partial\bm\varepsilon}\\
&\mathbb{D}=\frac{\partial\bm\sigma^+}{\partial\bm\varepsilon}+\frac{\partial\bm\sigma^-}{\partial\bm\varepsilon}\\
\end{aligned}\right.
\end{equation*}& $H = \max \limits_{\tau\in[0,t]}\psi_\varepsilon^+\left(\bm\varepsilon(\bm x,\tau)\right)$, $\bm\sigma^+=(1-\phi)^2\frac {\partial\psi_\varepsilon^+}{\partial\bm\varepsilon}$, $\bm\sigma^-=\frac {\partial\psi_\varepsilon^-}{\partial\bm\varepsilon}$, $\psi_{\varepsilon}^{\pm}(\bm \varepsilon) = \frac{\lambda}{2}\langle \mathrm{tr}(\bm\varepsilon)\rangle_{\pm}^2+\mu \mathrm{tr} \left(\bm\varepsilon_{\pm}^2\right)$\\
	Anisotropic \citep{amor2009regularized} &\begin{equation*}
\left\{\begin{aligned}
&\nabla\cdot\bm\sigma+\bm b =0\\
&-l_0^2\Delta \phi+\phi=\frac {2l_0}{G_c}(1-\phi)H\\
&\bm\sigma(\bm u,\phi)=(1-\phi)^2\frac {\partial\psi_0^+}{\partial\bm\varepsilon}+\frac {\partial\psi_0^-}{\partial\bm\varepsilon}\\
&\mathbb{D}=\frac{\partial\bm\sigma^+}{\partial\bm\varepsilon}+\frac{\partial\bm\sigma^-}{\partial\bm\varepsilon}\\
\end{aligned}\right.
\end{equation*}& $H = \max \limits_{\tau\in[0,t]}\psi_0^+\left(\bm\varepsilon(\bm x,\tau)\right)$, $\bm\sigma^+=(1-\phi)^2\frac {\partial\psi_0^+}{\partial\bm\varepsilon}$, $\bm\sigma^-=\frac {\partial\psi_0^-}{\partial\bm\varepsilon}$, $\psi_0^+=\frac 1 2 K_n\langle(\mathrm{tr}(\bm\varepsilon)\rangle_+^2+\mu(\bm\varepsilon^{\mathrm{dev}}:\bm\varepsilon^{\mathrm{dev}})$,  $\psi_0^-=\frac 1 2 K_n\langle(\mathrm{tr}(\bm\varepsilon)\rangle_-^2$ with $K_n$ a physical parameter\\
	Hybrid \citep{ambati2015review} &\begin{equation*}
\left\{\begin{aligned}
&\nabla\cdot\bm\sigma+\bm b =0\\
&-l_0^2\Delta \phi+\phi=\frac {2l_0}{G_c}(1-\phi)H\\
&\bm\sigma(\bm u,\phi)=(1-\phi)^2\frac {\partial\psi_0}{\partial\bm\varepsilon}\\
&\mathbb{D}=\frac{\partial\bm\sigma}{\partial\bm\varepsilon}\\
&\forall \bm x:\hspace{0.1cm}\Psi_0^+<\Psi_0^-\rightarrow\phi=0\\
\end{aligned}\right.
\end{equation*}& $H = \max \limits_{\tau\in[0,t]}\psi_\varepsilon^+\left(\bm\varepsilon(\bm x,\tau)\right)$, $\psi_0=\frac \lambda 2 (\mathrm{tr}(\bm\varepsilon))^2+\mu \mathrm{tr}(\bm\varepsilon^2)$, $\psi_{\varepsilon}^{+}(\bm \varepsilon) = \frac{\lambda}{2}\langle \mathrm{tr}(\bm\varepsilon)\rangle_{+}^2+\mu \mathrm{tr} \left(\bm\varepsilon_{+}^2\right)$\\
	Hybrid \citep{zhang2017modification} &\begin{equation*}
\left\{\begin{aligned}
&\nabla\cdot\bm\sigma+\bm b =0\\
&(1-\phi)\left(\frac {H_1}{G_{cI}}+\frac {H_2}{G_{cII}}\right)-\frac 1 {2l_0}\phi + \frac {l_0} 2 \Delta\phi =0\\
&\bm\sigma(\bm u,\phi)=(1-\phi)^2\frac {\partial\psi_0}{\partial\bm\varepsilon}\\
&\mathbb{D}=\frac{\partial\bm\sigma}{\partial\bm\varepsilon}\\
\end{aligned}\right.
\end{equation*}& $H_1 = \max \limits_{\tau\in[0,t]}\lambda\langle\mathrm{tr}(\bm\varepsilon)\rangle_+^2$, $H_2 = \max \limits_{\tau\in[0,t]}\mu\mathrm{tr}[\langle\bm\varepsilon\rangle_+^2]$, $G_{cI}$ and $G_{cII}$ are the critical energy release rate of modes I and II\\
	%\hline
	\bottomrule
	\end{tabular}
\flushleft{\scriptsize *Note: in this table the stability parameter $k$ is ignored for simplicity because it can be introduced by different forms in different models.} 
	\end{table}

\subsection{Modified phase field model for compressive-shear fractures in rock-like solids}

For rock-like materials, one way to include the effects of cohesion and internal friction angle is to apply the well-known Mohr-Coulomb criterion. However, this criterion is more suitable for macroscopic compressive-shear failure and it is a stress based failure criterion. In linear elastic fracture mechanics, the stress at crack tip is singular and is prone to exceed the strength defined by the Mohr-Coulomb criterion even when the external loads are quite small. Therefore, the macroscopically suitable Mohr-Coulomb criterion cannot be directly applied to model fracture initiation and propagation. It is also not applicable to the phase field model, which is simply an energy based method. However, many improved Mohr-Coulomb type formulations have been developed and applied in prediction of strength and damage initiation. Among these improvements are the stress level used in the statistical damage model for rocks \citep{li2012statistical, zhou2017statistical}, and the energy based driving force to damage $w_s$ \citep{li2017damage}:
	\begin{equation}
	w_s=\sum_{i=j+1}^{3}\sum_{j=1}^{2}\frac 1 {2G} \left(\frac {\sigma_i-\sigma_j}{2\mathrm{cos}\varphi_{ij}}+\frac{\sigma_i+\sigma_j}{2}\mathrm{tan}{\varphi_{ij}} \right)^2
	\label{li2017damagemodel}
	\end{equation}

\noindent where $G=\mu$ is the shear modulus, $\sigma_i$ and $\sigma_j$ are principal stresses, $\varphi_{ij}$ is the calculated angle from the Mohr diagram, and $\varphi_{13}$ is the internal friction angle. Obviously, the driving force proposed by \citet{li2017damage} only includes the internal friction angle while the effect of cohesion is neglected.

Given that the phase field method to fracture is an energy-driven approach, we reconstruct a new driving energy $\psi_p$ based on Eq. \eqref{li2017damagemodel}. For the purpose of simplicity and preventing unrealistic fractures from small shear stress, the cohesion in rock-like solids and a positive operator $\langle\centerdot\rangle_{+}$ are introduced to Eq. \eqref{li2017damagemodel}, and the energy $\psi_p$ is thereby written as
	\begin{equation}
	\psi_p=\sum_{i=j+1}^{3}\sum_{j=1}^{2}\frac 1 {2G} \left\langle\frac {\sigma_{i0}-\sigma_{j0}}{2\mathrm{cos}\varphi}+\frac{\sigma_{i0}+\sigma_{j0}}{2}\mathrm{tan}{\varphi}-c \right\rangle_+^2
	\label{new energy1}
	\end{equation}

\noindent where $c$ is the cohesion, $\varphi$ is the internal friction angle, and $\sigma_{i0}$ and $\sigma_{j0}$ are the principal stresses without including the phase field. In this study, we focus on the compressive-shear fractures. That is, the strain decomposition of \citet{miehe2010phase} is still applicable but only the compressive part of the principal strain is used to construct the energy in Eq. \eqref{new energy1}. Therefore, the principal stresses $\sigma_{i0}$ and $\sigma_{j0}$ are then determined by the following equation:
	\begin{equation}
	\left\{\begin{aligned}
	&\sigma_{i0}= \lambda\left(\varepsilon_{1p}+\varepsilon_{2p}+\varepsilon_{3p}\right)+2\mu\varepsilon_{ip}\\ &\sigma_{j0}= \lambda\left(\varepsilon_{1p}+\varepsilon_{2p}+\varepsilon_{3p}\right)+2\mu\varepsilon_{jp}
	\end{aligned}
	\right.
	\label{principals stress}
	\end{equation}

\noindent where $\varepsilon_{ip} = \langle\varepsilon_i\rangle_-$ and $\varepsilon_{jp} = \langle\varepsilon_j\rangle_-$.

By substituting Eq. \eqref{principals stress} into Eq. \eqref{new energy1}, the energy $\psi_p$ is rewritten as 
	\begin{equation}
	\psi_p=\sum_{i=j+1}^{3}\sum_{j=1}^{2}\frac 1 {2G} \left\langle \frac {\mu(\varepsilon_{ip}-\varepsilon_{jp})}{\mathrm{cos}\varphi}+\left[\lambda(\varepsilon_{1p}+\varepsilon_{2p}+\varepsilon_{3p})+\mu(\varepsilon_{ip}+\varepsilon_{jp})\right]\mathrm{tan}{\varphi}-c \right\rangle_+^2
	\label{new energy2}
	\end{equation}

To ensure the fracture irreversibility, the new driving force is defined as $H_p=\max \limits_{\tau\in[0,t]}\psi_p$. We then replace the driving force $H$ in \citet{ambati2015review} by the new driving force $H_p$, and the governing equation set of the hybrid phase field model for simulating compressive-shear fractures in rock-like materials are thereby denoted as follows:
	\begin{equation}
	\left\{\begin{aligned}
	&\nabla\cdot\bm\sigma+\bm b =0\\
	&\left[\frac{2l_0(1-k)H_p}{G_c}+1\right]\phi-l_0^2\nabla^2\phi=\frac{2l_0(1-k)H_p}{G_c}\\
	&\bm\sigma(\bm u,\phi)=\left[(1-k)(1-\phi)^2+k\right]\frac {\partial\psi_0}{\partial\bm\varepsilon}\\
	&\mathbb{D}=\frac{\partial\bm\sigma}{\partial\bm\varepsilon}\\
	\end{aligned}\right.
	\label{phase field model of compressive-shear fractures}
	\end{equation}

\noindent in which $G_c$ accordingly refers to the critical energy release rate of mode II. It should be noted that the proposed hybrid phase field model for compressive-shear fractures can account for the influence of internal friction angle and cohesion on initiation and propagation of fractures as well as the load-displacement curves of rock-like specimens. This important feature will be further verified in the numerical examples presented in Section \ref{Numerical examples}.

\section{Numerical implementation}\label{Numerical implementation}

Most of the PFMs are solved within the framework of conventional finite element methods (FEMs), and there are two different approaches at moment for coupling the displacement or other physical field with the phase field, namely the monolithic \citep{miehe2010thermodynamically, liu2016abaqus} and staggered ones \citep{kuhn2008phase, amor2009regularized, miehe2010phase,  bourdin2012variational, hesch2014thermodynamically, liu2016abaqus, wu2017unified, zhou2018phase, zhou2018phase3}. In this study, the weak forms of the governing equations \eqref{phase field model of compressive-shear fractures} are first given by 
	\begin{equation}
	\int_{\Omega}\bm\sigma:\delta \bm {\varepsilon} \mathrm{d}\Omega = \int_{\Omega}\bm b \cdot \delta \bm u  \mathrm{d}\Omega +\int_{\Omega_{h}}\bm t \cdot \delta \bm u  \mathrm{d}S
	\label{weak form 1}
	\end{equation}

\noindent and
	\begin{equation}
	\int_{\Omega}-2(1-k)H_p(1-\phi)\delta\phi\mathrm{d}\Omega+\int_{\Omega}G_c\left(l_0\nabla\phi\cdot\nabla\delta\phi+\frac{1}{l_0}\phi\delta\phi\right)\mathrm{d}\Omega=0
	\label{weak form 2}
	\end{equation}

We use the standard vector-matrix notation and the displacement and phase field are discretized as
	\begin{equation}
	\bm u = \mathbf {N_u} \mathbf d,\hspace{0.5cm} \phi = \mathbf{\bm N_{\phi}} \hat{\bm\phi}
	\end{equation}

\noindent where $\mathbf d$ and $\hat{\bm\phi}$ are the constructed vectors comprising the node values $\bm u_i$ and $\phi_i$. In addition, the shape function matrices $\mathbf {N_u}$ and $\mathbf {N_{\phi}}$ in 2D are defined as
	\begin{equation}		
			\mathbf {N_u} = \left[ \begin{array}{ccccc}
			N_{1}&0&\dots&N_{n}&0\\
			0&N_{1}&\dots&0&N_{n}
			\end{array}\right], \hspace{0.5cm}
			\mathbf {N_\phi} = \left[ \begin{array}{cccc}
			N_{1}&N_{2}&\dots&N_{n}
			\end{array}\right]
	\end{equation}

\noindent where $n$ is the node number in one element and $N_i$ is the shape function at node $i$. The test functions are then assumed to have the same discretization:
	\begin{equation}
	\delta \bm u = \mathbf {N_u} \delta \mathbf d,\hspace{0.5cm} \delta \phi = \mathbf {N_{\phi}} \delta \hat{\bm\phi}
	\end{equation}

\noindent where $\delta \mathbf d$ and $\delta \hat{\bm\phi}$ are the vectors composed of the node values of the test functions.

The gradients of the trial and test functions are defined as follows,
	\begin{equation}
	\bm \varepsilon =  \mathbf {B_u} \mathbf d,\hspace{0.5cm} \nabla\phi = \mathbf {B_\phi} \hat{\bm\phi}, \hspace{0.5cm}\bm \delta \varepsilon =  \mathbf {B_u} \delta \mathbf d,\hspace{0.5cm} \nabla\phi = \mathbf{B_\phi}  \delta \hat{\bm\phi}
	\end{equation}

\noindent where $\mathbf {B_u}$ and $\mathbf {B_\phi}$ are the derivatives of the shape functions defined by
	\begin{equation}
	\mathbf {B_u}=\left[
		\begin{array}{ccccc}
		N_{1,x}&0&\dots&N_{n,x}&0\\
		0&N_{1,y}&\dots&0&N_{n,y}\\
		N_{1,y}&N_{1,x}&\dots&N_{n,y}&N_{n,x}
		\end{array}\right],\hspace{0.2cm}
		\mathbf {B_\phi}=\left[
		\begin{array}{cccc}
		N_{1,x}&N_{2,x}&\dots&N_{n,x}\\
		N_{1,y}&N_{2,y}&\dots&N_{n,y}
		\end{array}\right]
		\label{BiBu}
	\end{equation}

The equations of weak form \eqref{weak form 1} and \eqref{weak form 2} are subsequently written as
	\begin{equation}
	(\delta\mathbf d)^\mathrm{T} \left[\int_{\Omega} \mathbf {B_u}^\mathrm{T} \mathbf {D_e} \mathbf {B_u} \mathrm{d}\Omega \mathbf d \right]= 
	(\delta\mathbf d)^\mathrm{T} \left[\int_{\Omega}\mathbf {N_u}^\mathrm{T}\bm b \mathrm{d}\Omega+\int_{\Omega_{h}} \mathbf {N_u}^\mathrm{T} \bm t\mathrm{d}S \right]
	\label{discrete equation 1}
	\end{equation}
	\begin{equation}
	-(\delta\hat{\bm \phi})^{\mathrm{T}} \int_{\Omega}\left\{\mathbf {B_\phi}^{\mathrm{T}} G_c l_0 \mathbf {B_\phi} +\mathbf {N_\phi}^{\mathrm{T}} \left [ \frac{G_c}{l_0} + 2(1-k)H_p \right ]  \mathbf {N_\phi} \right \} \mathrm{d}\Omega \hat{\bm \phi}+ (\delta\hat{\bm \phi})^\mathrm{T} \int_{\Omega}2(1-k)H_p\mathbf {N_\phi}^{\mathrm{T}} \mathrm{d}\Omega = 0
	\label{discrete equation 2}      
	\end{equation}

Equations \eqref{discrete equation 1} and \eqref{discrete equation 2} always hold for arbitrary admissible test functions, therefore, the discrete weak form equations read
	\begin{equation}
	-\underbrace{\int_{\Omega} \mathbf {B_u}^\mathrm{T} \mathbf {D_e} \mathbf {B_u} \mathrm{d}\Omega \mathbf d}_{\bm F_u^{int}=\bm K_u \mathbf d} + \underbrace{\int_{\Omega}\mathbf {N_u}^\mathrm{T}\bm b \mathrm{d}\Omega+\int_{\Omega_{h}} \mathbf {N_u}^\mathrm{T} \bm t\mathrm{d}S}_{\bm F_u^{ext}}=0
\label{discrete equation 3}
	\end{equation}
	\begin{equation}
	-\underbrace{ \int_{\Omega}\left\{\mathbf {B_\phi}^{\mathrm{T}} G_c l_0 \mathbf {B_\phi} +\mathbf {N_\phi}^{\mathrm{T}} \left [ \frac{G_c}{l_0} + 2(1-k)H_p \right ] \mathbf {N_\phi} \right \} \mathrm{d}\Omega \hat{\bm \phi}}_{\bm F_\phi^{int}=\bm K_\phi \hat{\bm \phi}}+ \underbrace{\int_{\Omega}2(1-k)H_p\mathbf {N_\phi}^{\mathrm{T}} \mathrm{d}\Omega}_{\bm F_\phi^{ext}} = 0 
\label{discrete equation 4}
	\end{equation}

\noindent where $\bm F_u^{int}$ and $\bm F_u^{ext}$ are the internal and external forces for the displacement field, and $\mathbf {D_e}$ is the degraded elasticity matrix. $\bm F_\phi^{int}$ and $\bm F_\phi^{ext}$ are the internal and external force terms of the phase field \citep{zhou2018phase}. In addition, the stiffness matrices of the displacement and phase fields read
	\begin{equation}
	\left\{\begin{aligned}\bm K_u &= \int_{\Omega} \mathbf {B_u}^\mathrm{T} \mathbf {D_e} \mathbf {B_u} \mathrm{d}\Omega\\
	\bm K_\phi &= \int_{\Omega}\left\{\mathbf {B_\phi}^{\mathrm{T}} G_c l_0 \mathbf {B_\phi} +\mathbf {N_\phi}^{\mathrm{T}} \left [ \frac{G_c}{l_0} + 2(1-k)H_p \right ] \mathbf {N_\phi} \right \} \mathrm{d}\Omega
	\end{aligned}\right .
	\end{equation}

Finally, Eqs. \eqref{discrete equation 3} and \eqref{discrete equation 4} are solved by using a staggered scheme because it provides greater flexibility as well as stability in solving the displacement and phase field compared with the monolithic scheme. The phase field and displacement are thereby solved sequentially with iterations in between. More details about the widely used staggered scheme for phase field modeling can be referred to \citet{miehe2010phase, miehe2010thermodynamically}. We use the implicit Generalized-$\alpha$ method for time integration and the Anderson acceleration technology for increasing the convergence rate in the simulations  \citep{zhou2018phase, zhou2018phase2, zhou2018phase3, zhou2018propagation}.

\section{Numerical examples}\label{Numerical examples}

In this section, some benchmark examples are tested to show the capability of the proposed PFM for brittle compressive-shear fractures in rock-like materials. These examples include uniaxial compression tests on intact rock-like specimen as well as those specimens with a single or double inclined pre-existing flaws. 

\subsection{An intact specimen under uniaxial compression}

In this example, we test the compressive-shear fractures in an intact prism specimen subjected to axial compression, and 2D plane strain and 3D simulations are performed. Geometry and boundary conditions of the specimen are shown in Fig. \ref{An intact rectangular specimen subjected to uniaxial compression. (a) Geometry and boundary conditions, (b) distribution of Young's modulus in 2D (Unit: Pa), and (c) distribution of Young's modulus in 3D (Unit: Pa)}a. We apply vertical displacement on the upper end of the specimen. To reduce complexity of this numerical test, we only introduce a random distribution on the Young's modulus of the specimen for the purpose of inducing a non-uniform stress field, which will further drives the fracture propagation in part of the specimen without producing everywhere identical phase field in the specimen. In this example, the Young's modulus $E(\bm x)$ is assumed to follow Weibull distribution \citep{zhou2017statistical, zhou2018longterm} and the corresponding probability density function $p$ is expressed as follows,
	\begin{equation}
	p(E)=\frac{m}{E_0}\left(\frac{E}{E_0}\right)^{m-1}\mathrm{exp}\left[-\left(\frac{E}{E_0}\right)^m\right]
	\end{equation}

\noindent where $E_0$ is the mean value of the Young's modulus and $m$ is a parameter that defines the shape of Weibull distribution (see also different density curves about Weibull distribution under different $m$ in \citet{zhou2017statistical}).

The resulting Young's modulus distribution is shown in Fig. \ref{An intact rectangular specimen subjected to uniaxial compression. (a) Geometry and boundary conditions, (b) distribution of Young's modulus in 2D (Unit: Pa), and (c) distribution of Young's modulus in 3D (Unit: Pa)}b and c. The mean value of $E$ is 90 GPa for both 2D and 3D while the maximum and minimum values are 903 GPa and 0.45 MPa in 2D, and the maximum and minimum values for 3D are 883 GPa and 0.431 MPa, respectively. The other parameters are chosen as $\nu=0.3$, $G_c = 6$ N/m, $m=1$, $l_0=1$ mm, $\varphi=20^\circ$, and $c=100$ kPa; these parameters are used for both 2D and 3D calculation for an intact rock specimen and are obtained from the experimental results (stress-strain curve of the first sample) in \citet{xia2015strength}. Note that we apply the initial Young's modulus of the rock specimen in the actual experimental test in the phase field modeling. In 2D simulation, a total of 16,000 regular Q4 elements with characteristic size of $h=0.5$ mm are used to discretize the specimen while 6-node prism elements with  maximum element size $h=1$ mm are applied for 3D and $k = 1 \times 10^{-9}$ is selected to prevent numerical singularity. In each time step, we set the displacement increment as $\Delta u=2\times10^{-6}$ mm. 

	\begin{figure}[htbp]
	\centering
	\subfigure[]{\includegraphics[height = 6cm]{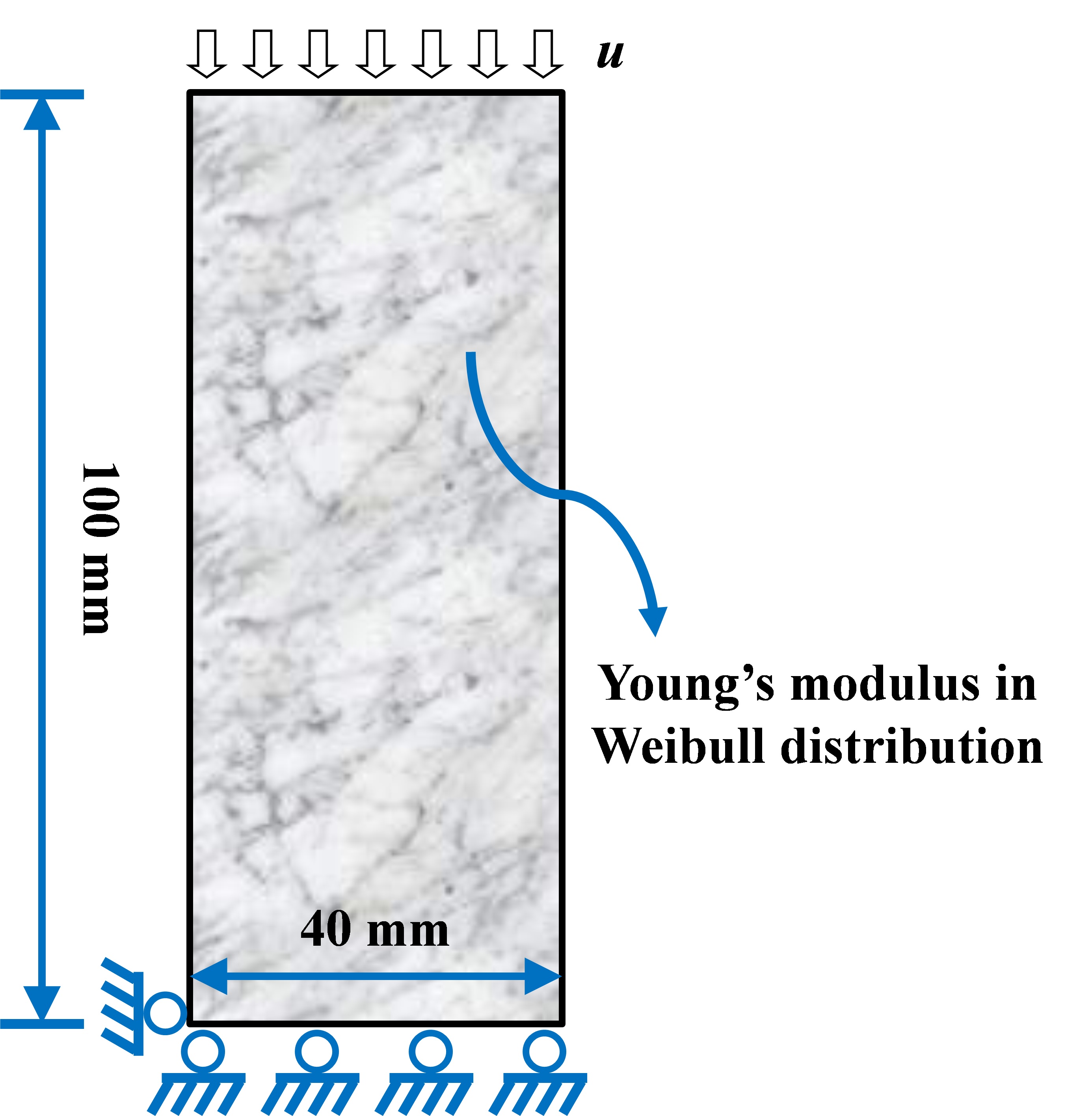}}
	\subfigure[]{\includegraphics[height = 6cm]{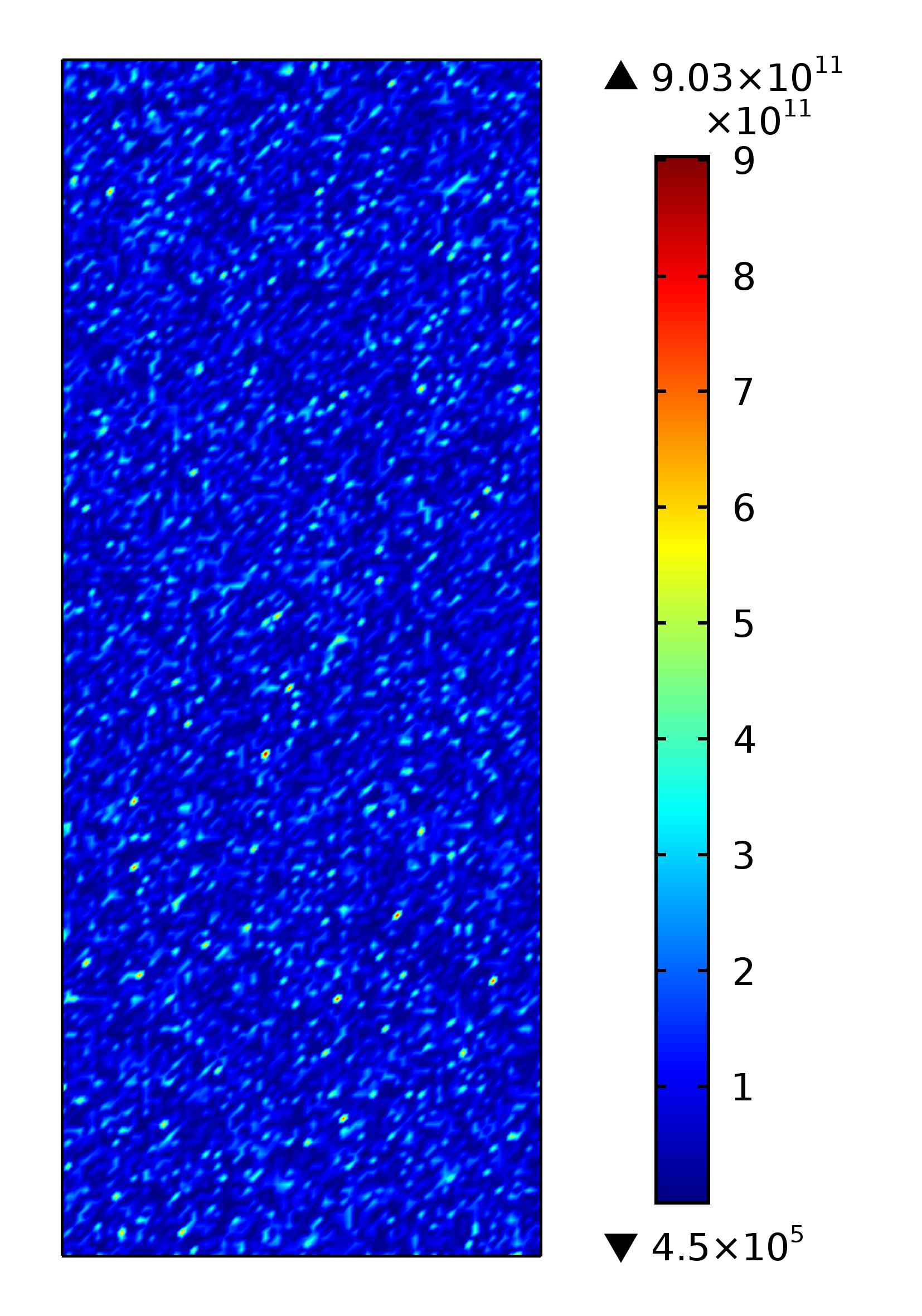}}
	\subfigure[]{\includegraphics[height = 6cm]{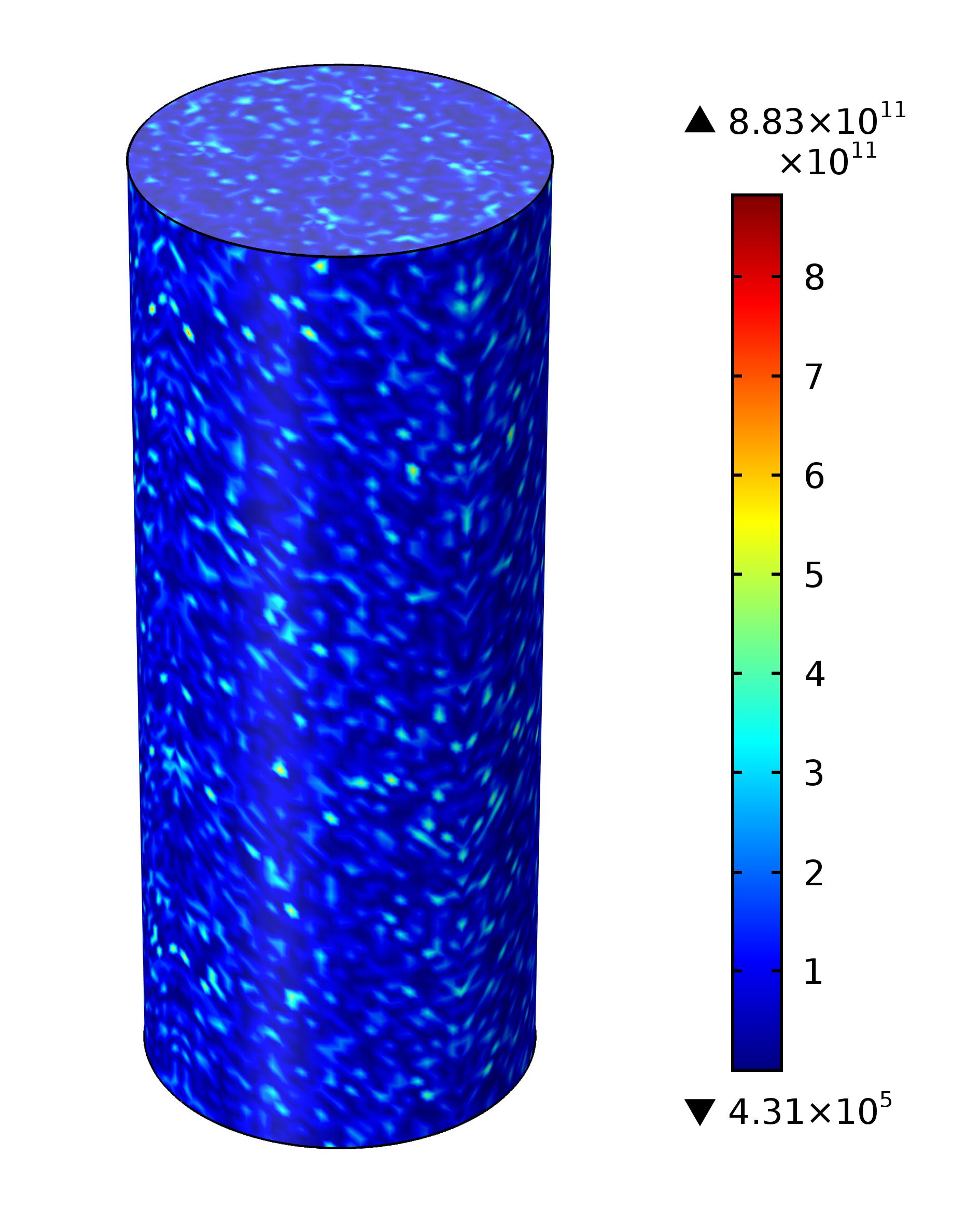}}\\
	\caption{An intact rectangular specimen subjected to uniaxial compression. (a) Geometry and boundary conditions, (b) distribution of Young's modulus in 2D (Unit: Pa), and (c) distribution of Young's modulus in 3D (Unit: Pa)}
	\label{An intact rectangular specimen subjected to uniaxial compression. (a) Geometry and boundary conditions, (b) distribution of Young's modulus in 2D (Unit: Pa), and (c) distribution of Young's modulus in 3D (Unit: Pa)}
	\end{figure}

\begin{figure}[htbp]
	\centering
	\subfigure[$u=0.92$ mm, 2D]{\includegraphics[height = 6.5cm]{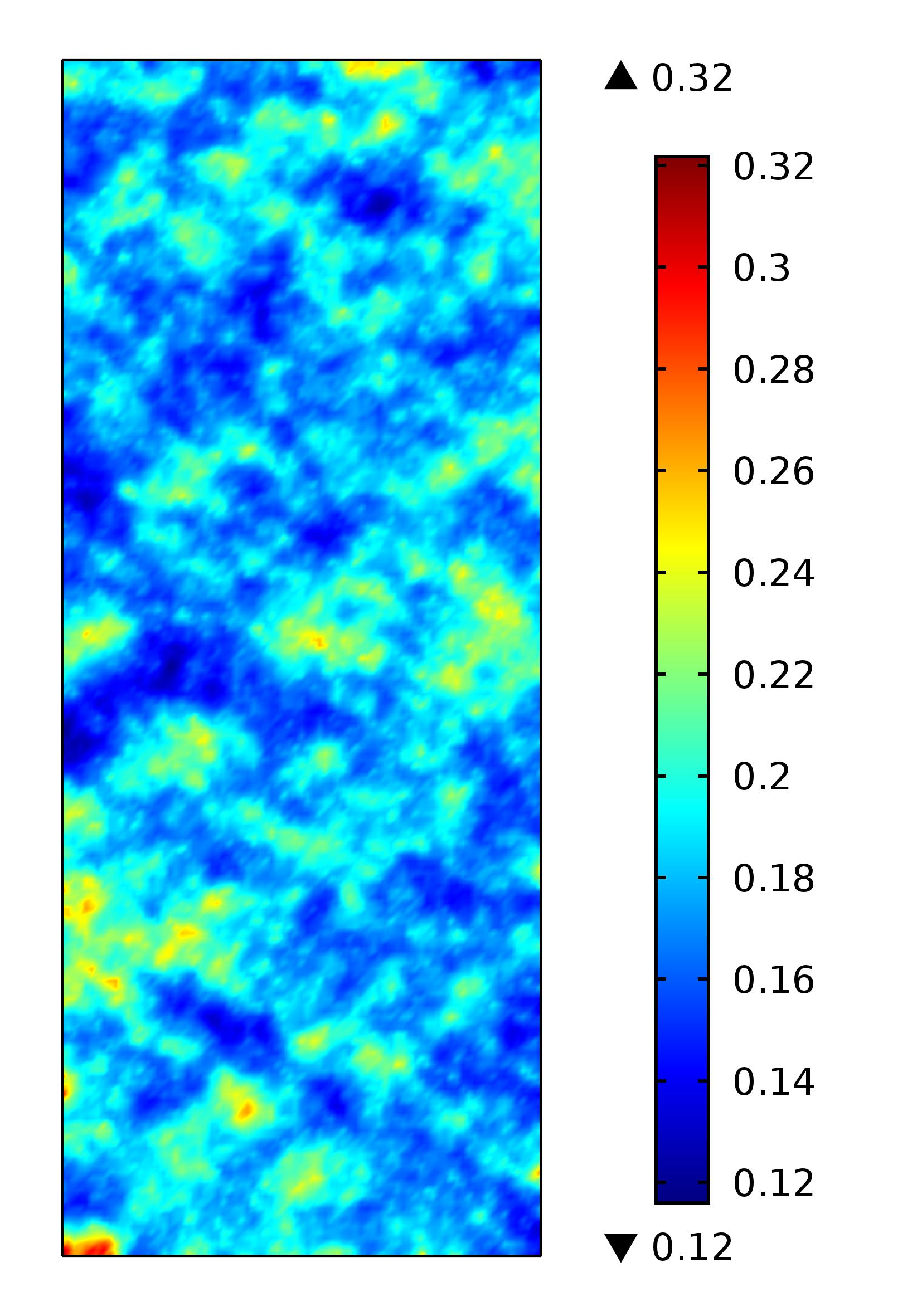}}
	\subfigure[$u=0.9215$ mm, 2D]{\includegraphics[height = 6.5cm]{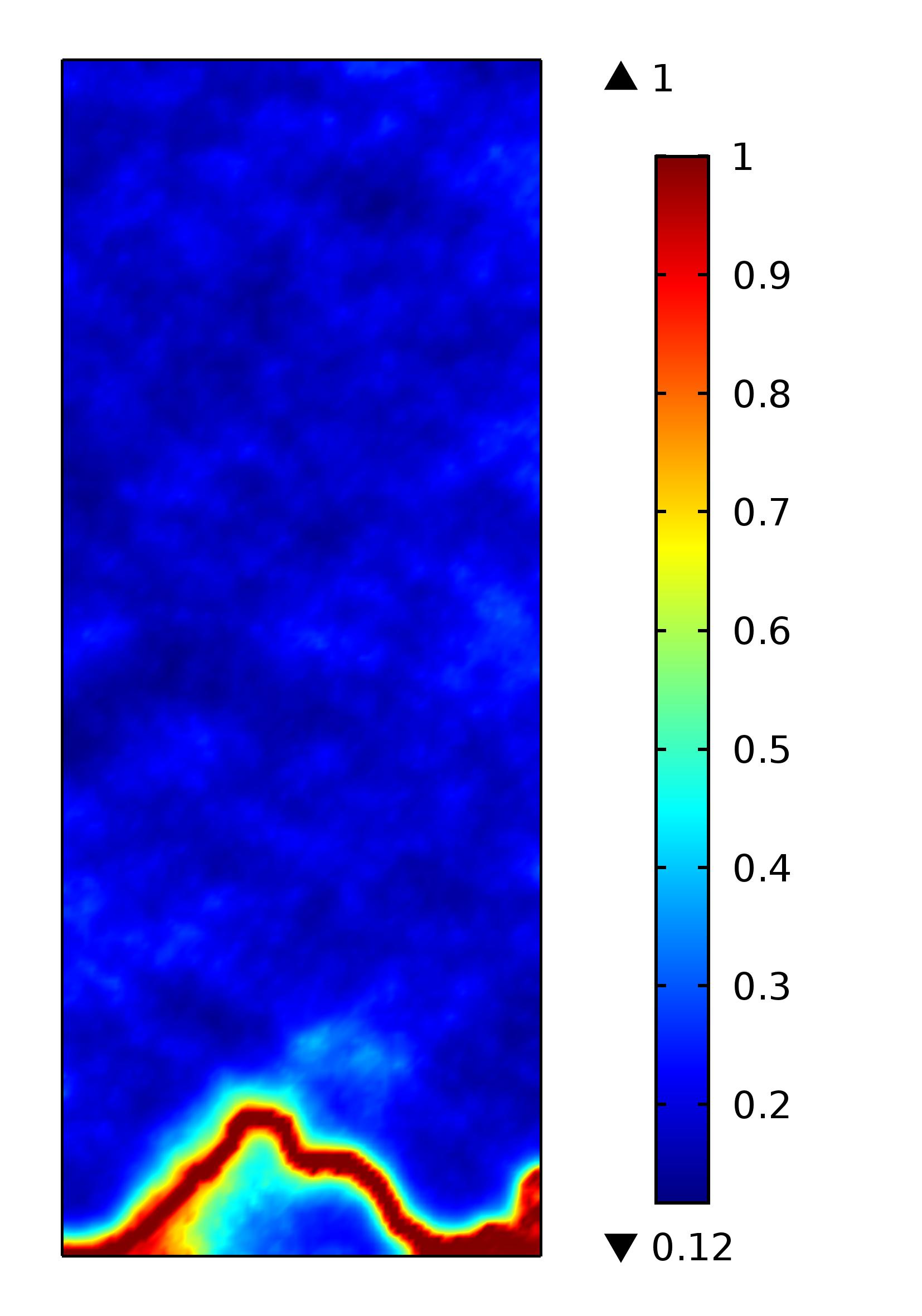}}
	\subfigure[$u=0.104$ mm, 3D]{\includegraphics[height = 6.5cm]{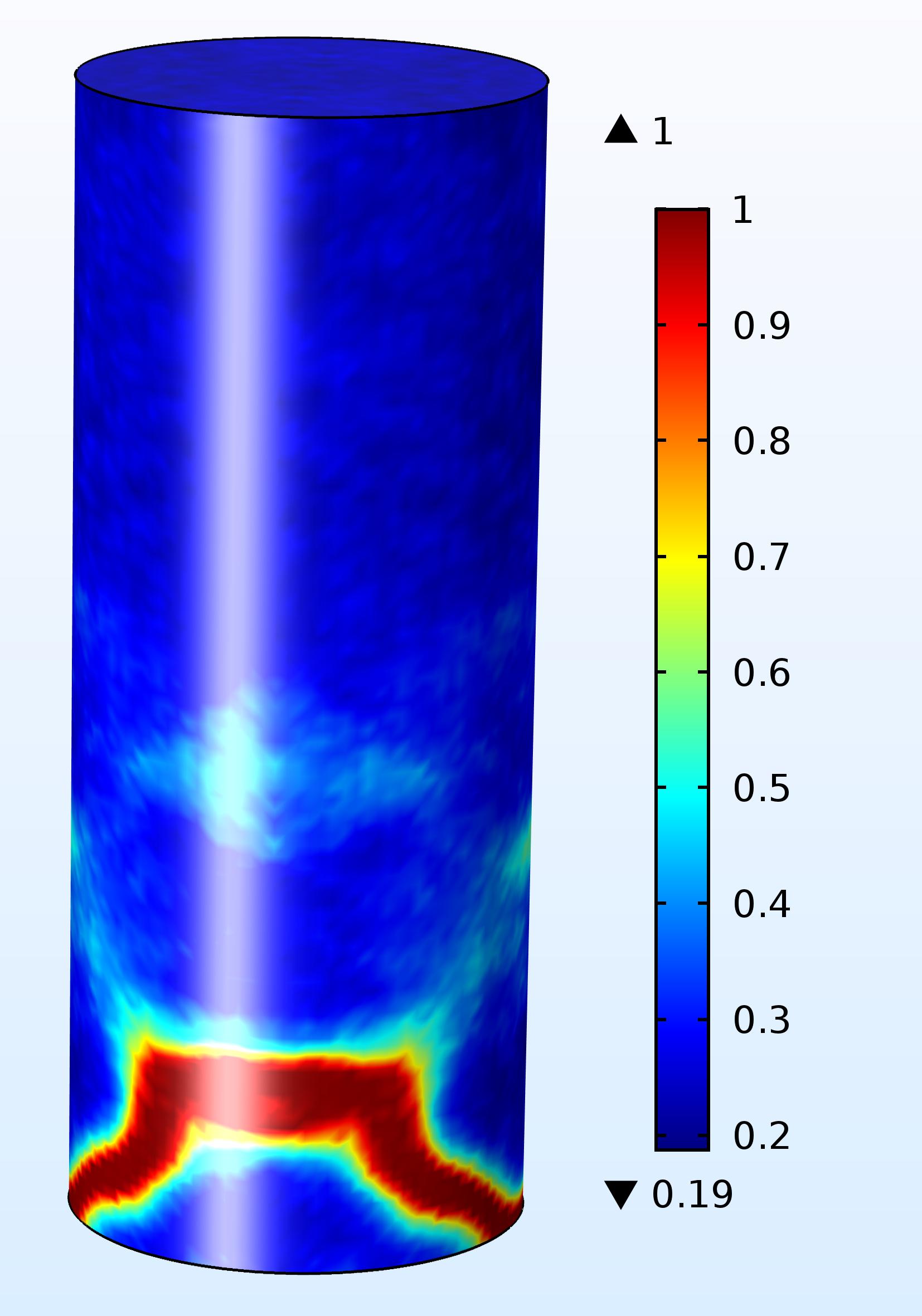}}\\
	\caption{Phase field of the intact specimen at different displacements}
	\label{Phase field of the intact specimen at different displacements}
	\end{figure}

\begin{figure}[htbp]
	\centering
	\subfigure[Basalt \citep{xia2015strength}]{\includegraphics[height = 6cm]{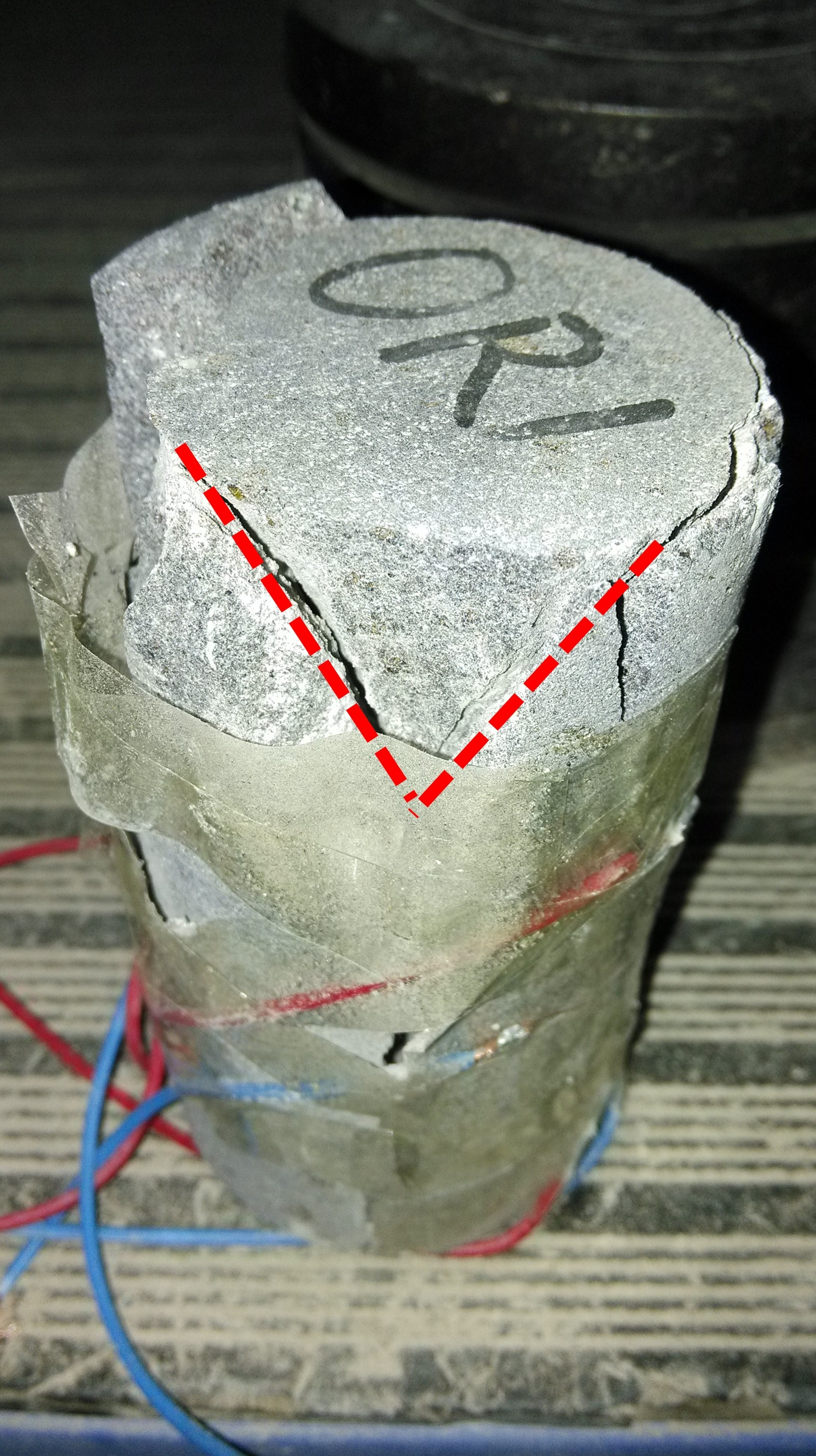}}
	\subfigure[Cement mortar]{\includegraphics[height = 6cm]{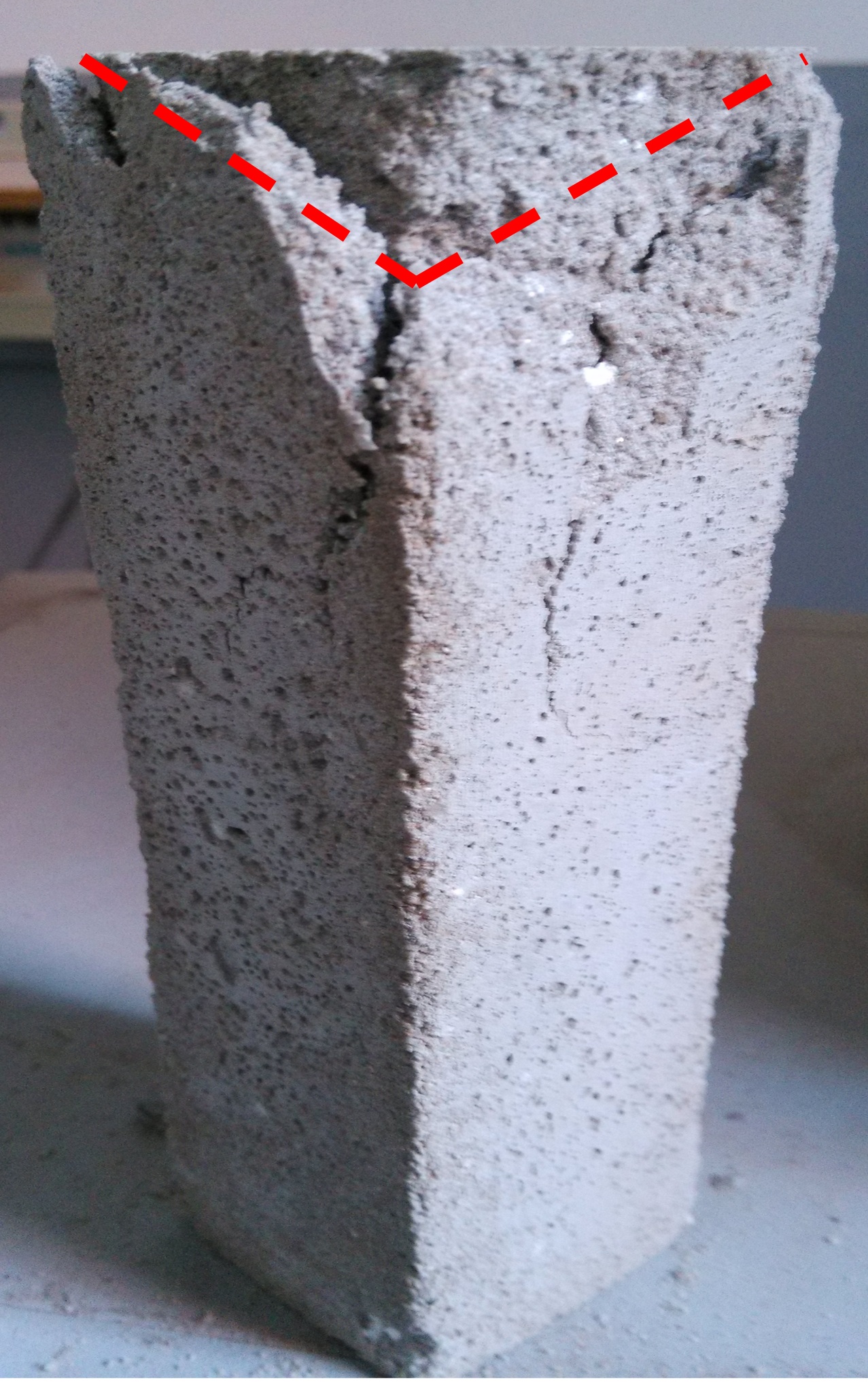}}
	\subfigure[Sandstone \citep{basu2013rock}]{\includegraphics[height = 6cm]{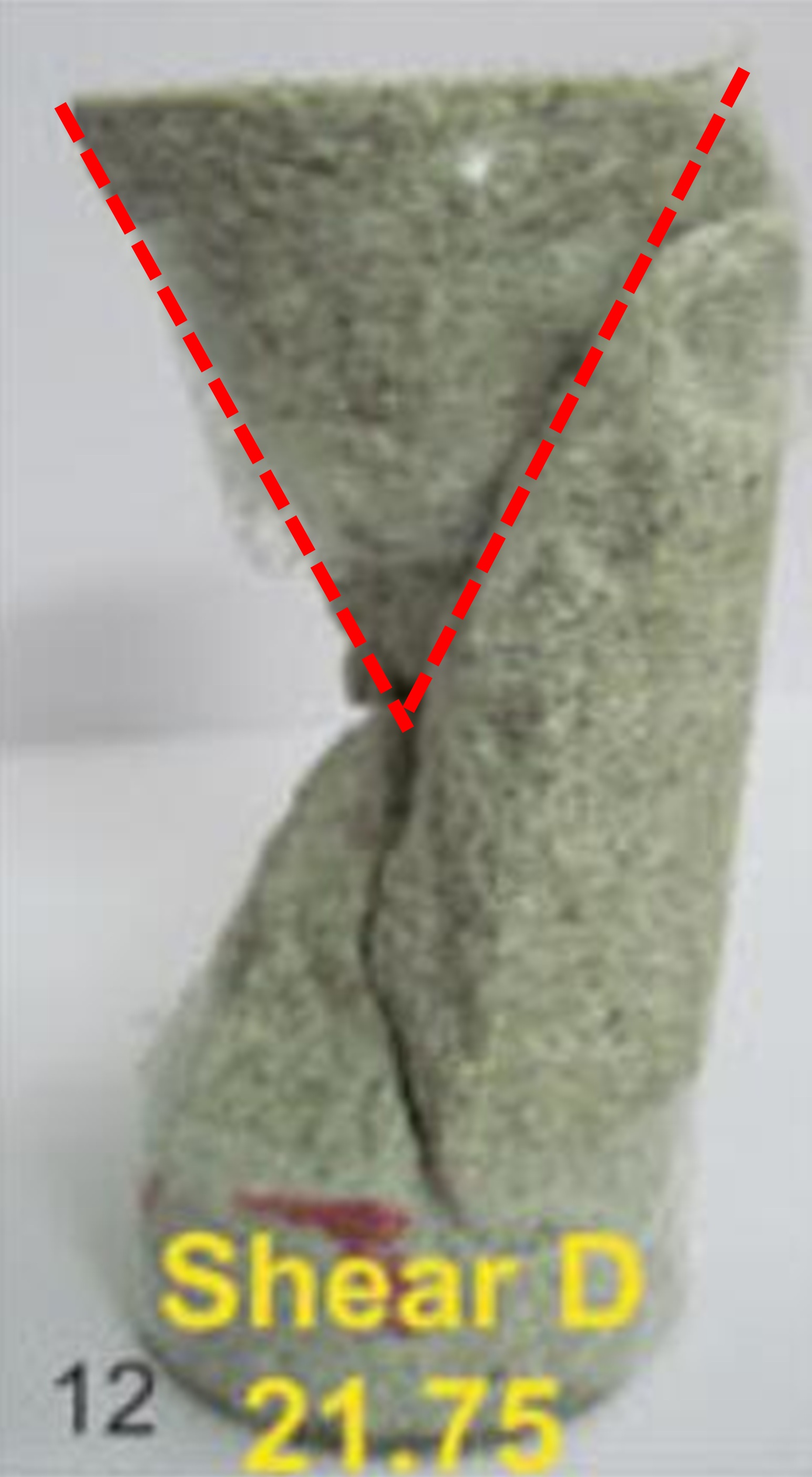}}
	\subfigure[Sandstone \citep{basu2013rock}]{\includegraphics[height = 6cm]{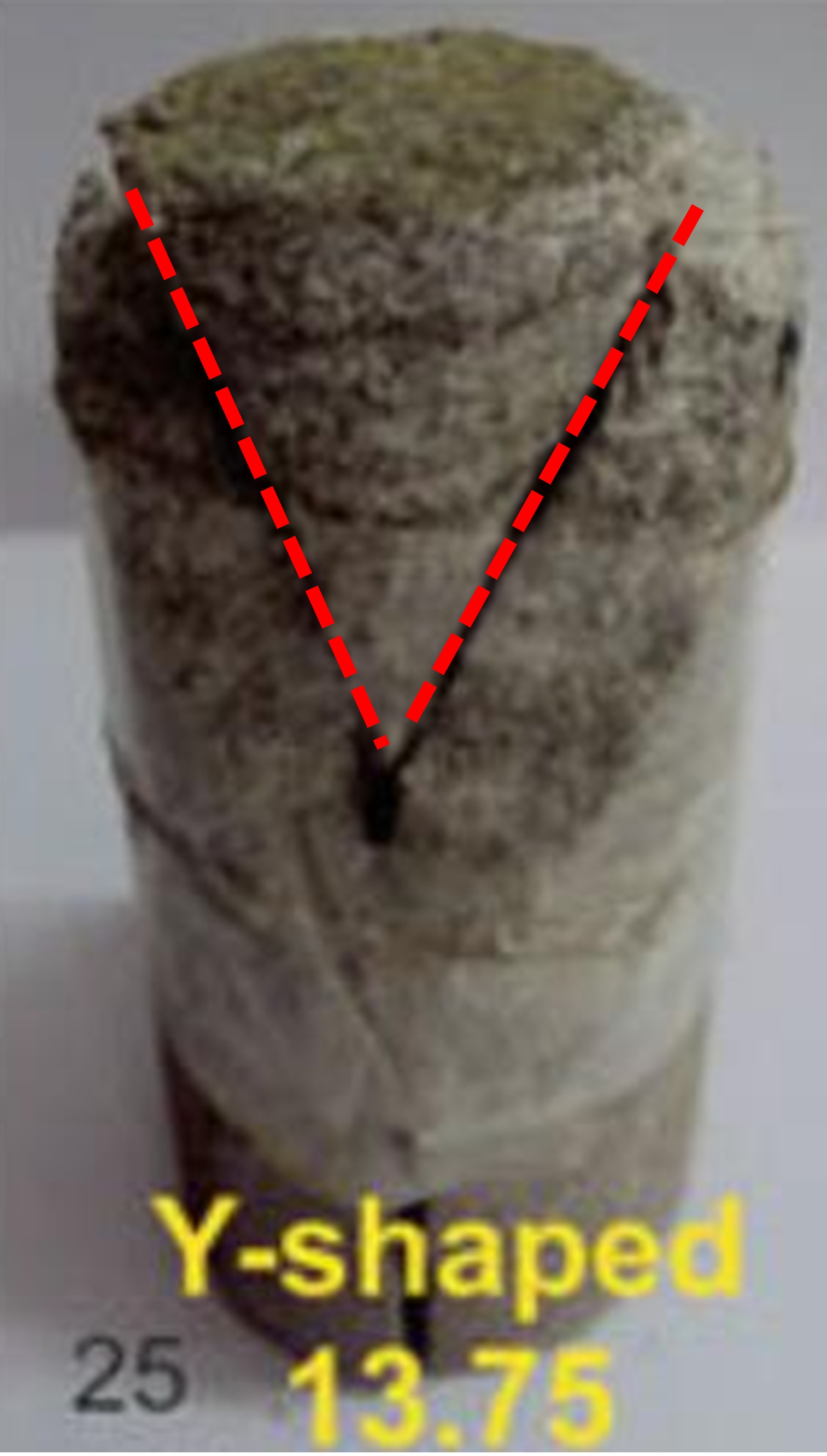}}\\
	\caption{Experimental observations of the V-shaped compressive-shear fractures}
	\label{Experimental observations of the V-shaped compressive-shear fractures}
	\end{figure}

Figure \ref{Phase field of the intact specimen at different displacements} shows the phase field contour plot of the intact specimen at different displacements. As can be seen in Fig. \ref{Phase field of the intact specimen at different displacements}a, the phase field is randomly distributed and its maximum value occurs at the lower left corner of the specimen. On the other hand, Fig. \ref{Phase field of the intact specimen at different displacements}b displays the final fracture pattern for 2D. A V-shaped compressive-shear fracture forms in Fig. \ref{Phase field of the intact specimen at different displacements}b with the increasing axial compressive load. For 3D, Fig. \ref{Phase field of the intact specimen at different displacements}c also shows a V-shaped fracture pattern. The 2D and 3D phase field simulations in this study agree well with the experimental observations where the V-shaped compressive-shear fracture pattern is the most frequently observed in experimental tests. Some representative examples of the V-shaped shear fracture can be seen in Fig. \ref{Experimental observations of the V-shaped compressive-shear fractures}. Note that the fracture type in basalt (Fig. \ref{Experimental observations of the V-shaped compressive-shear fractures}a) comes from the sample OR1 in \citet{xia2015strength} and the compressive-shear fractures were observed in cement mortar (Fig. \ref{Experimental observations of the V-shaped compressive-shear fractures}b) while the fractures in sandstones in Fig. \ref{Experimental observations of the V-shaped compressive-shear fractures}c and d are obtained in \citet{basu2013rock}.

\begin{figure}[htbp]
	\centering
	\includegraphics[width = 9cm]{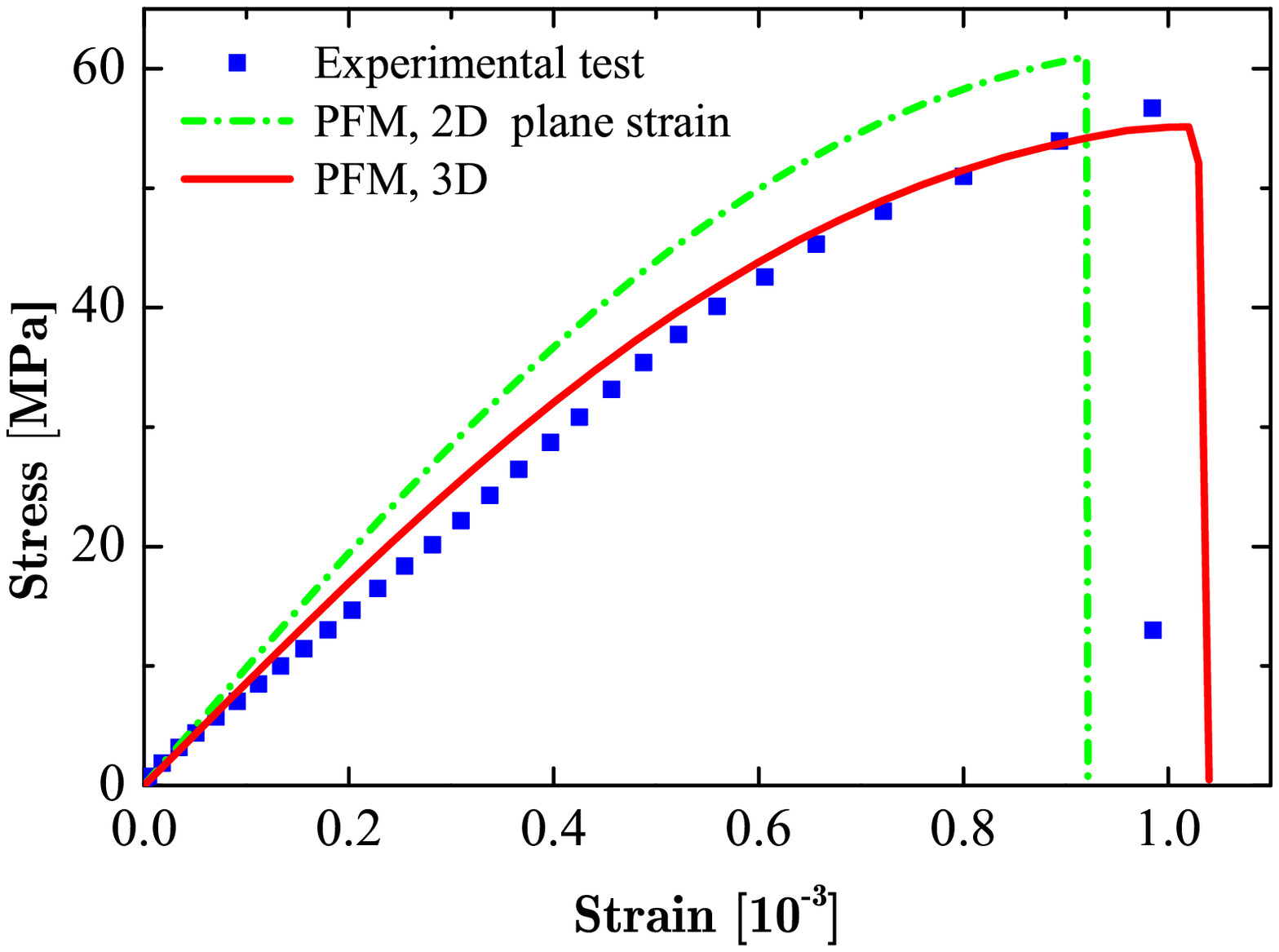}
	\caption{Comparison of the stress-strain curves of the phase field simulation and experimental tests}
	\label{Comparison of the stress-strain curves of the phase field simulation and experimental tests}
	\end{figure}

Figure \ref{Comparison of the stress-strain curves of the phase field simulation and experimental tests} compares the stress-strain curves obtained by the phase field simulation and the experimental tests (sample OR1 in \citet{xia2015strength}). In this figure, the stress refers to the averaged pressure on the upper boundary of the specimen and the strain refers to the ratio of the displacement on the upper end to the height of the specimen. As observed, the proposed PFM achieves a close stress-strain curve to the experimental test and as expected the 3D PFM agrees better to the experiment than 2D plane strain model. The 2D PFM obtains a stress-strain curve with a slight deviation from the experimental curve because 2D plane strain assumption provides extra stiffness for the vertical direction due to the displacement restriction perpendicular to the plane. Therefore, for an intact specimen, the proposed phase field model can exactly reproduce the fracture pattern and stress-strain curve, which preliminarily indicates the effectiveness of the proposed PFM for compressive-shear fractures.

\subsection{Uniaxial compression of a specimen with an inclined flaw}

\begin{figure}[htbp]
	\centering
	\includegraphics[width = 6cm]{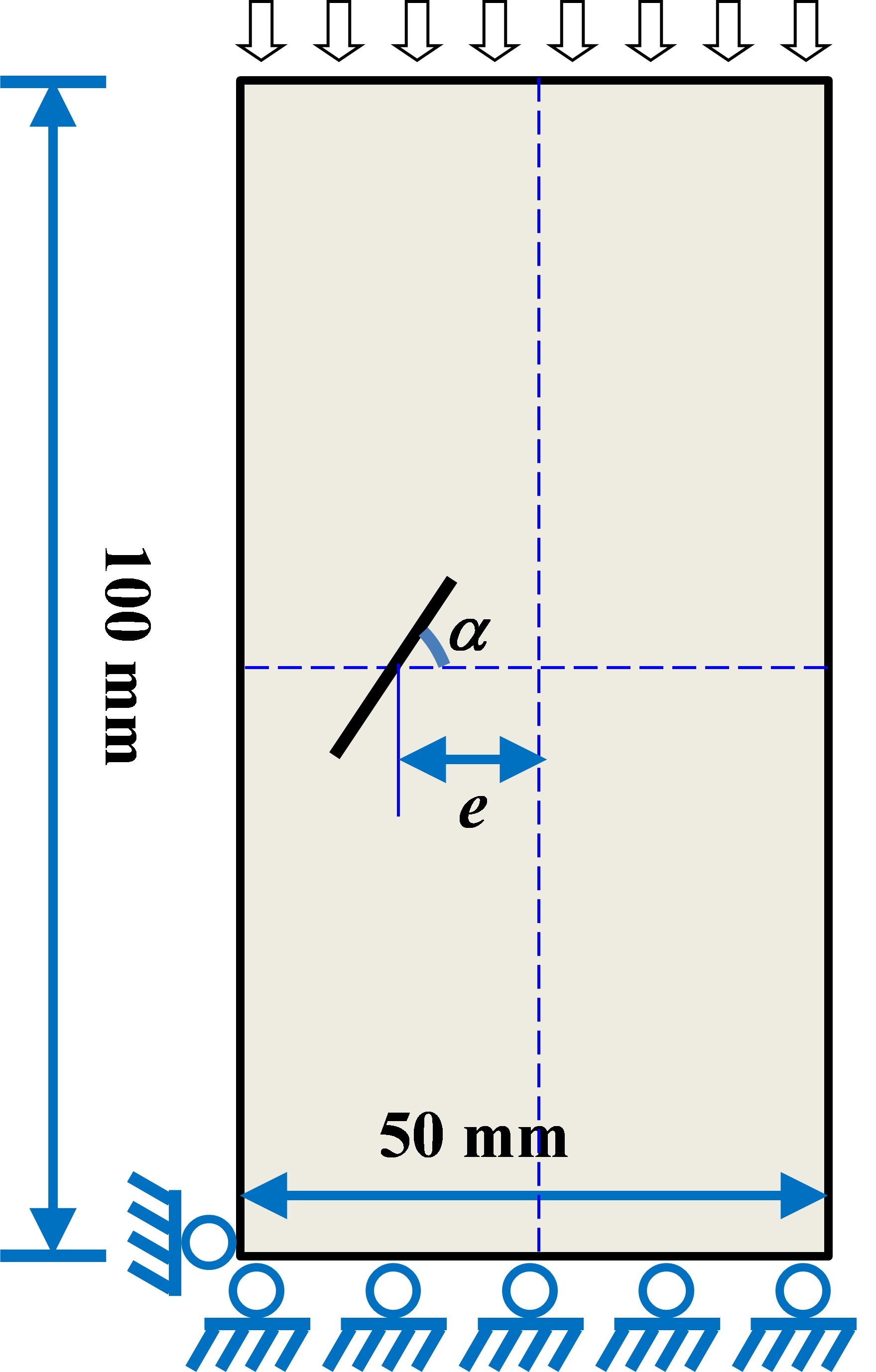}
	\caption{Setup of a rectangular specimen with an inclined flaw subjected to uniaxial compression}
	\label{Setup of a rectangular specimen with an inclined flaw subjected to uniaxial compression}
	\end{figure}

The fractures caused by uniaxial compression on rock-like materials with an inclined flaw have been widely investigated both experimentally and numerically (see some typical contributions in \citet{lajtai1974brittle, ingraffea1980finite, wong2009systematic, yang2011strength, zhuang2014comparative, zhang2017modification} and these literature applied cuboid or cubic rock specimens). Therefore, in this example, we apply the proposed PFM to simulate the compressive-shear fractures in a rectangular specimen with an inclined flaw subjected to axial compression. The setup of the problem including the geometry and boundary conditions is illustrated in Fig. \ref{Setup of a rectangular specimen with an inclined flaw subjected to uniaxial compression} along with the width and height of the rock-like specimen being 50 mm and 100 mm, respectively. The flaw is 5 mm in length and 1 mm in width; its center is located in the axis of symmetry and deviates from the center of the specimen by an eccentricity $e$. Moreover, the pre-existing flaw is inclined at an angle $\alpha$ to the horizontal direction, and a vertical downward displacement is applied on the upper end of the specimen.

These base parameters are used for calculation: inclination angle $\alpha = 45^\circ$, Young's modulus $E$ = 60 GPa, Poisson's ratio $\nu$ = 0.3, critical energy release rate $G_c$ = 100 N/m, length scale parameter $l_0$ = 1 mm, $k=1\times10^{-9}$, cohesion $c$ = 100 kPa, and internal friction angle $\varphi=15^\circ$. According to \citet{zhang2017numerical}, the length scale used in this example can capture accurate crack paths since the ratio of the length scale to the specimen size is sufficiently small. A total of 51906 triangular elements are used to discretize the domain with the maximum element size $h$ = 0.5 mm. In addition, we adopt a displacement increment $\Delta u = 1\times 10^{-4}$ mm in the simulations.

\begin{figure}[htbp]
	\centering
	\subfigure[$u = 0.1540$ mm]{\includegraphics[width = 4cm]{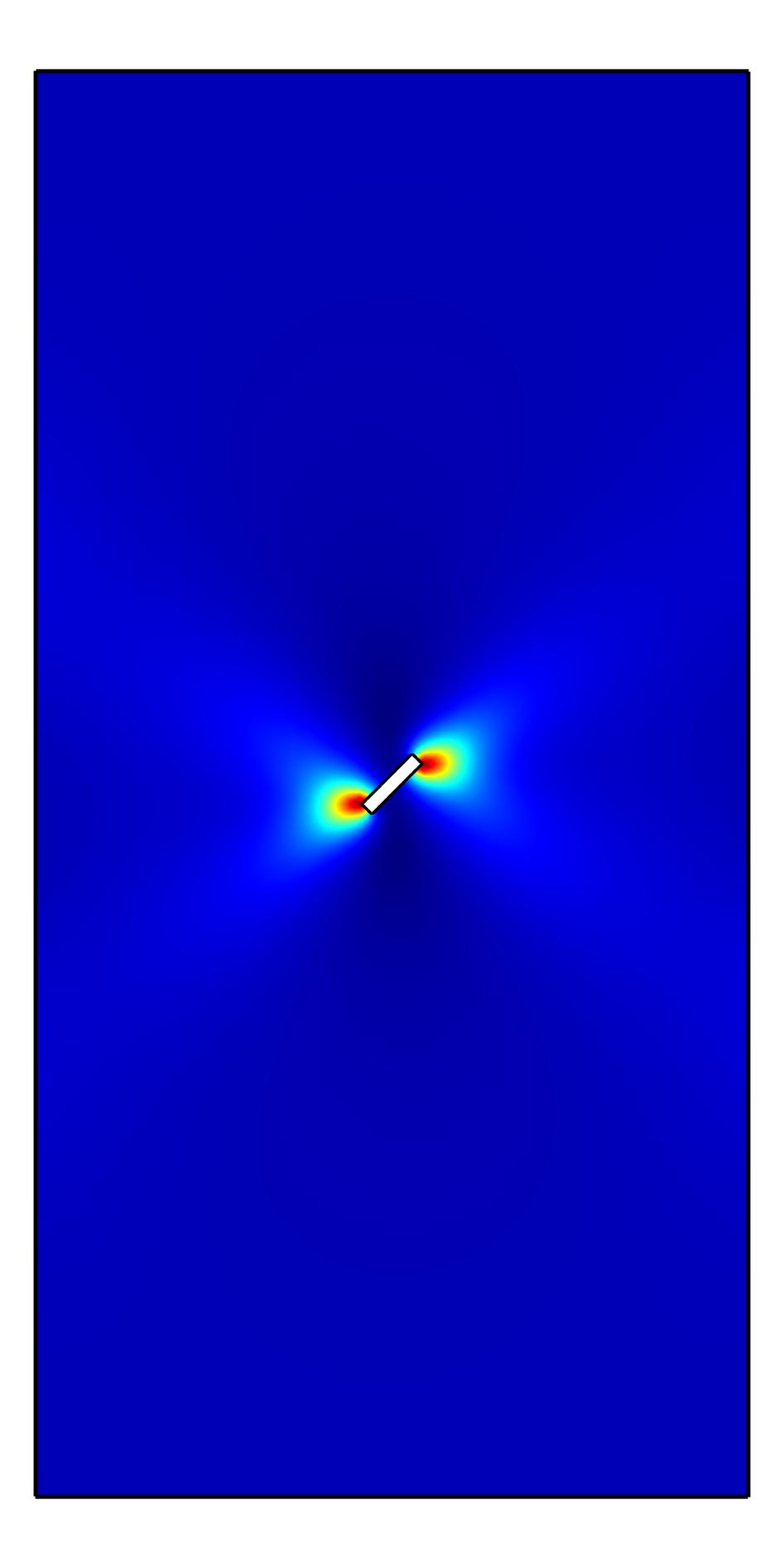}}
	\subfigure[$u = 0.1554$ mm]{\includegraphics[width = 4cm]{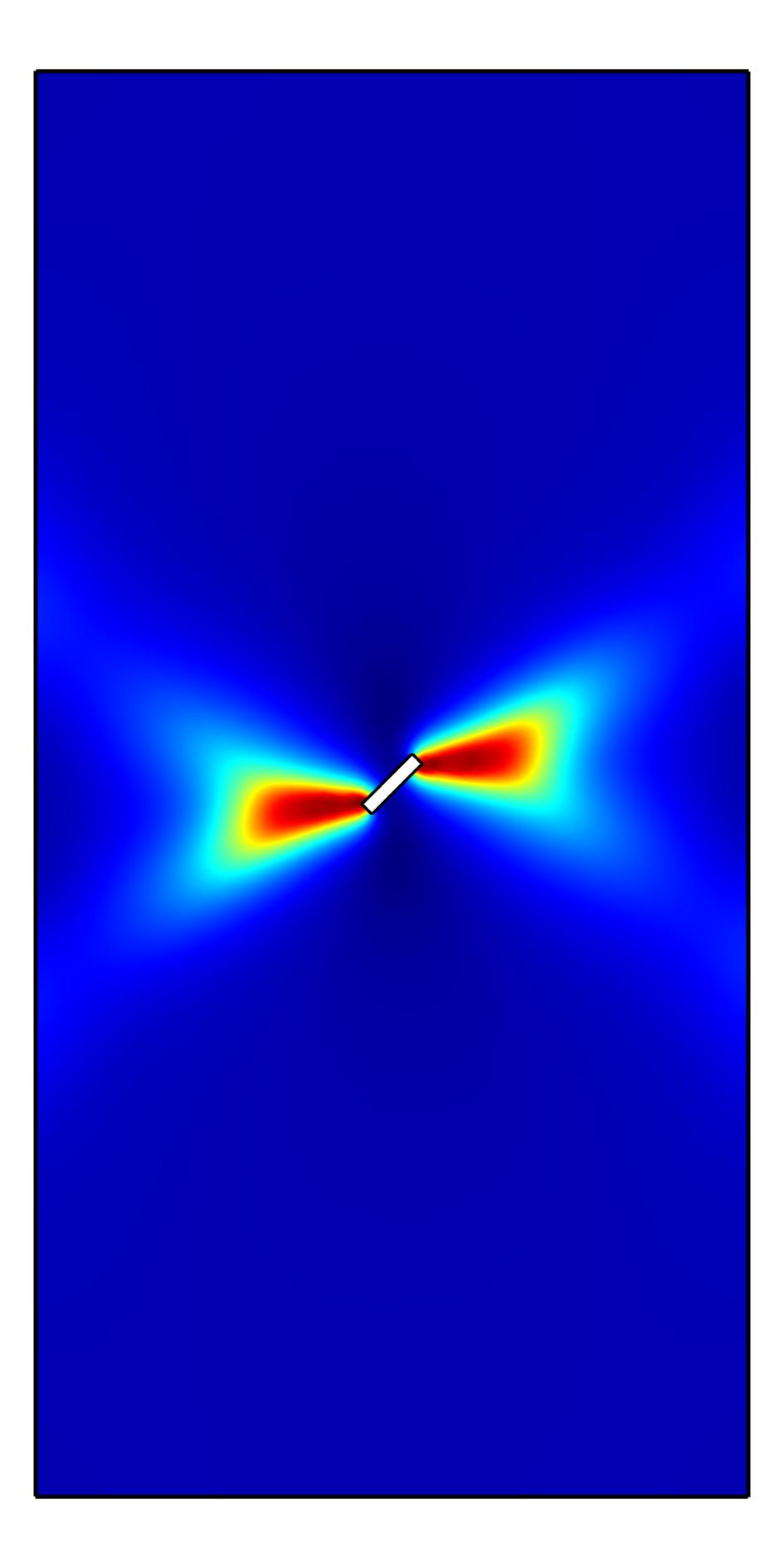}}
	\subfigure[$u = 0.1558$ mm]{\includegraphics[width = 4cm]{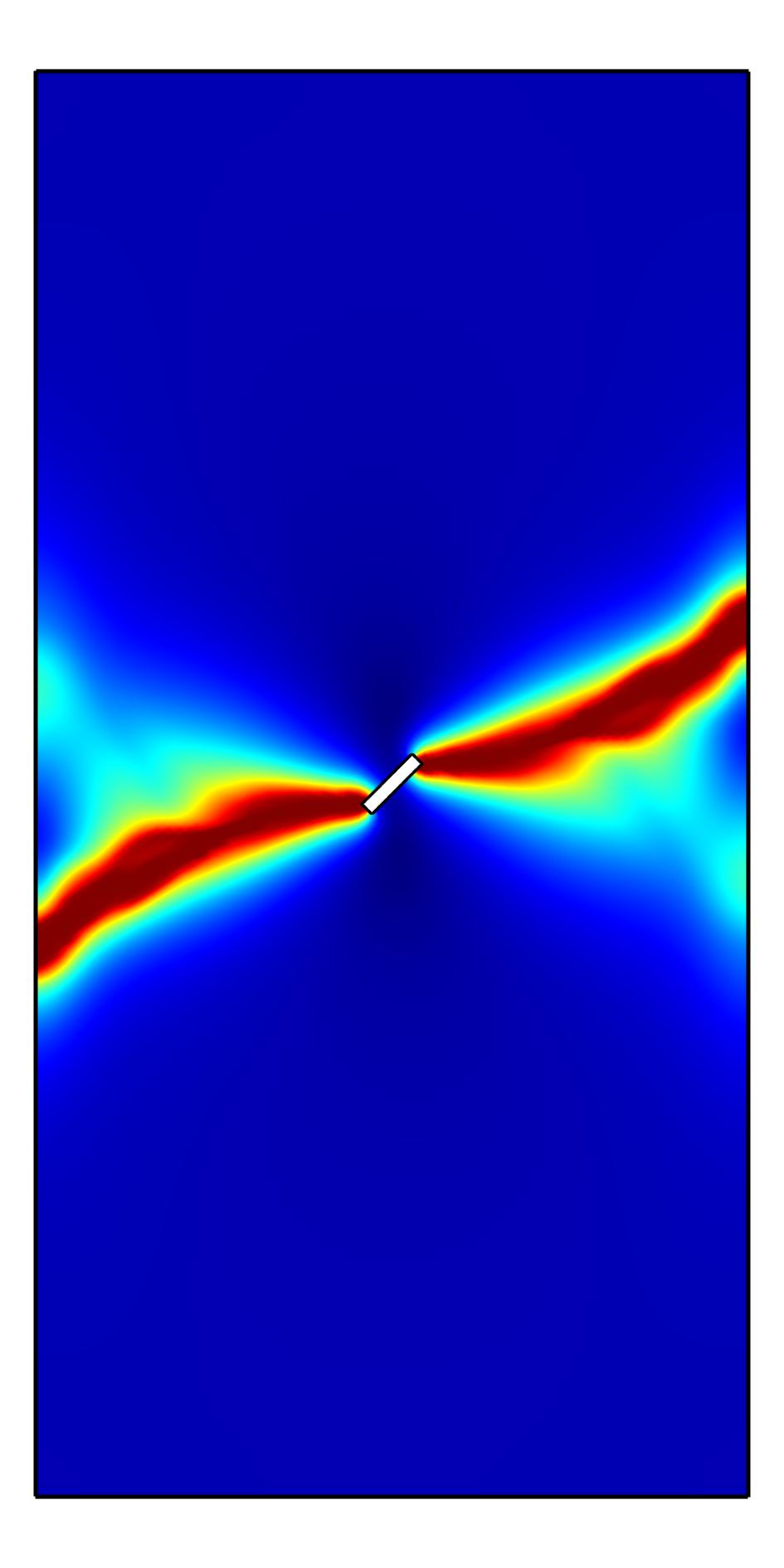}}\\
	\caption{Fracture patterns of a specimen with an inclined flaw ($\alpha=45^\circ$)}
	\label{Fracture patterns of a specimen with an inclined flaw alpha = 45}
	\end{figure}

\begin{figure}[htbp]
	\centering
	\subfigure[$u = 0.1540$ mm]{\includegraphics[width = 5cm]{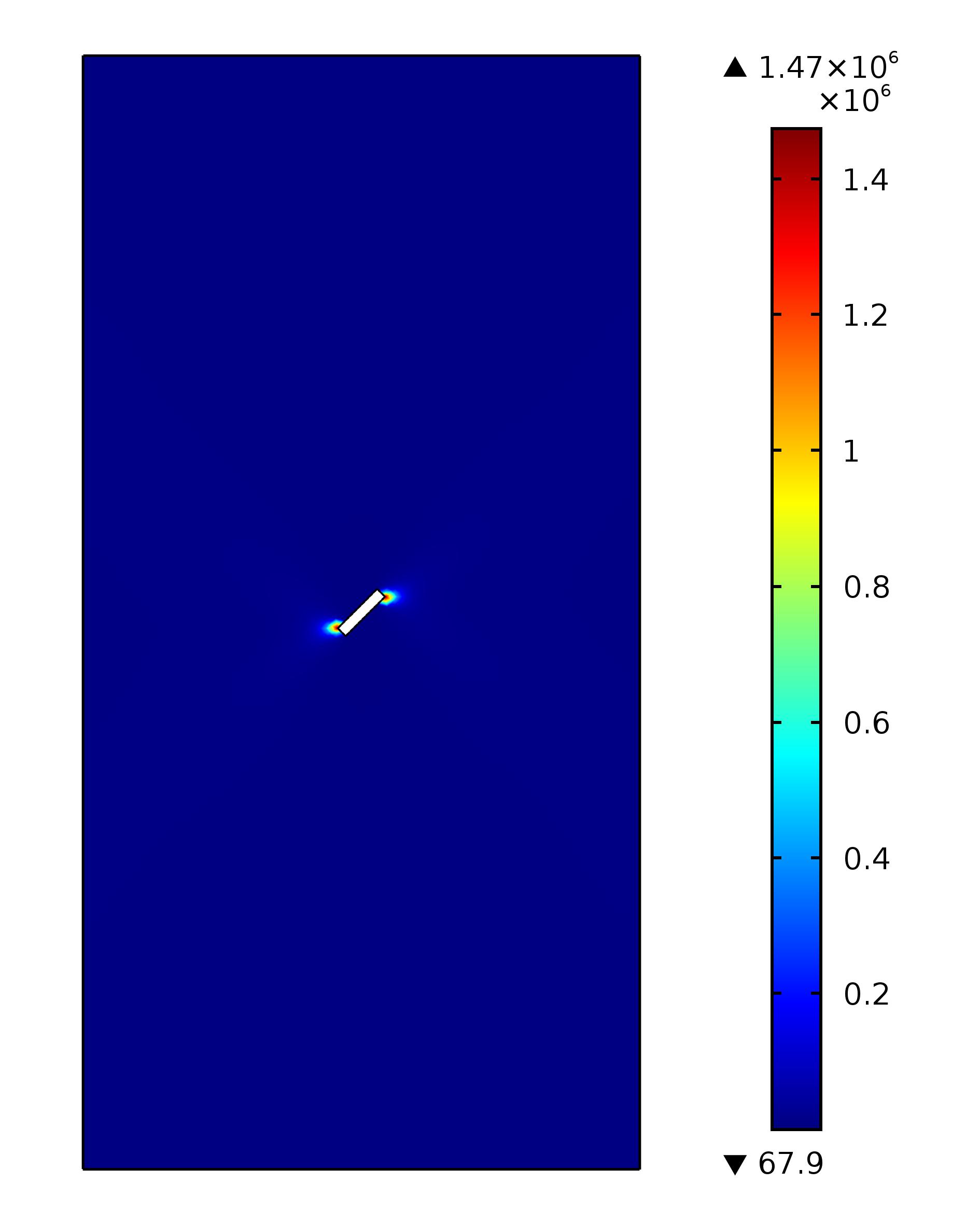}}
	\subfigure[$u = 0.1554$ mm]{\includegraphics[width = 5cm]{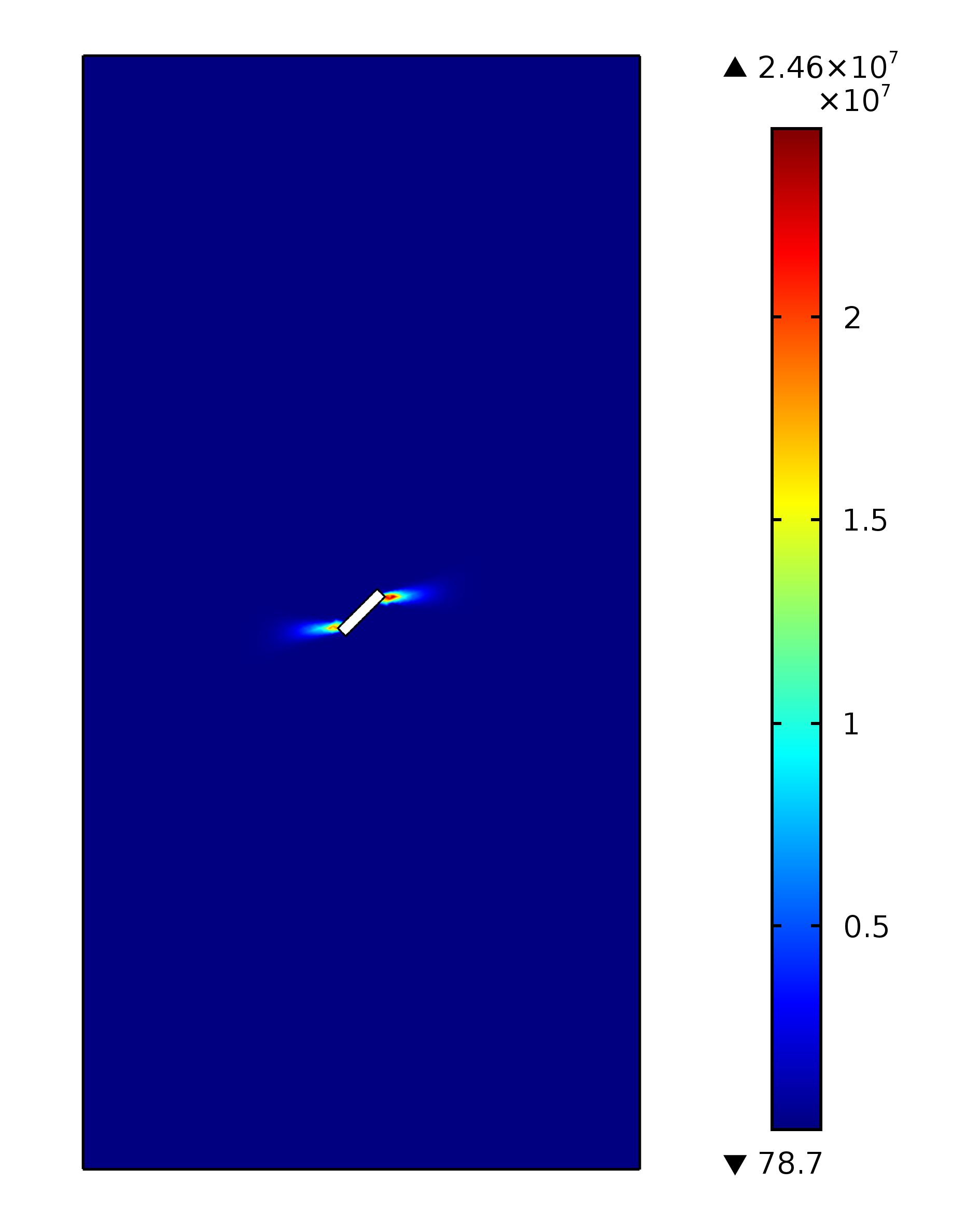}}
	\subfigure[$u = 0.1558$ mm]{\includegraphics[width = 5cm]{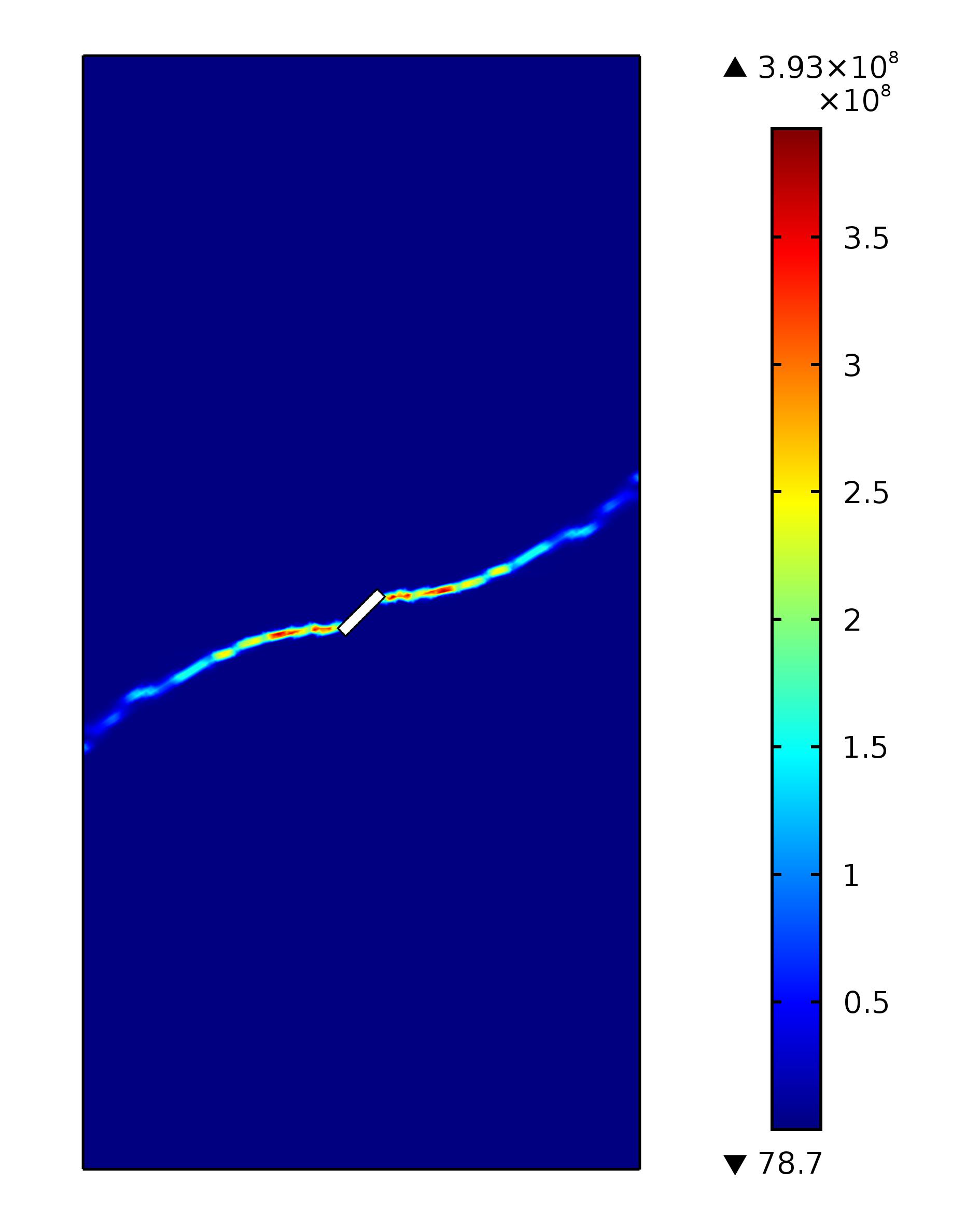}}\\
	\caption{Evolution of $H_p$ for a specimen with an inclined flaw ($\alpha=45^\circ$) (Unit:J/m$^3$)}
	\label{Evolution of the history reference of a specimen with an inclined flaw alpha = 45}
	\end{figure}

For an inclination angle of the flaw $\alpha=45^\circ$, the progressive fracture patterns obtained by the proposed phase field model are shown in Fig. \ref{Fracture patterns of a specimen with an inclined flaw alpha = 45}. It can be seen from Fig. \ref{Fracture patterns of a specimen with an inclined flaw alpha = 45}a that with the increase in the compressive loads, shear zones occur around the tips of the inclined flaw, and that two new fractures initiate at the flaw tips due to the increasing energy reference $H_p$, the evolution of which is shown in Fig \ref{Evolution of the history reference of a specimen with an inclined flaw alpha = 45}. As shown in Fig. \ref{Fracture patterns of a specimen with an inclined flaw alpha = 45}b, when the displacement on the upper boundary of the specimen increases to $0.1554$ mm, the fractures propagate obliquely and become wider because of complex compressive-shear zones around the fractures. In addition, Fig. \ref{Fracture patterns of a specimen with an inclined flaw alpha = 45}c indicates that when $u= 0.1558$ mm the fractures continue propagating at an increasing angle to the horizontal direction; finally an inclined fracture pattern is observed.

\begin{figure}[htbp]
	\centering
	\includegraphics[width = 12cm]{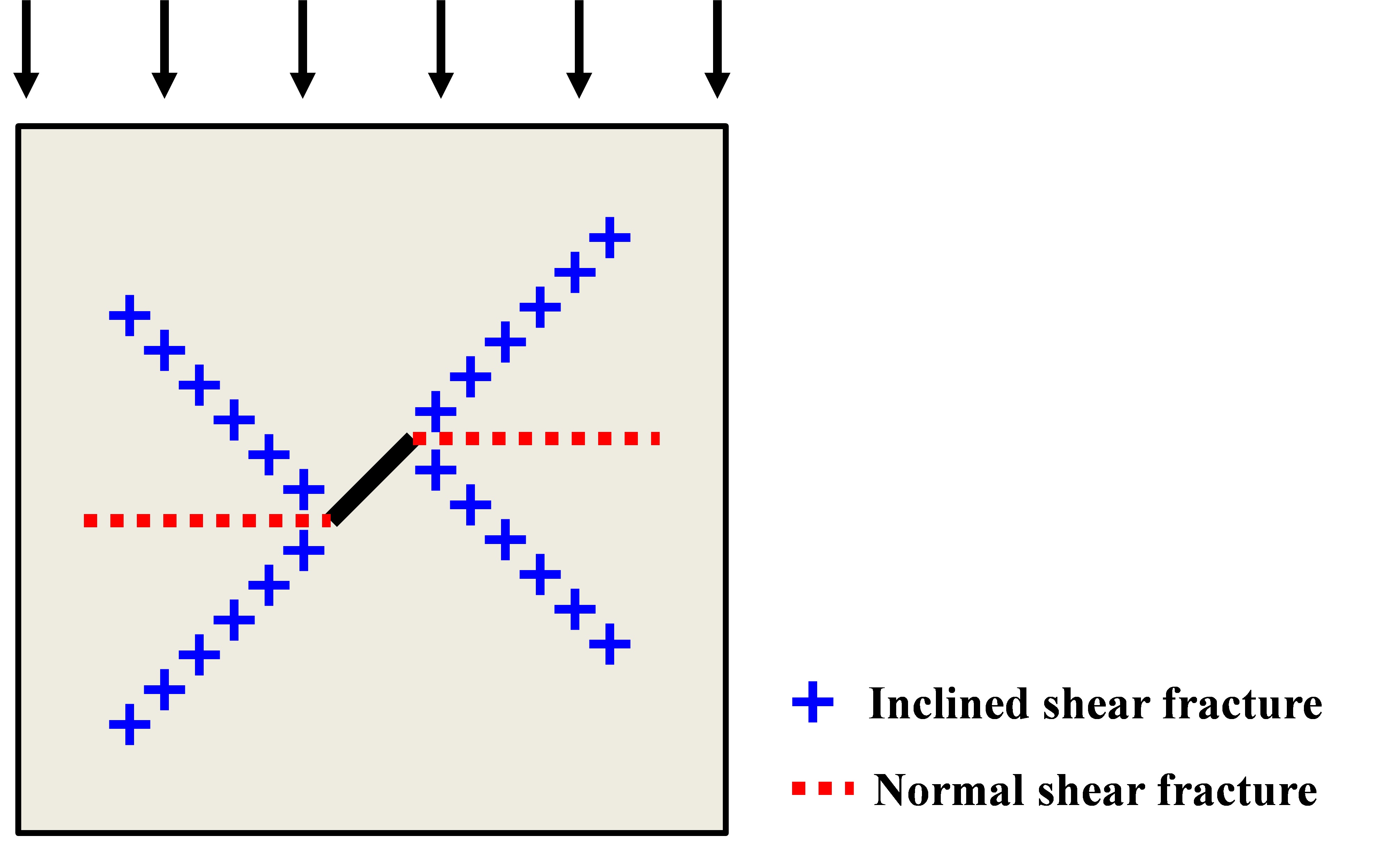}
	\caption{A specimen with an inclined flaw ($\alpha=45^\circ$). Experimental observations of the compressive-shear fractures in \citet{lajtai1974brittle, wong2009systematic}}
	\label{A specimen with an inclined flaw (alpha=45) Experimental observations of the compressive-shear fractures in lajtai1974brittle}
	\end{figure}

\begin{figure}[htbp]
	\centering
	\subfigure[Lateral fractures]{\includegraphics[width = 4cm]{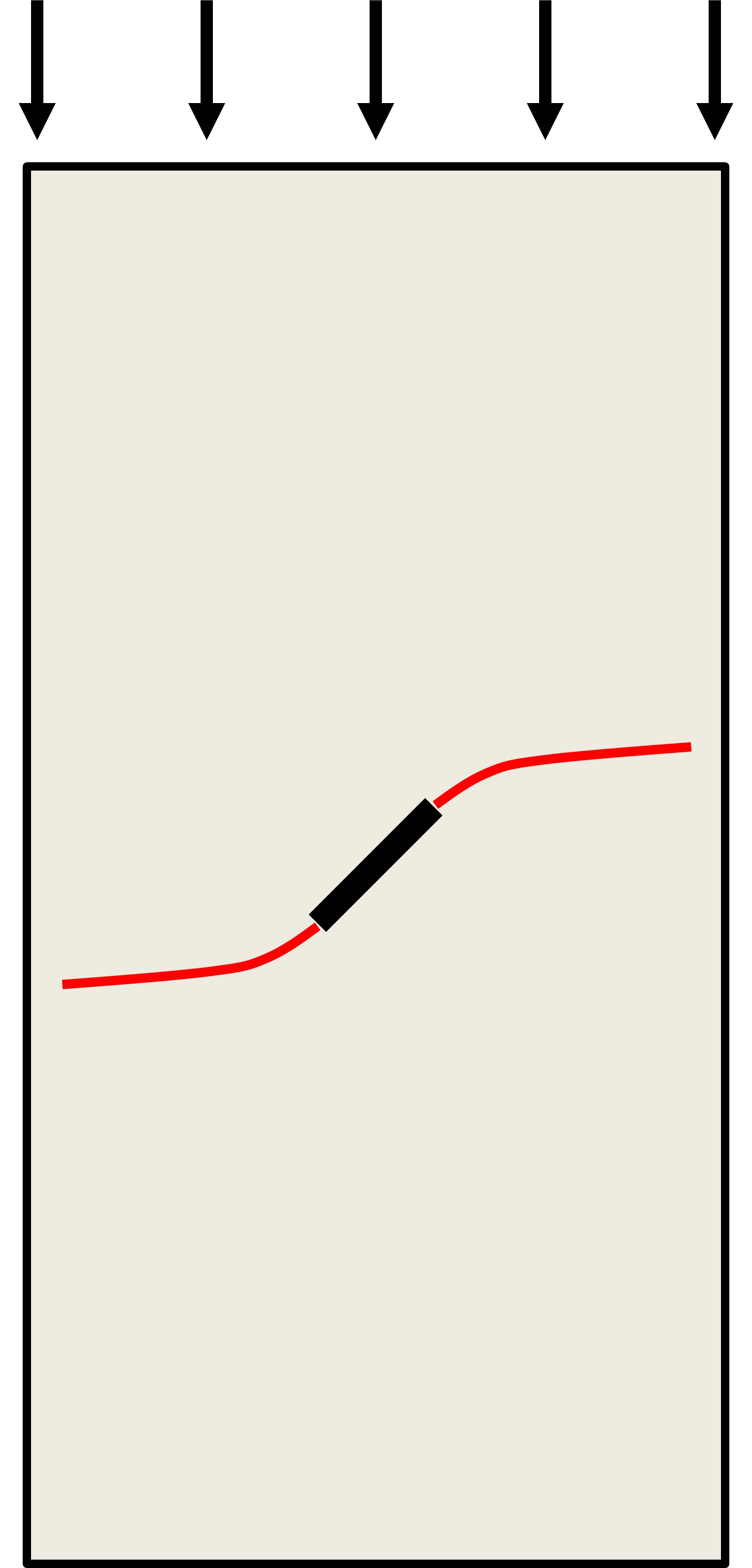}}
	\subfigure[Pure shear fractures]{\includegraphics[width = 4cm]{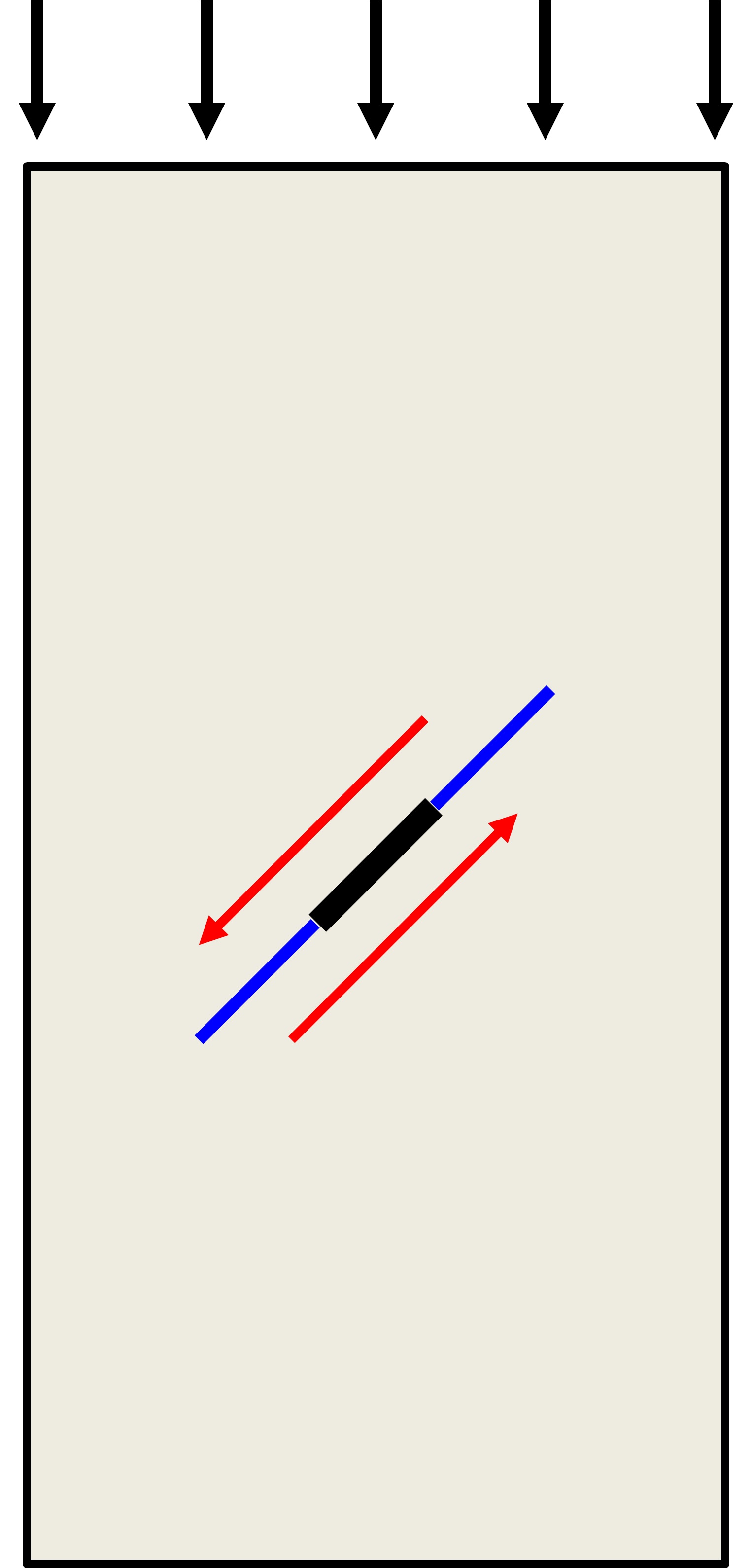}}
	\caption{Experimental observations of the compressive-shear fractures for a specimen with an inclined flaw ($\alpha=45^\circ$) in \citet{yang2011strength}}
	\label{A specimen with an inclined flaw (alpha=45) Experimental observations of the compressive-shear fractures in yang2011strength}
	\end{figure}

In an experimental test of uniaxial compression on rock-like materials with a flaw of $\alpha=45^\circ$, some salient features about the compressive-shear fracture responses can be observed and further classified into two categories: the normal and inclined shear fractures \citep{lajtai1974brittle, wong2009systematic}. These two types can be seen in Figs. \ref{A specimen with an inclined flaw (alpha=45) Experimental observations of the compressive-shear fractures in lajtai1974brittle} and \ref{A specimen with an inclined flaw (alpha=45) Experimental observations of the compressive-shear fractures in yang2011strength}, which interpret the experimental results of the compressive-shear fractures in square and rectangular specimens respectively, although the fractures are named a lateral fracture and a pure shear fracture by \citet{yang2011strength}. Comparing Figs. \ref{Fracture patterns of a specimen with an inclined flaw alpha = 45}, \ref{A specimen with an inclined flaw (alpha=45) Experimental observations of the compressive-shear fractures in lajtai1974brittle}, and \ref{A specimen with an inclined flaw (alpha=45) Experimental observations of the compressive-shear fractures in yang2011strength} show that the compressive-shear fractures modeled by the proposed PFM are consistent with the experimental results.

\begin{figure}[htbp]
	\centering
	\includegraphics[width = 12cm]{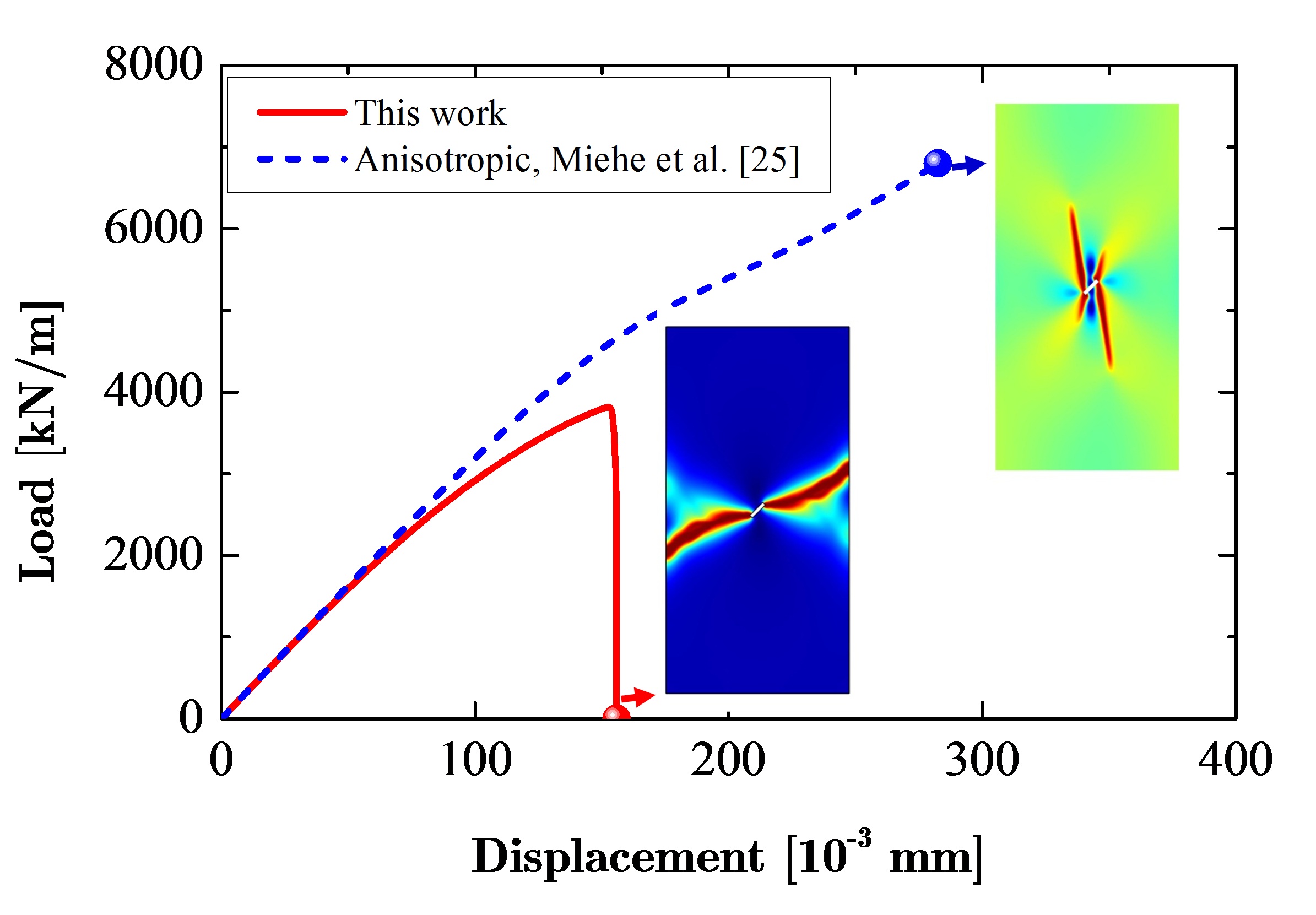}
	\caption{Comparison of fracture pattern and load-displacement curve obtained by the proposed PFM and the anisotropic model of \citep{miehe2010thermodynamically}}
	\label{Comparison of fracture pattern and load-displacement curve obtained by the proposed PFM and the anisotropic model of}
	\end{figure}

We also compare the fracture pattern and load-displacement curves obtained by our model and the anisotropic model \citep{miehe2010thermodynamically} in Fig. \ref{Comparison of fracture pattern and load-displacement curve obtained by the proposed PFM and the anisotropic model of}. Note that both models are tested under the same parameters. In Fig. \ref{Comparison of fracture pattern and load-displacement curve obtained by the proposed PFM and the anisotropic model of}, the anisotropic model of \citep{miehe2010thermodynamically} is found to not have a drop stage in the load-displacement curve while the drop stage in our model is obvious. Moreover, only wing and anti-wing tensile cracks are simulated in the model of \citep{miehe2010thermodynamically}. Therefore, the comparison in Fig. \ref{Comparison of fracture pattern and load-displacement curve obtained by the proposed PFM and the anisotropic model of} convinces the finding that the anisotropic PFM is not suitable for predicting compressive-shear fracture in rock-like materials.

\begin{figure}[htbp]
	\centering
	\subfigure[$u = 0.1378$ mm]{\includegraphics[width = 4cm]{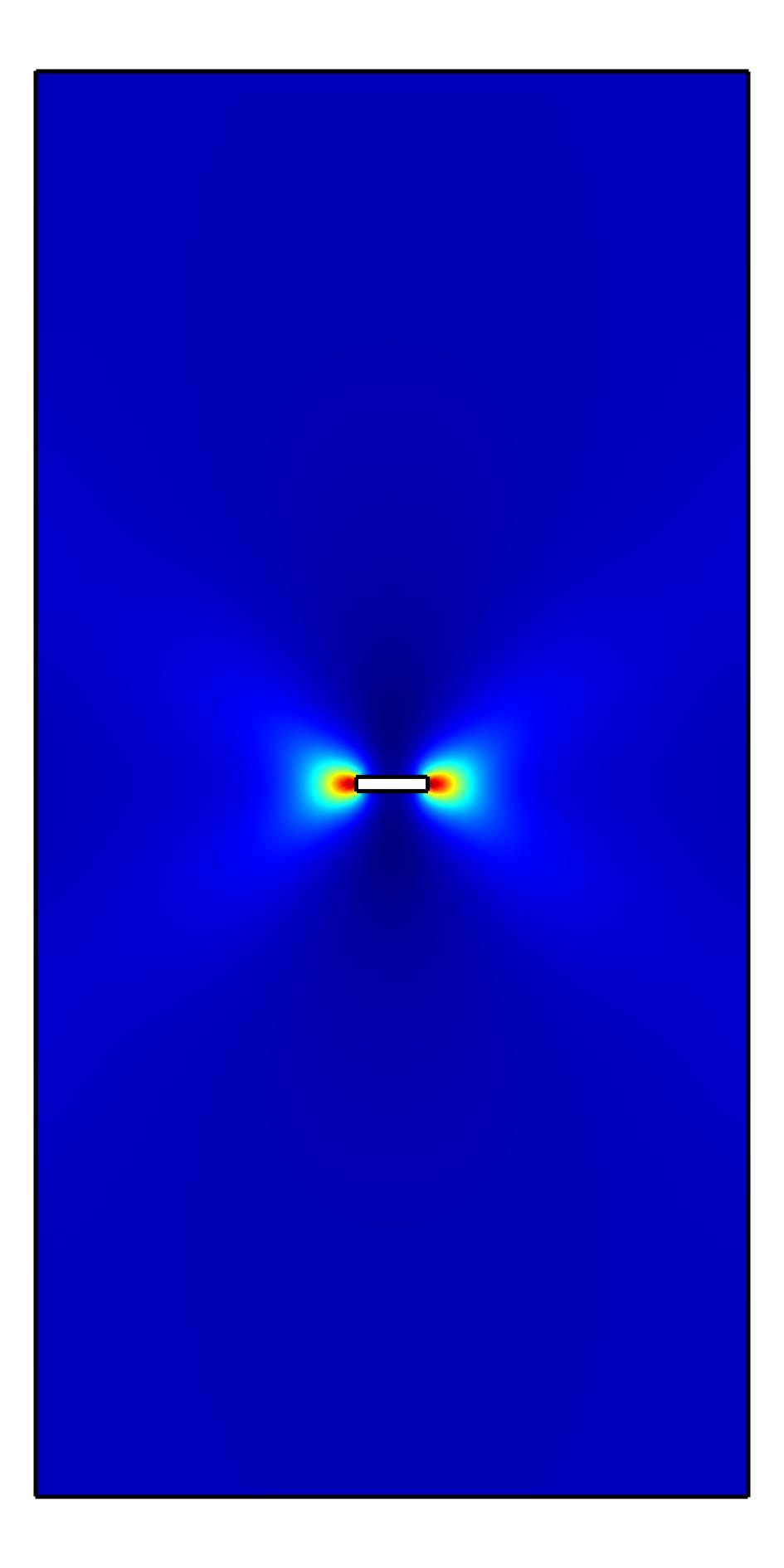}}
	\subfigure[$u = 0.1400$ mm]{\includegraphics[width = 4cm]{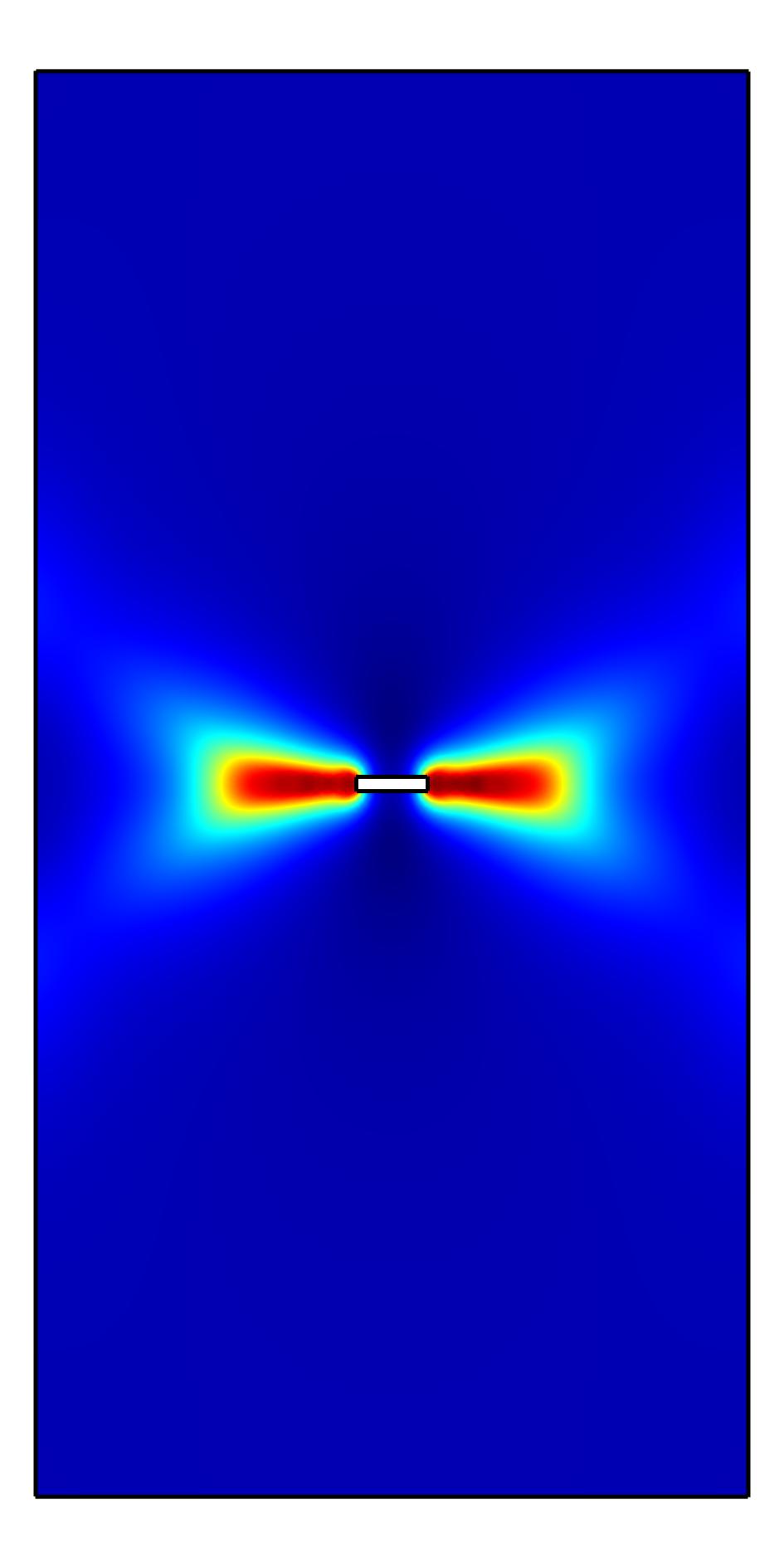}}
	\subfigure[$u = 0.1406$ mm]{\includegraphics[width = 4cm]{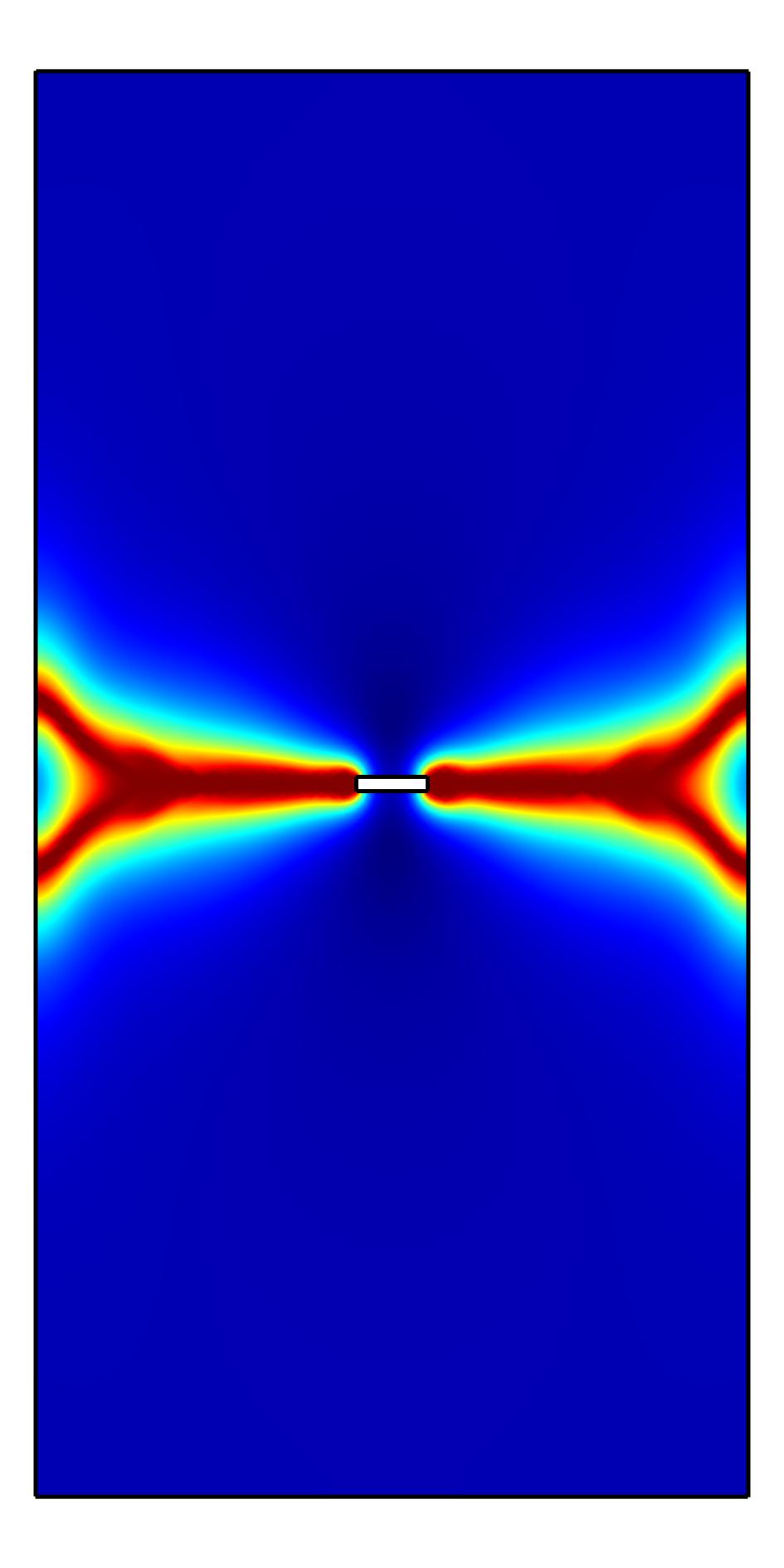}}\\
	\caption{Fracture patterns (phase field evolution) of a specimen with an inclined flaw of $\alpha=0^\circ$}
	\label{Fracture patterns of a specimen with an inclined flaw alpha = 0}
	\end{figure}

\begin{figure}[htbp]
	\centering
	\subfigure[$u = 0.1976$ mm]{\includegraphics[width = 4cm]{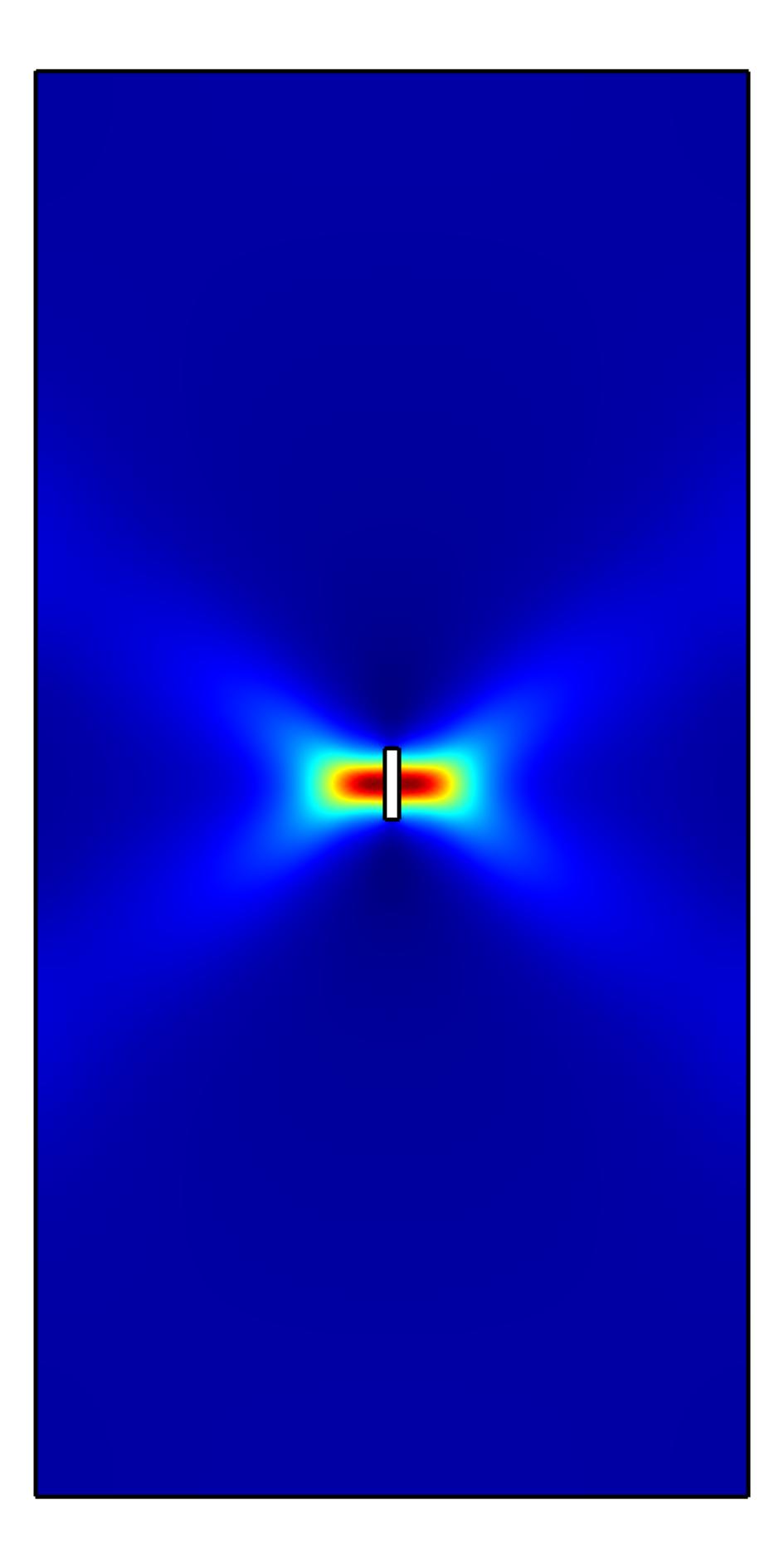}}
	\subfigure[$u = 0.1980$ mm]{\includegraphics[width = 4cm]{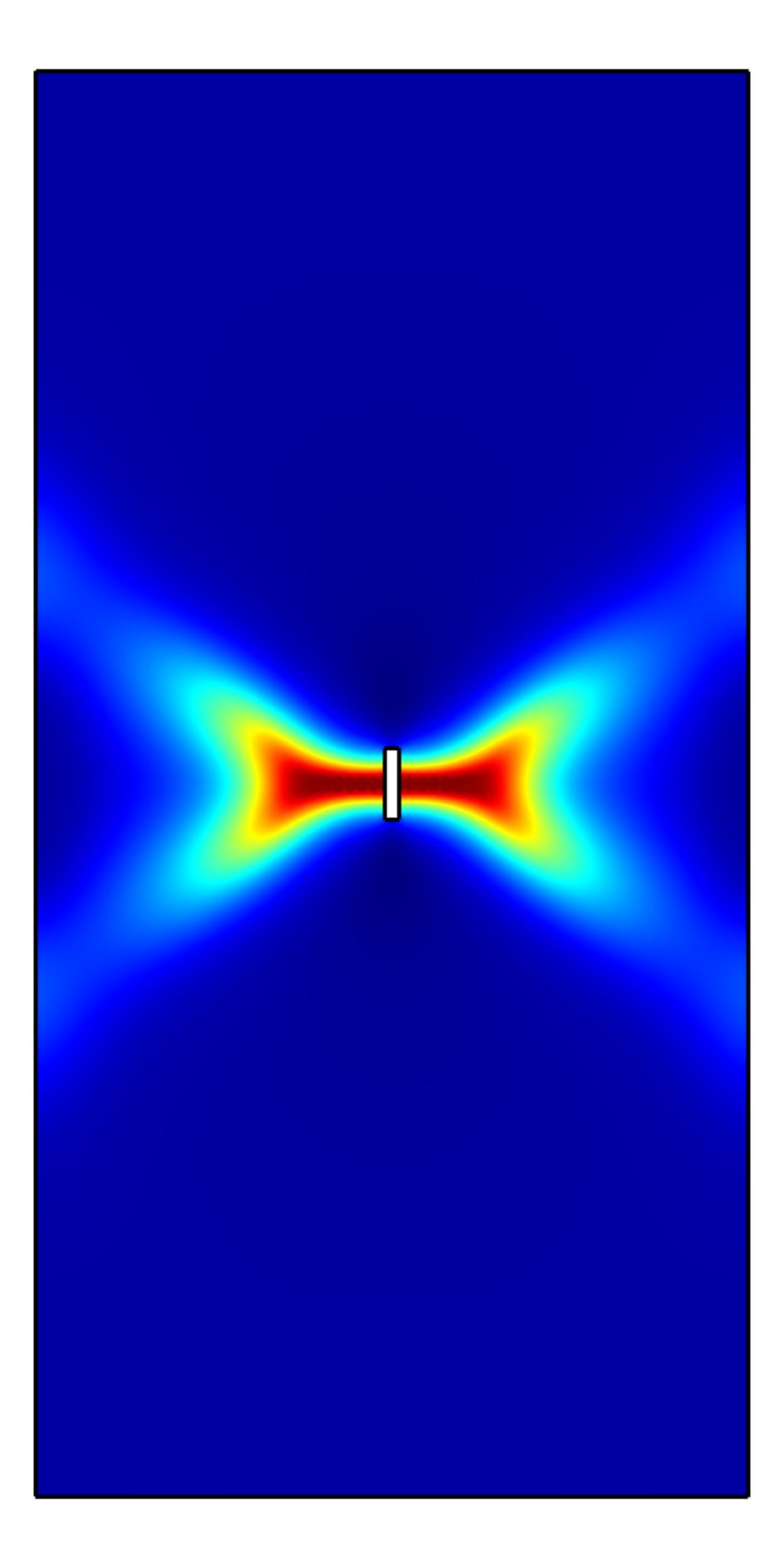}}
	\subfigure[$u = 0.1981$ mm]{\includegraphics[width = 4cm]{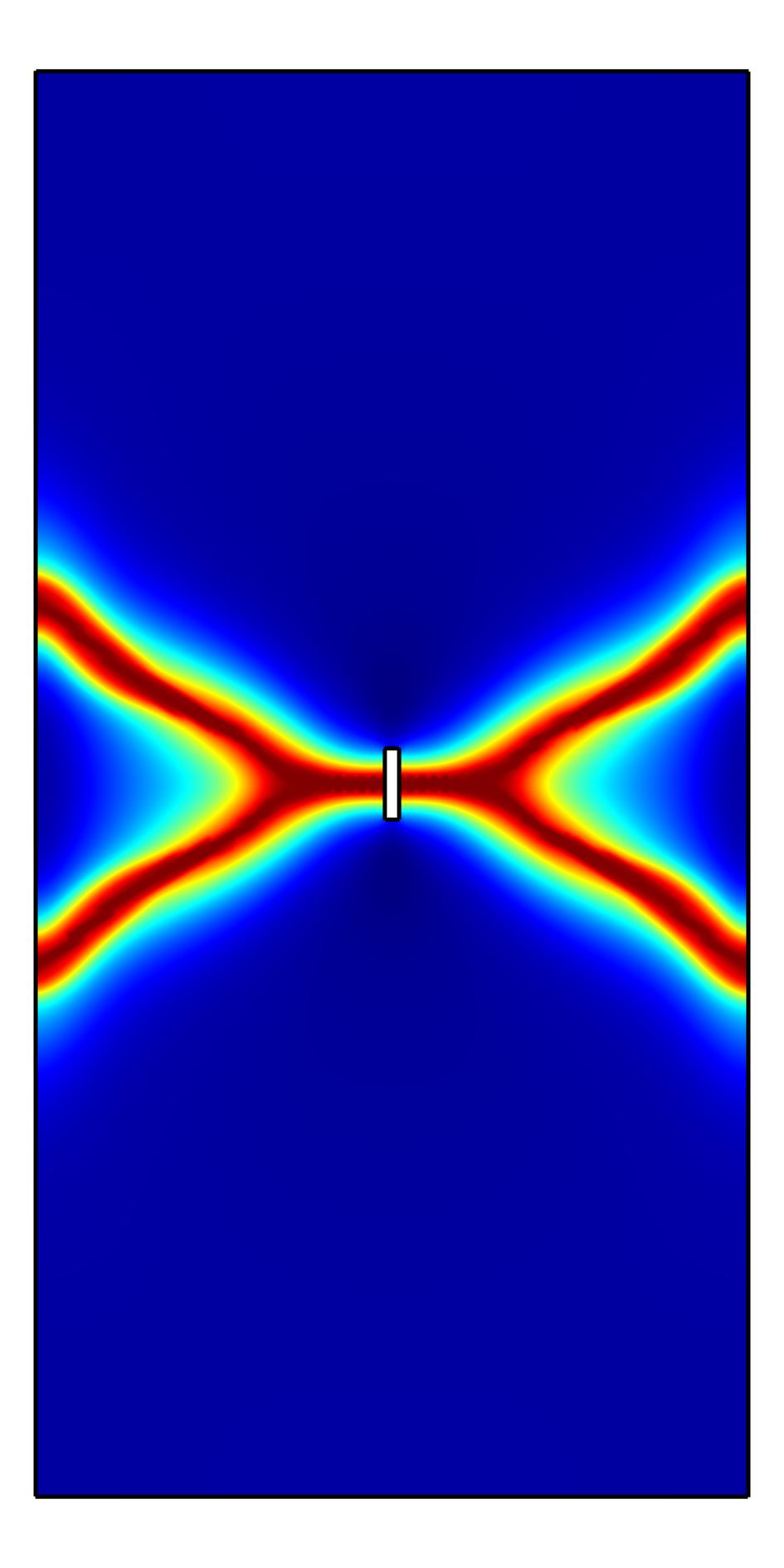}}\\
	\caption{Fracture patterns of a specimen with an inclined flaw of $\alpha=90^\circ$}
	\label{Fracture patterns of a specimen with an inclined flaw alpha = 90}
	\end{figure}

\begin{figure}[htbp]
	\centering
	\includegraphics[width = 5cm]{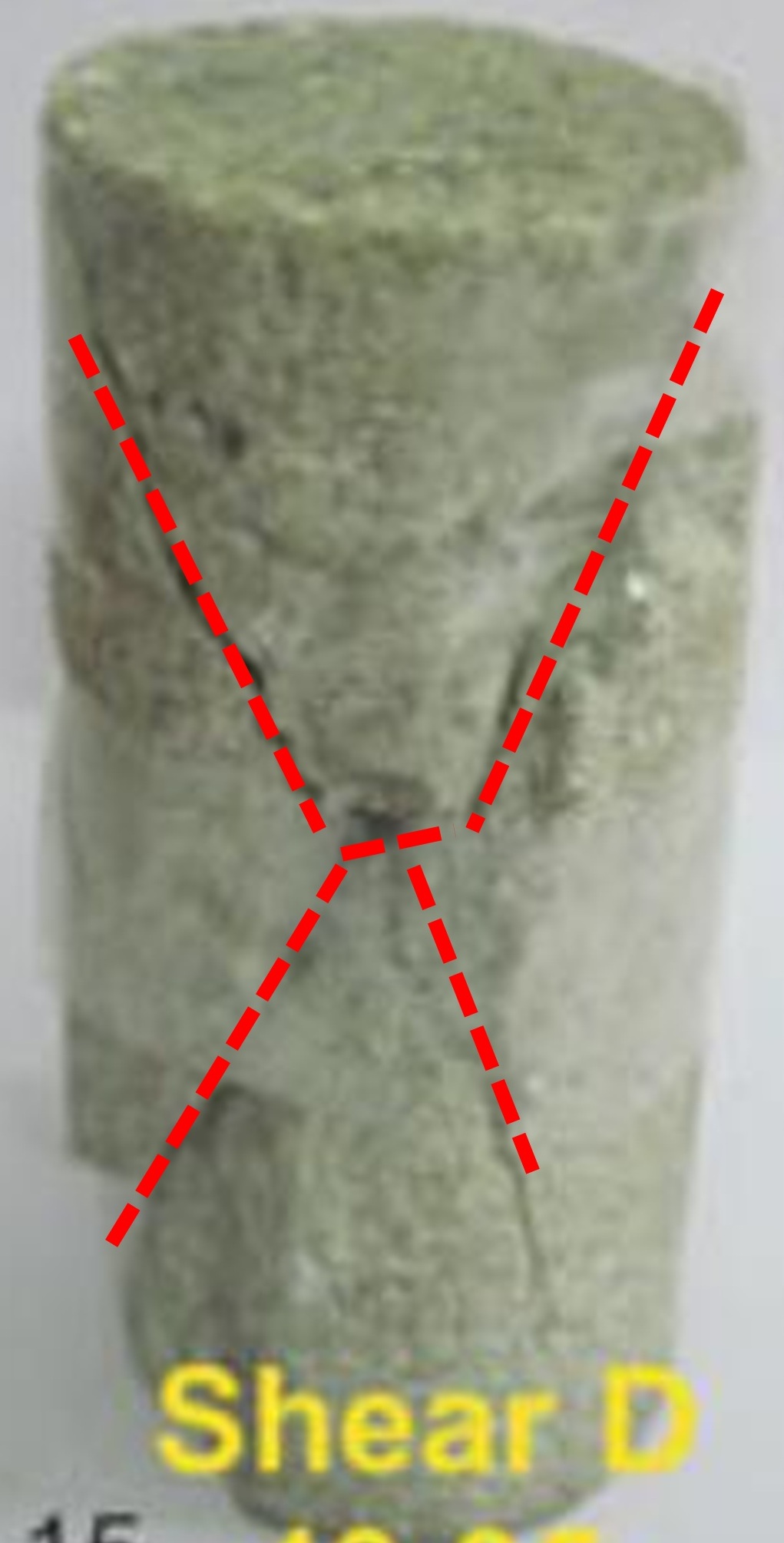}
	\caption{Experimental observation of the double shear fractures in rocks \citep{basu2013rock}}
	\label{Experimental observation of the double shear fractures in rocks}
	\end{figure}

We then test the influence of the inclination angle $\alpha$ on the compressive-shear fracture patterns. The inclination angle varies from $0^\circ$, $15^\circ$, $30^\circ$, $60^\circ$, $75^\circ$, to $90^\circ$ while the other parameters remain unchanged. For $\alpha$ = $15^\circ$, $30^\circ$, $60^\circ$, and $75^\circ$, the progressive fracture process is similar to that for $\alpha = 45^\circ$. All the modeled fractures in these cases belong to the inclined shear fractures. However, for $\alpha = 0^\circ$ and $90^\circ$, the fracture patterns are different from $\alpha = 45^\circ$ because of the presence of symmetry. Figures \ref{Fracture patterns of a specimen with an inclined flaw alpha = 0} and \ref{Fracture patterns of a specimen with an inclined flaw alpha = 90} show the fracture process when $\alpha = 0^\circ$ and $90^\circ$. For $\alpha = 0^\circ$, fractures initiate at the flaw tips when $u = 0.1378$ mm. Subsequently, the fractures propagates horizontally and fall into the category of normal shear fracture when $u = 0.1400$ mm. However, when $u = 0.1406$ mm, the propagating fractures approach the left and right boundaries and a slight fracture branching is shown in the specimen; this phenomenon is more obvious for $\alpha = 90^\circ$. Differently from $\alpha = 0^\circ$, fractures initiate in the middle of the flaw for $\alpha = 90^\circ$ when $u = 0.1976$ mm. Fractures start to branch when $u = 0.1980$ mm and the bifurcated fractures reach the left and right boundaries when $u = 0.1981$ mm. It should be noted that the modeled X-shaped fracture is in good agreement with the experimental observed pattern in Fig. \ref{Experimental observation of the double shear fractures in rocks}, and it is named the double shear fracture in \citet{basu2013rock}.

\begin{figure}[htbp]
	\centering
	\includegraphics[width = 10cm]{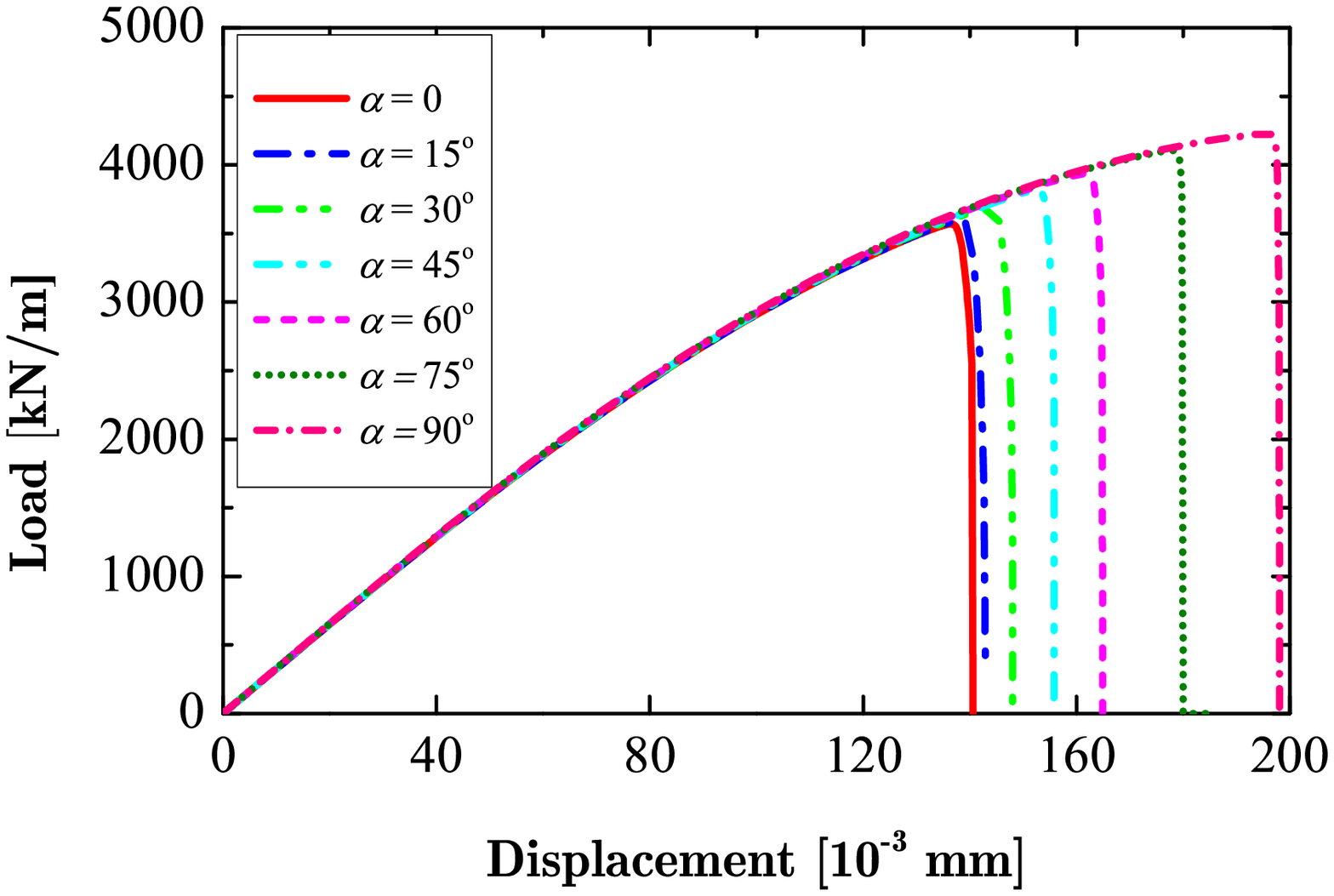}
	\caption{Load-displacement curves of the specimen with an inclined flaw under different angles}
	\label{Load-displacement curves of the specimen with an inclined flaw under different angles}
	\end{figure}

\begin{figure}[htbp]
	\centering
	\subfigure[$l_0 = 0.5$ mm]{\includegraphics[width = 4cm]{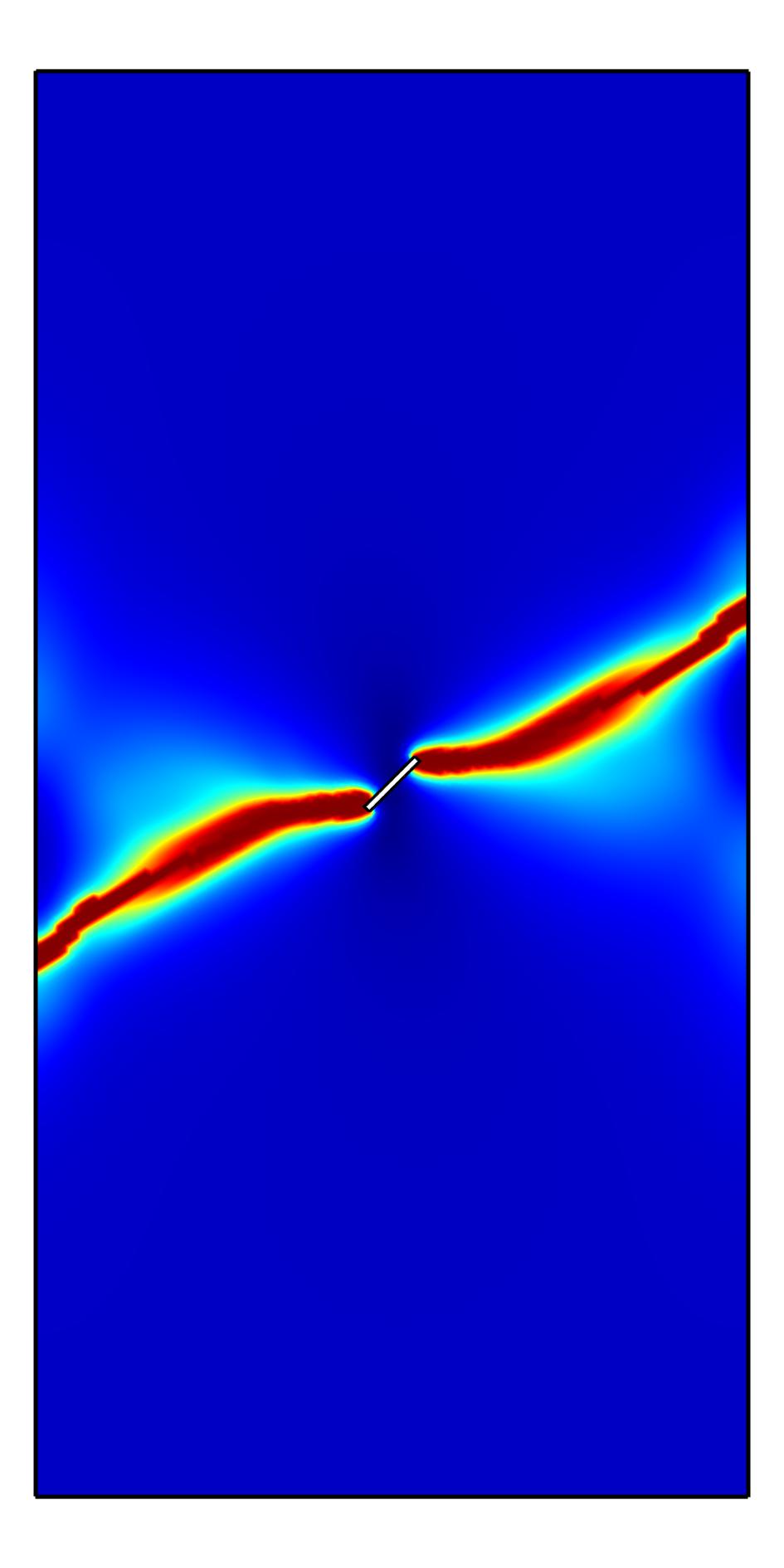}}
	\subfigure[$l_0 = 1$ mm]{\includegraphics[width = 4cm]{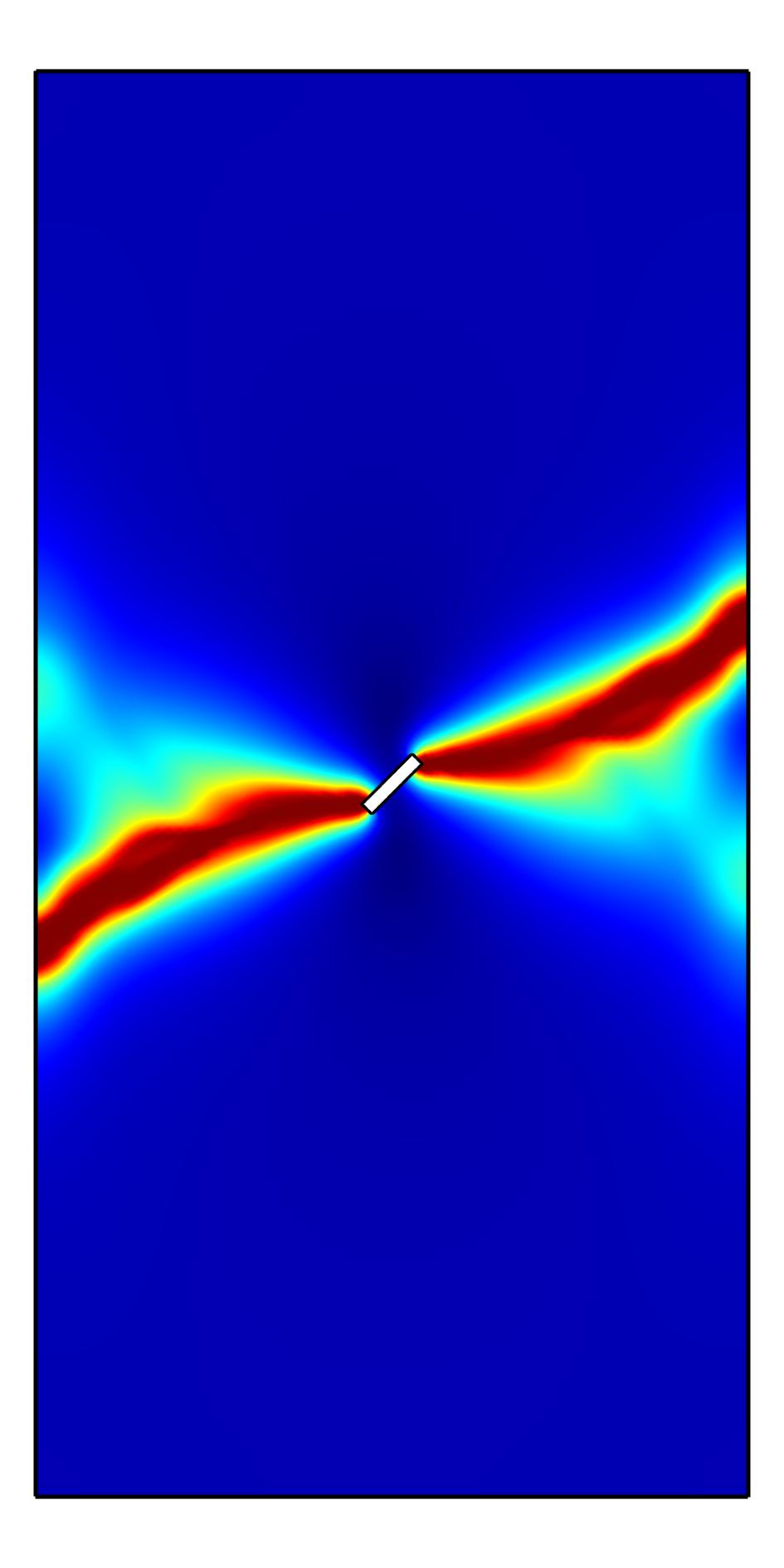}}\\
	\caption{Fracture patterns of a specimen with an inclined flaw ($\alpha=45^\circ$) under different $l_0$}
	\label{Fracture patterns of a specimen with an inclined flaw alpha under different l0}
	\end{figure}

\begin{figure}[htbp]
	\centering
	\includegraphics[width = 10cm]{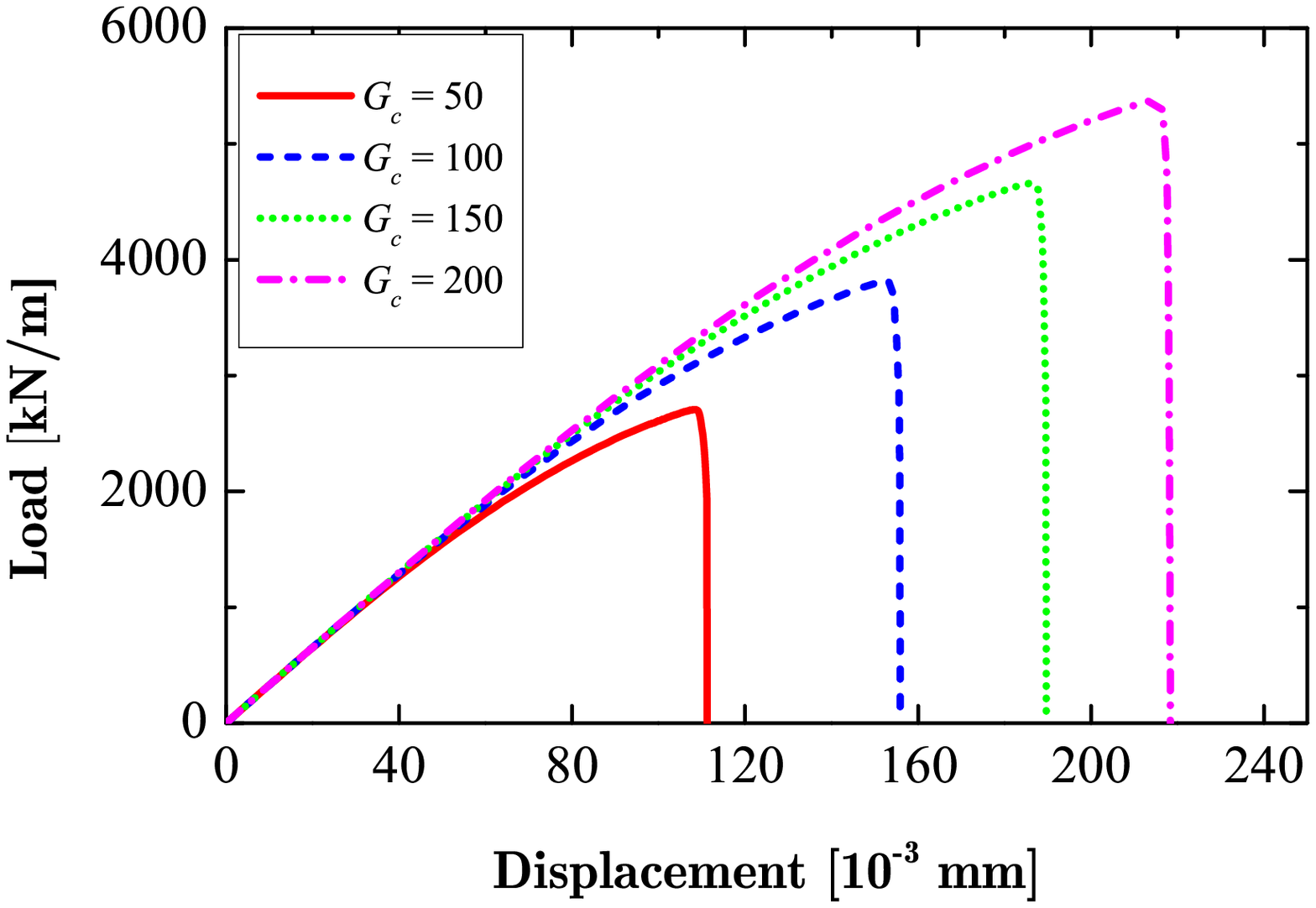}
	\caption{Load-displacement curves of the specimen with an inclined flaw under different $G_c$}
	\label{Load-displacement curves of the specimen with an inclined flaw under different energy release rate}
	\end{figure}

\begin{figure}[htbp]
	\centering
	\includegraphics[width = 10cm]{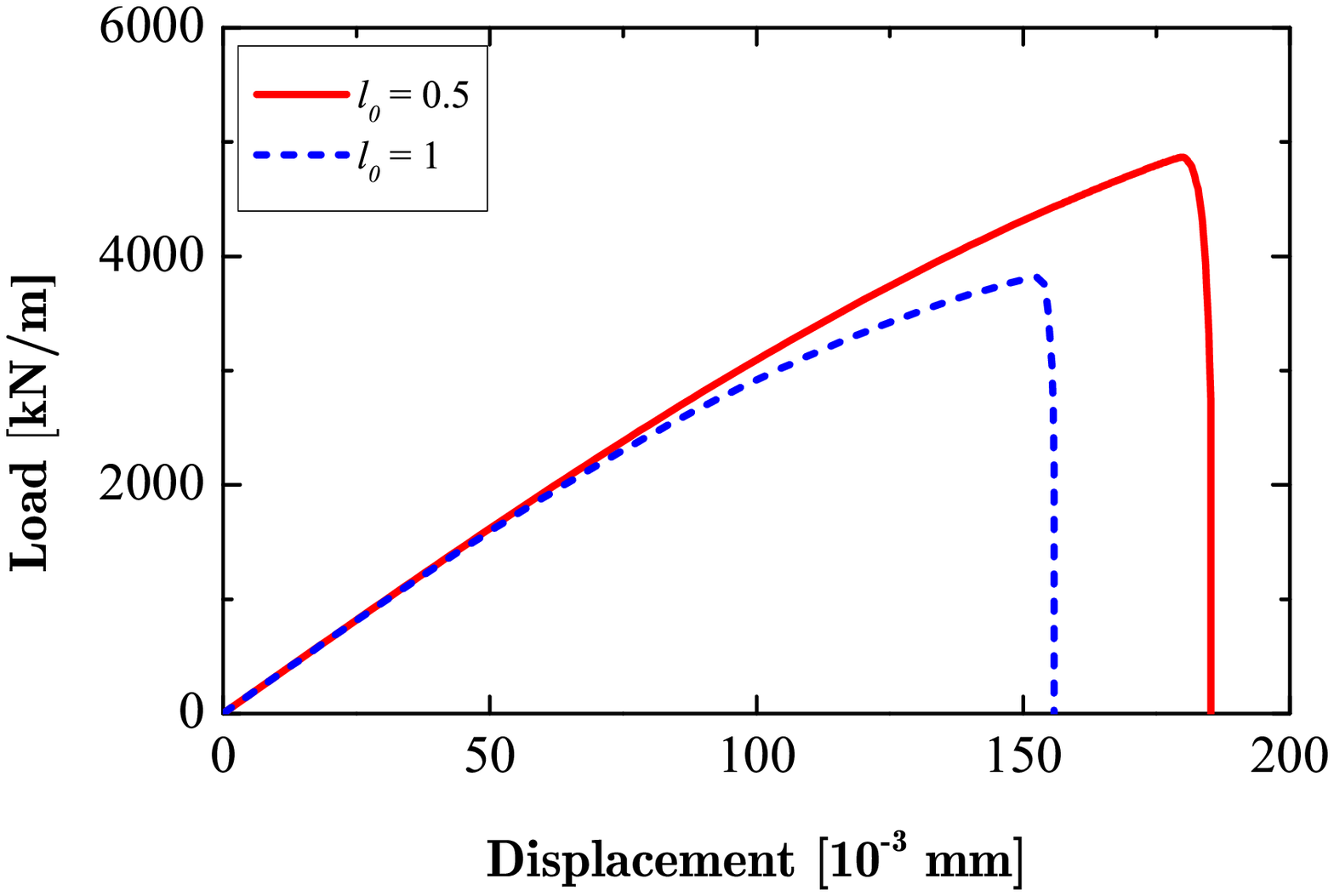}
	\caption{Load-displacement curves of the specimen with an inclined flaw under different $l_0$}
	\label{Load-displacement curves of the specimen with an inclined flaw under different length scale parameter}
	\end{figure}

The load-displacement curves of the specimen with an inclined flaw under different angles are shown in Fig. \ref{Load-displacement curves of the specimen with an inclined flaw under different angles}. As observed, the maximum load that the specimen can sustain increases with the increase in the inclination angle under the compressive-shear mode. Moreover, the maximum displacement of the specimen increases because the factual load-bearing area increases as the angle of inclination increases. The influence of the energy release rate $G_c$ and length scale parameter $l_0$ on the fracture propagation are also tested. Herein, we set $G_c$ = 100, 200, 300, and 400 N/m and $l_0$ = 0.5 mm and 1 mm while the other base parameters are fixed. It is found that the fracture patterns are not affected by the energy release rate and length scale. However, a larger length scale $l_0$ produces a larger fracture width, as shown in Fig. \ref{Fracture patterns of a specimen with an inclined flaw alpha under different l0}. In addition, the load-displacement curves under different $G_c$ and $l_0$ are shown in Figs. \ref{Load-displacement curves of the specimen with an inclined flaw under different energy release rate} and \ref{Load-displacement curves of the specimen with an inclined flaw under different length scale parameter}. As expected, the peak load of the specimen increases with an increasing $G_c$ but a decreasing $l_0$. Figure \ref{Load-displacement curves of the specimen with an inclined flaw under different length scale parameter} indicates that the proposed PFM is sensitive to the length scale parameter and in the future the new driving force proposed in this work can be coupled to the length-scale insensitive phase field framework proposed by \citet{mandal2019phase, wu2018length} to overcome this kind of sensitivity.

\begin{figure}[htbp]
	\centering
	\includegraphics[width = 10cm]{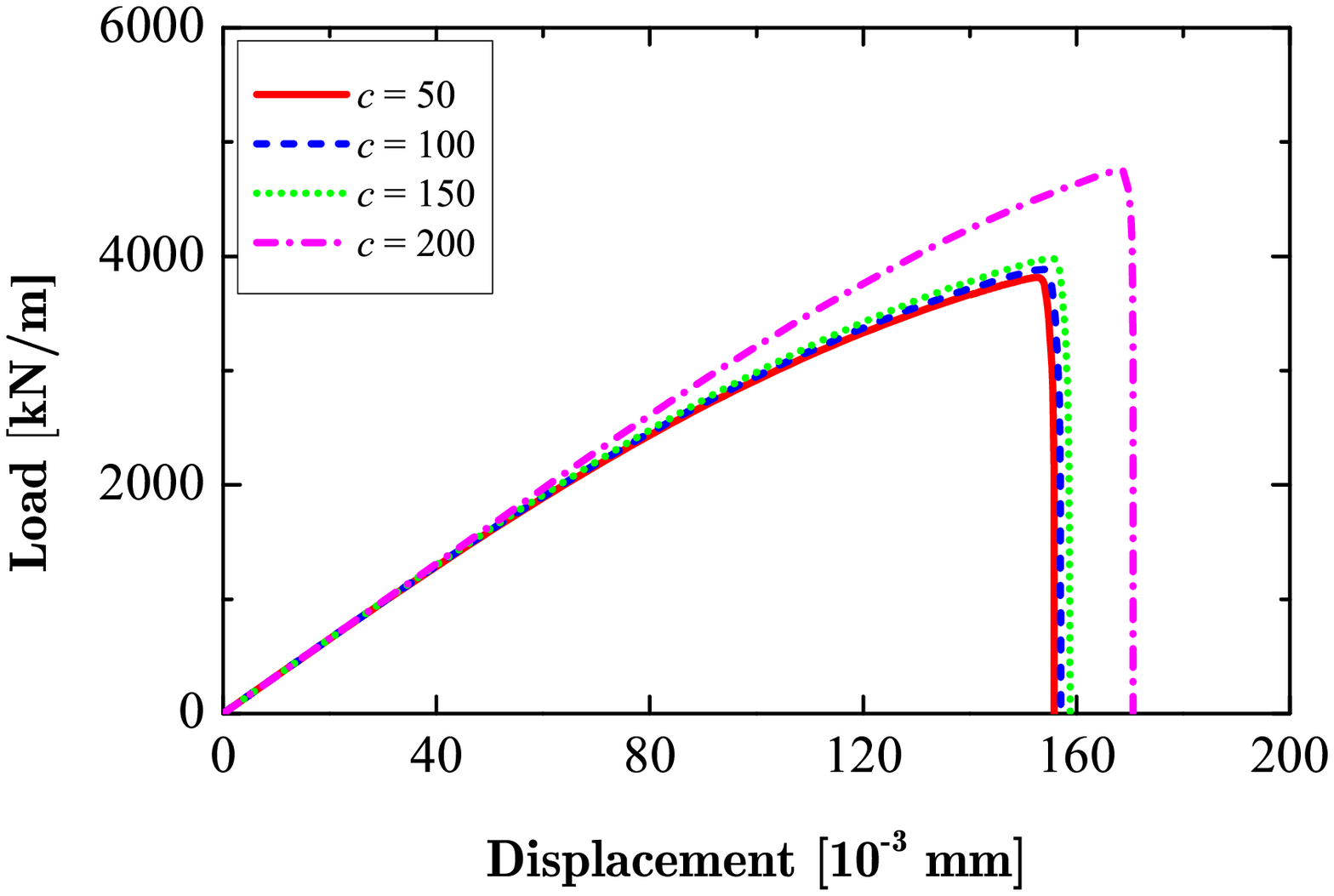}
	\caption{Load-displacement curves of the specimen with an inclined flaw under different $c$}
	\label{Load-displacement curves of the specimen with an inclined flaw under different cohesion}
	\end{figure}

\begin{figure}[htbp]
	\centering
	\includegraphics[width = 10cm]{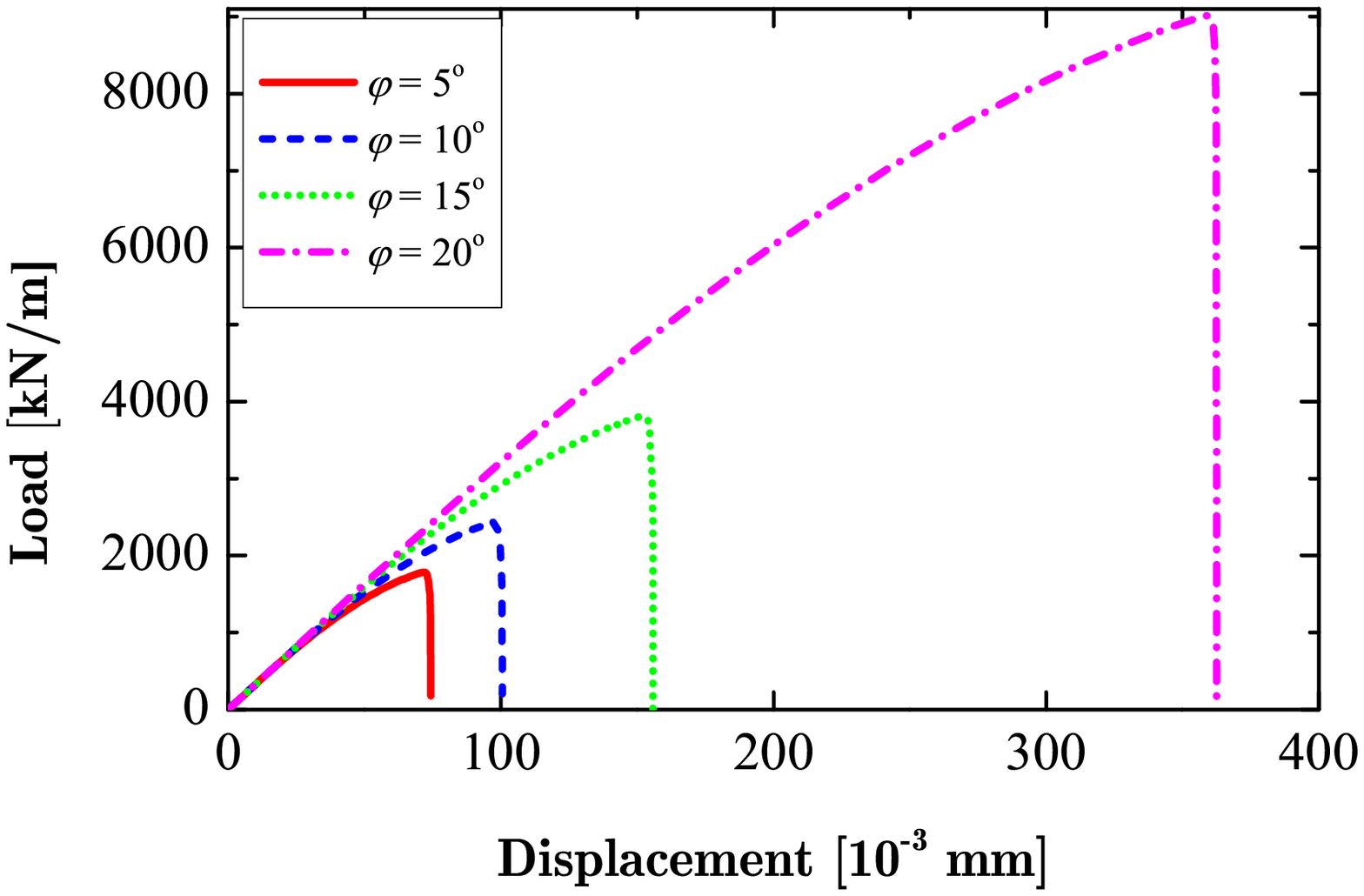}
	\caption{Load-displacement curves of the specimen with an inclined flaw under different $\varphi$}
	\label{Load-displacement curves of the specimen with an inclined flaw under different internal friction angle}
	\end{figure}

Finally, we test the influence of the cohesion $c$, internal friction angle $\varphi$, and eccentricity $e$ on the compressive-shear fracture pattern. We adopt $c=$ 100, 500, 1000 and 5000 kPa, $\varphi=5^\circ$, $10^\circ$, $15^\circ$ and $20^\circ$, and $e$ = 0, 5 and 10 mm while keeping the other base parameters unchanged. The numerical results show that the fracture pattern is not affected by the cohesion and internal friction angle. In addition, the load-displacement curves of the specimen under different $c$ and $\varphi$ are presented in Figs. \ref{Load-displacement curves of the specimen with an inclined flaw under different cohesion} and \ref{Load-displacement curves of the specimen with an inclined flaw under different internal friction angle}, respectively. As observed, the peak loads of the specimen increase with the increase in cohesion $c$ and internal friction angle $\varphi$, which accurately reflects the compressive-shear nature in rock fractures and corresponds to those experimental observations in rock tests. 

\begin{figure}[htbp]
	\centering
	\subfigure[$e = 0$]{\includegraphics[width = 4cm]{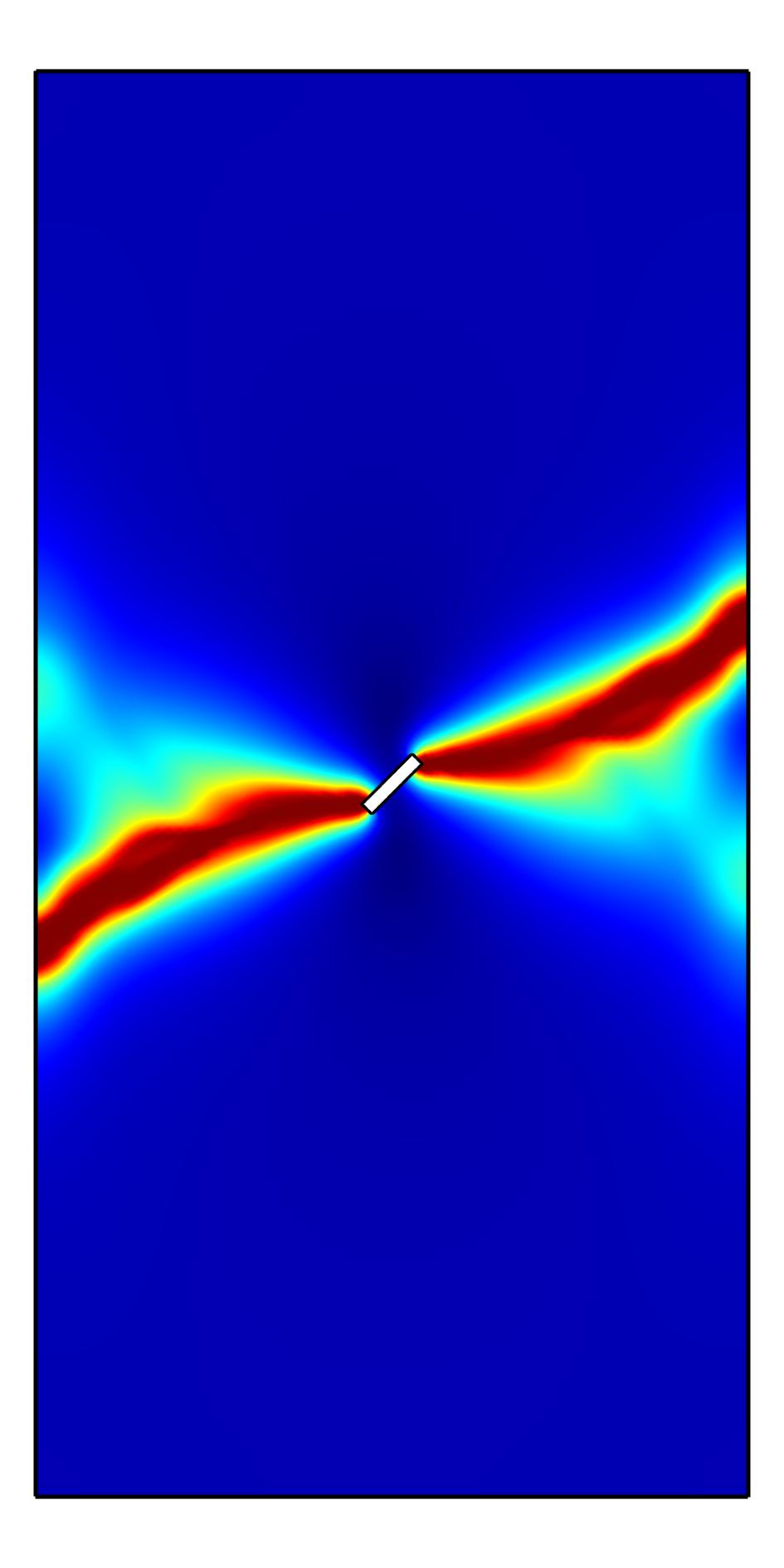}}
	\subfigure[$e = 5$ mm]{\includegraphics[width = 4cm]{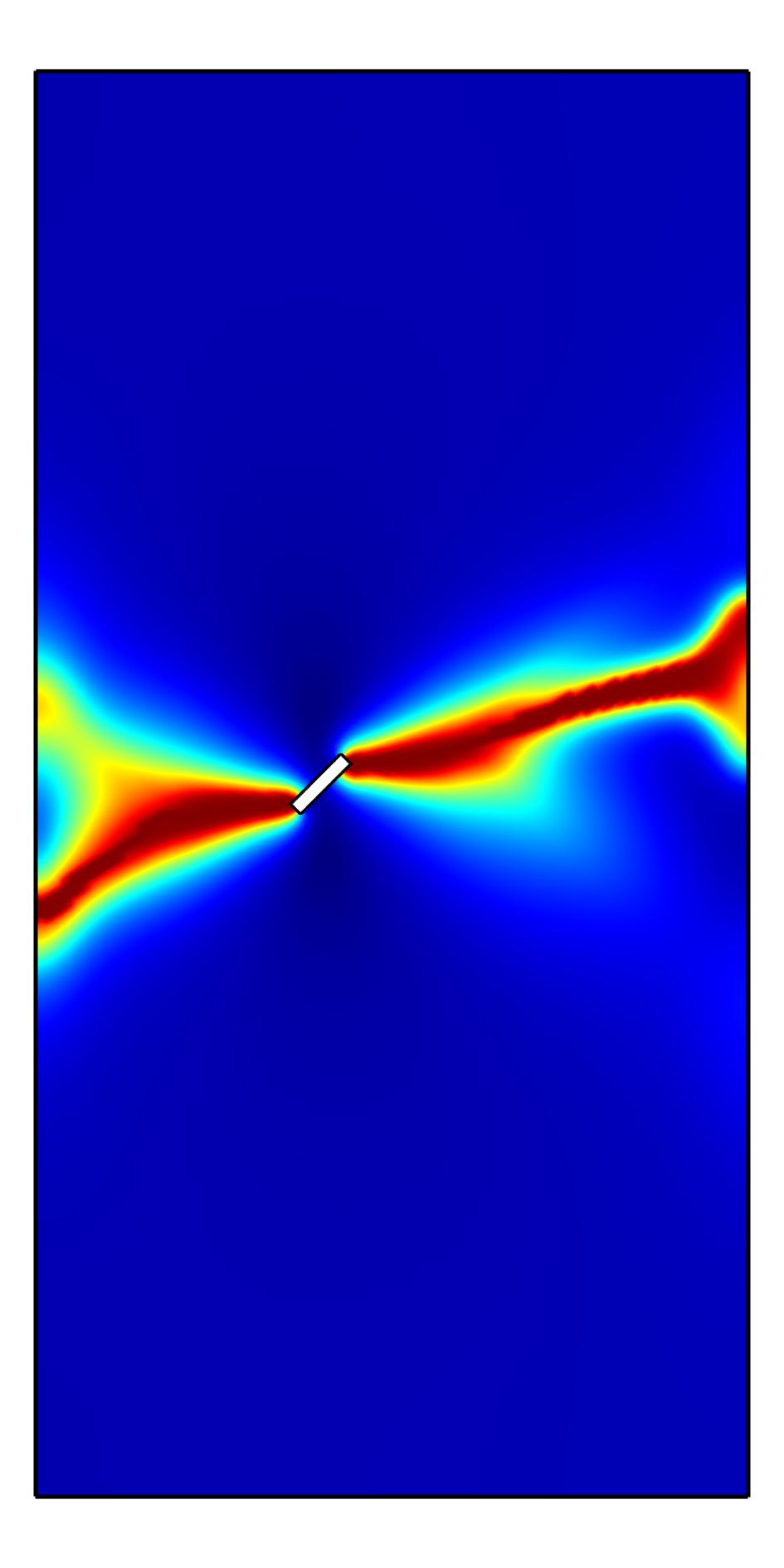}}
	\subfigure[$e = 10$ mm]{\includegraphics[width = 4cm]{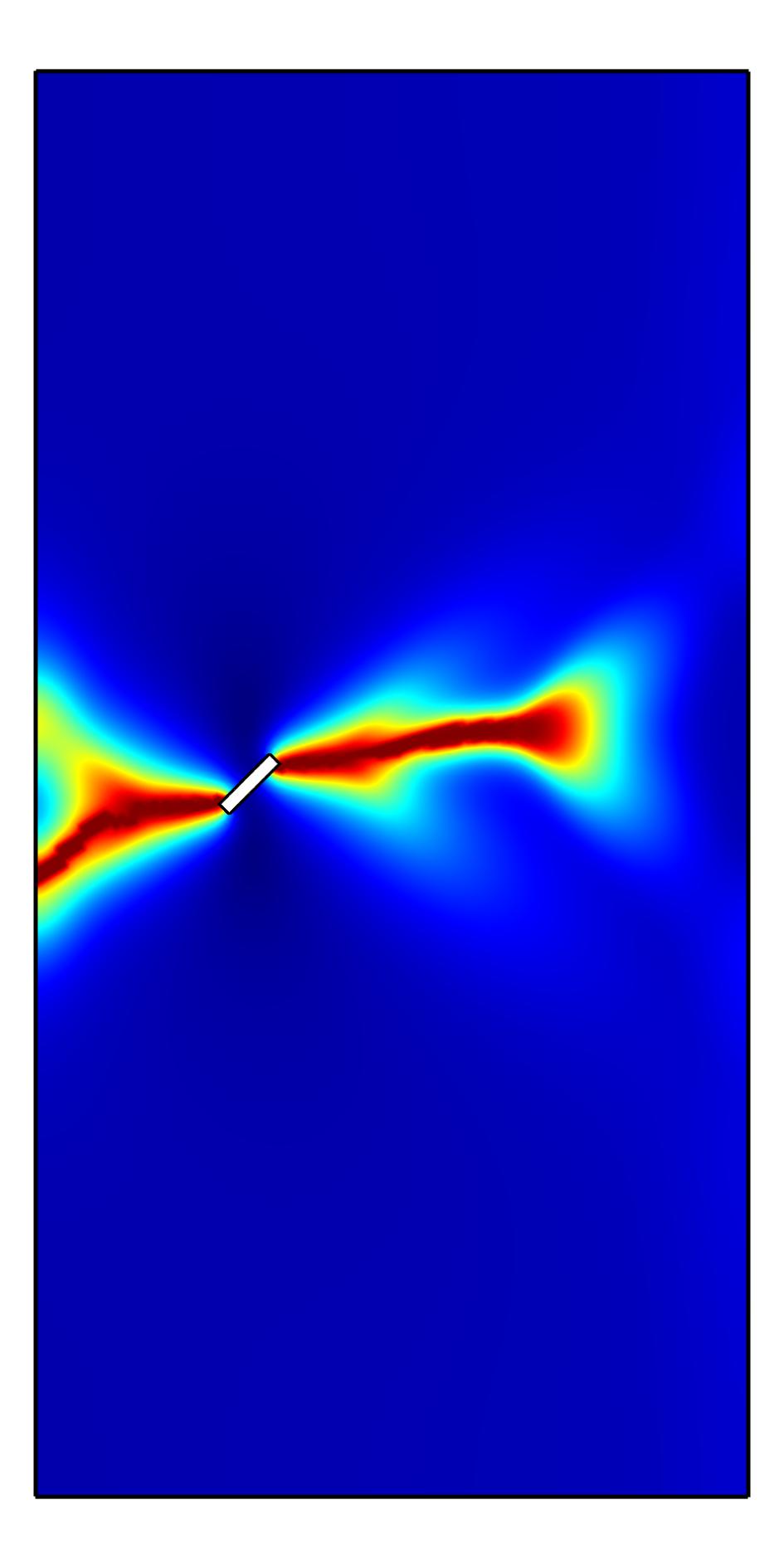}}\\
	\caption{Fracture patterns of a specimen with an inclined flaw ($\alpha=45^\circ$) under different $e$}
	\label{Fracture patterns of a specimen with an inclined flaw alpha under different eccentricity}
	\end{figure}

\begin{figure}[htbp]
	\centering
	\includegraphics[width = 10cm]{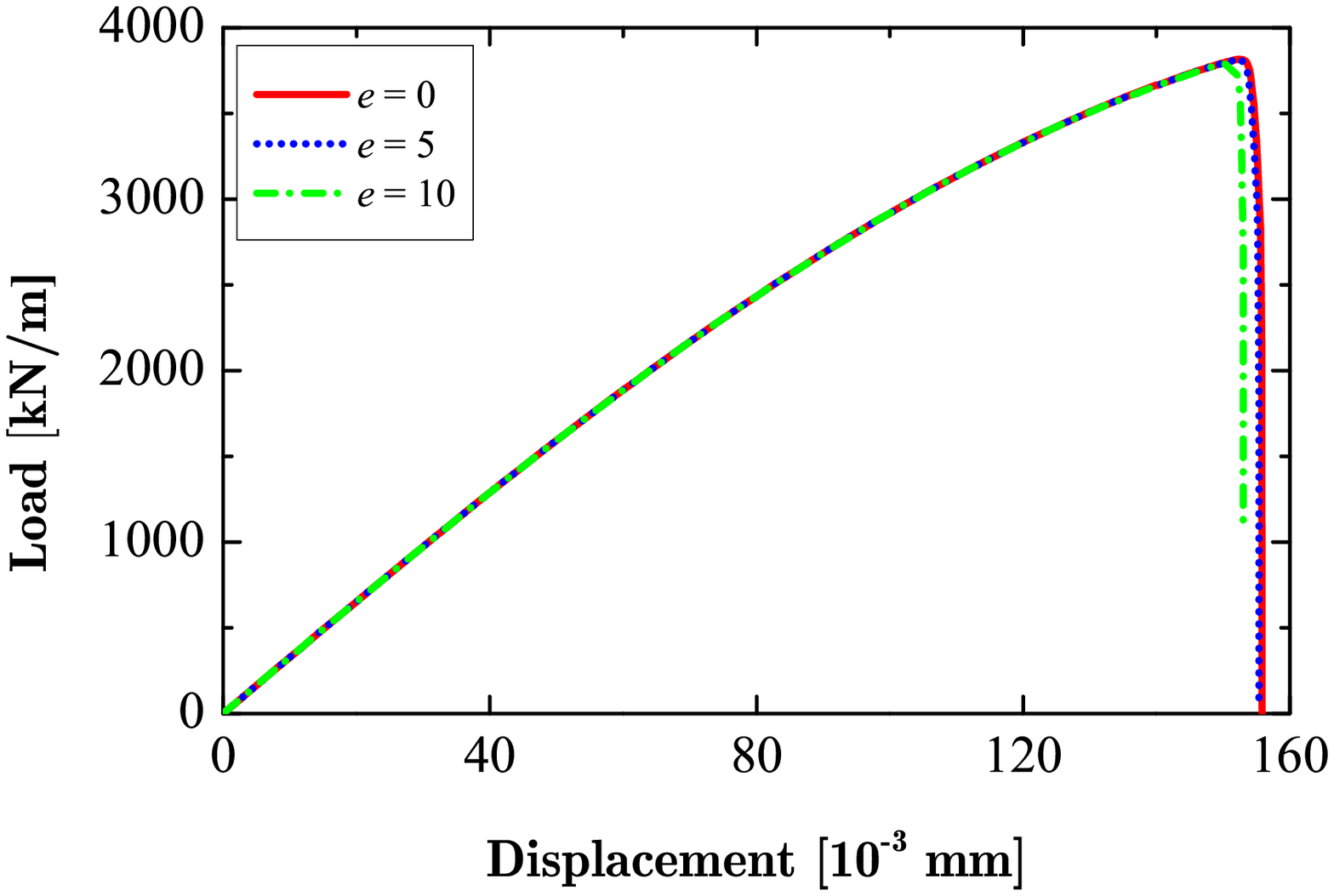}
	\caption{Load-displacement curves of the specimen with an inclined flaw under different $e$}
	\label{Load-displacement curves of the specimen with an inclined flaw under different eccentricity}
	\end{figure}

Figure \ref{Fracture patterns of a specimen with an inclined flaw alpha under different eccentricity} shows the fracture patterns of the specimen under different eccentricity $e$. With the increase in the eccentricity $e$, the compressive-shear fracture is found to propagate at a decreasing inclination angle to the horizontal direction. However, the load-displacement curves in Fig. \ref{Load-displacement curves of the specimen with an inclined flaw under different eccentricity} indicates an ignorable decrease in the peak load of the specimen because the factual load-bearing area under compression is almost unchanged under different $e$. In summary, the proposed phase field method can model well the compressive-shear fractures in the rock-like specimen with an inclined flaw, and can reflect the increase in the load-bearing capacity under compression as the intrinsic cohesion and internal friction angle of the materials increase.

\subsection{Uniaxial compression of a specimen with two parallel inclined flaws}

\begin{figure}[htbp]
	\centering
	\includegraphics[width = 12cm]{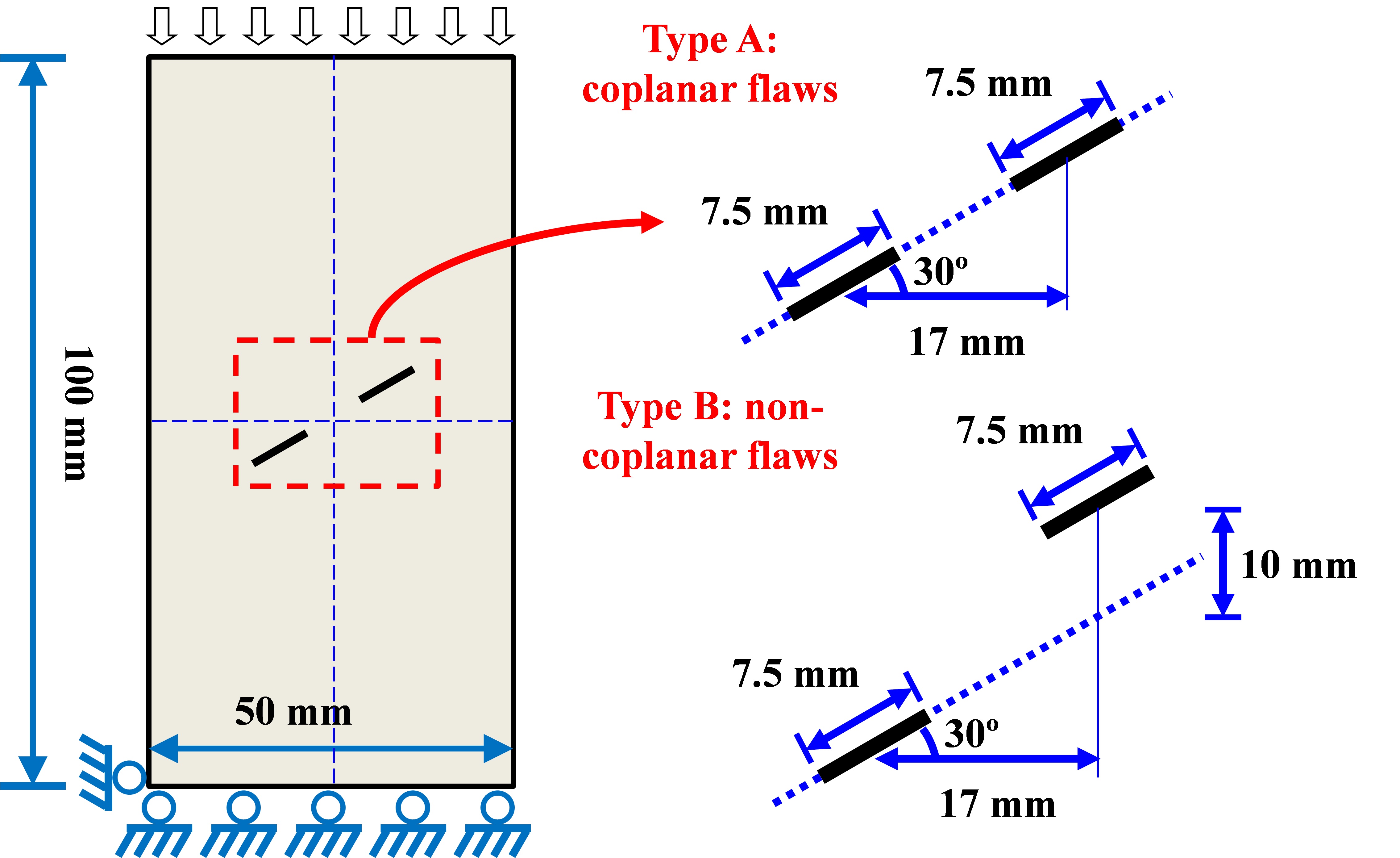}
	\caption{Geometry and boundary conditions of a rectangular specimen with two parallel inclined flaws subjected to uniaxial compression}
	\label{Geometry of a rectangular specimen with two parallel inclined flaws subjected to uniaxial compression}
	\end{figure}

In this final example, we test the compressive-shear fractures in a rock-like specimen with two parallel inclined flaws. Uniaxial compression is applied on the upper end of the specimen along with the geometry of the problem shown in Fig. \ref{Geometry of a rectangular specimen with two parallel inclined flaws subjected to uniaxial compression}. The inclination angle, length, and width of the flaws are $30^\circ$, 7.5 mm, and 1 mm, respectively. Moreover, we set two types of flaw arrangement in Fig. \ref{Geometry of a rectangular specimen with two parallel inclined flaws subjected to uniaxial compression}, namely, the coplanar (Type A) and non-coplanar (Type B) pre-existing flaws. These parameters are used for the phase field simulations: Young's modulus $E$ = 60 GPa, Poisson's ratio $\nu$ = 0.3, critical energy release rate $G_c$ = 100 N/m, length scale parameter $l_0$ = 1 mm, $k=1\times10^{-9}$, cohesion $c$ = 100 kPa, and internal friction angle $\varphi=15^\circ$. The specimen with two flaws are then discretized using triangular elements with the maximum element size $h$ = 0.5 mm. The simulation is proceeded under displacement control and we use a displacement increment $\Delta u = 1\times 10^{-4}$ mm in this example.

\begin{figure}[htbp]
	\centering
	\subfigure[$u = 0.1178$ mm]{\includegraphics[width = 4cm]{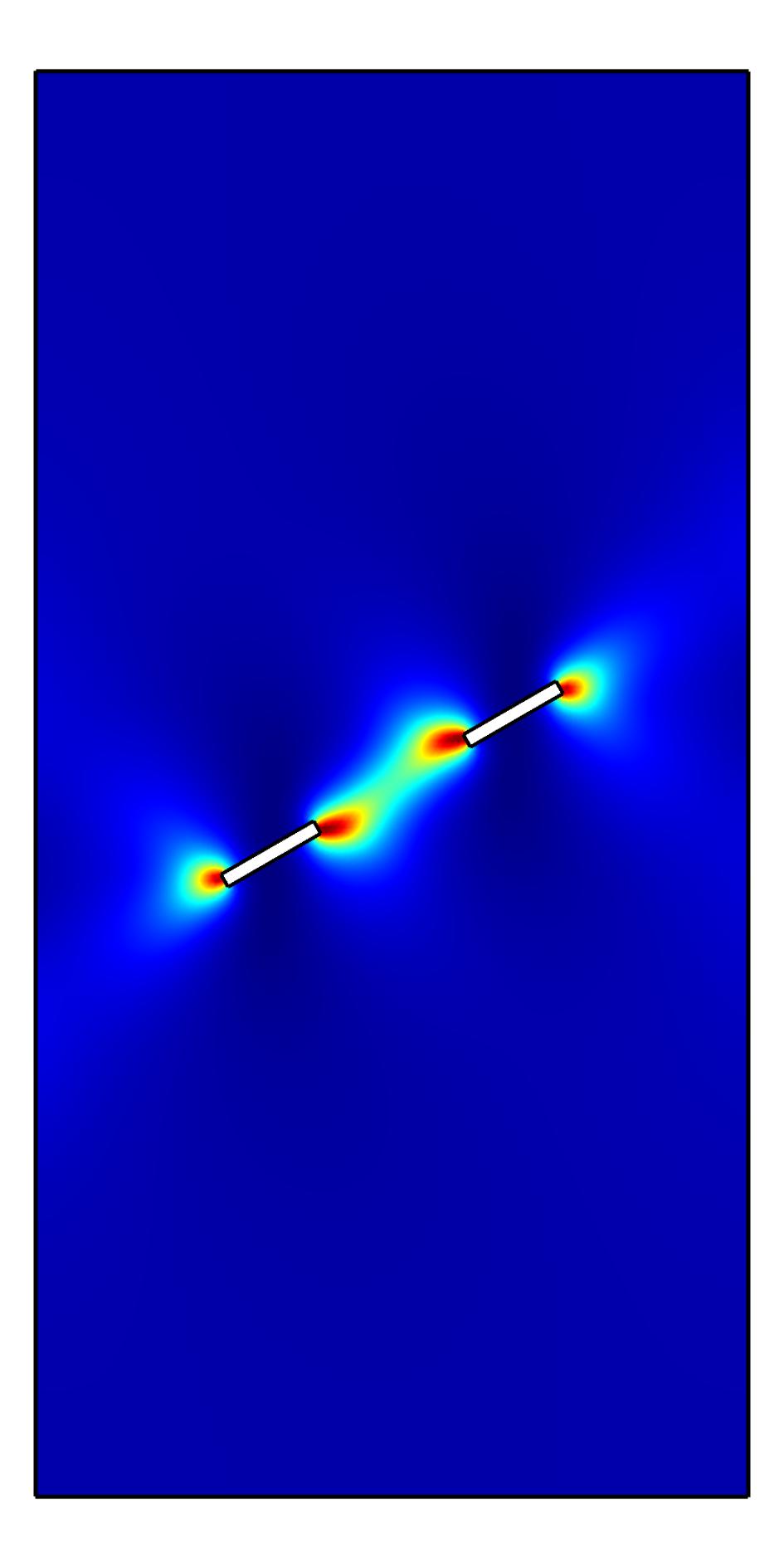}}
	\subfigure[$u = 0.1184$ mm]{\includegraphics[width = 4cm]{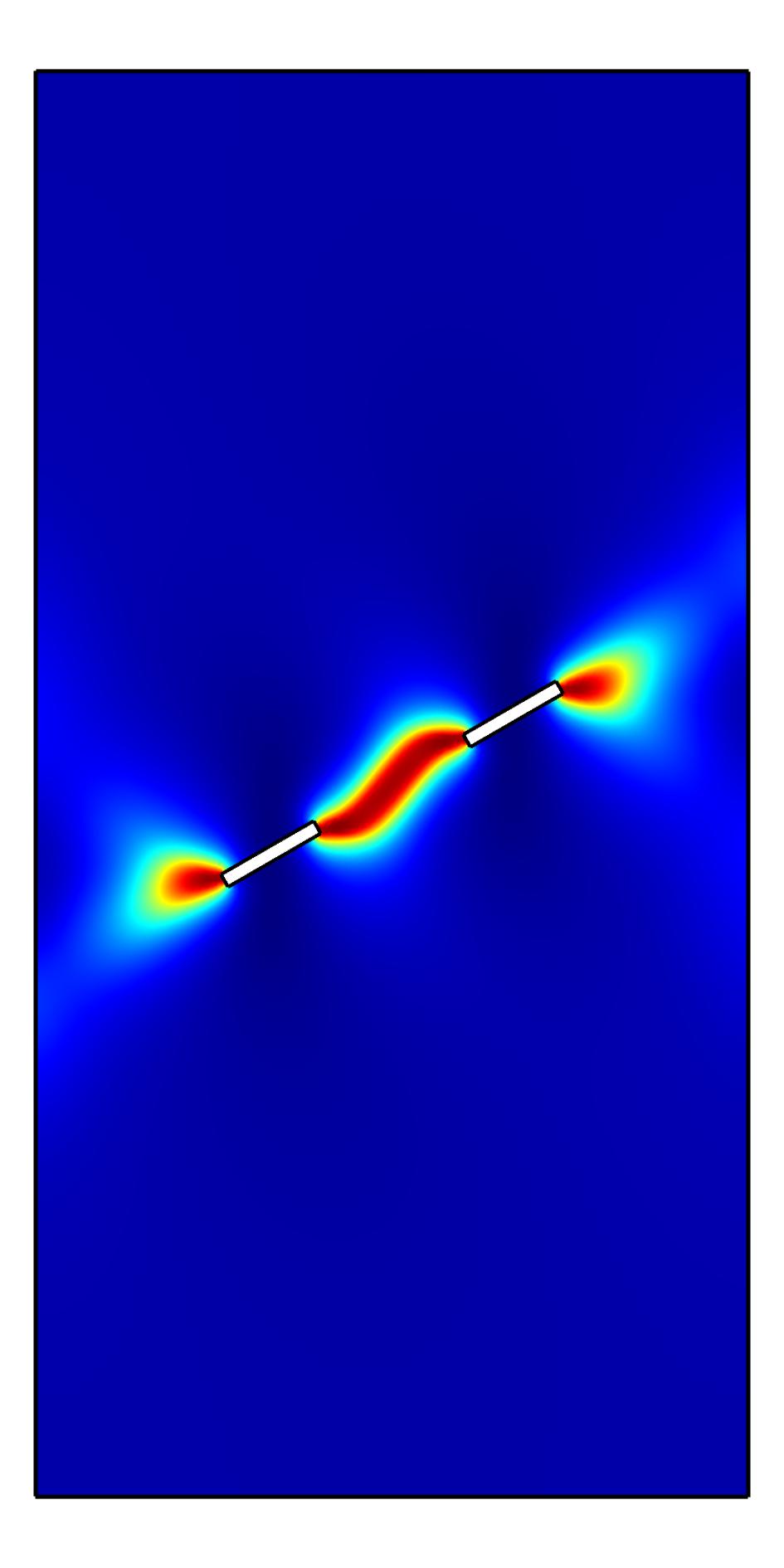}}
	\subfigure[$u = 0.1200$ mm]{\includegraphics[width = 4cm]{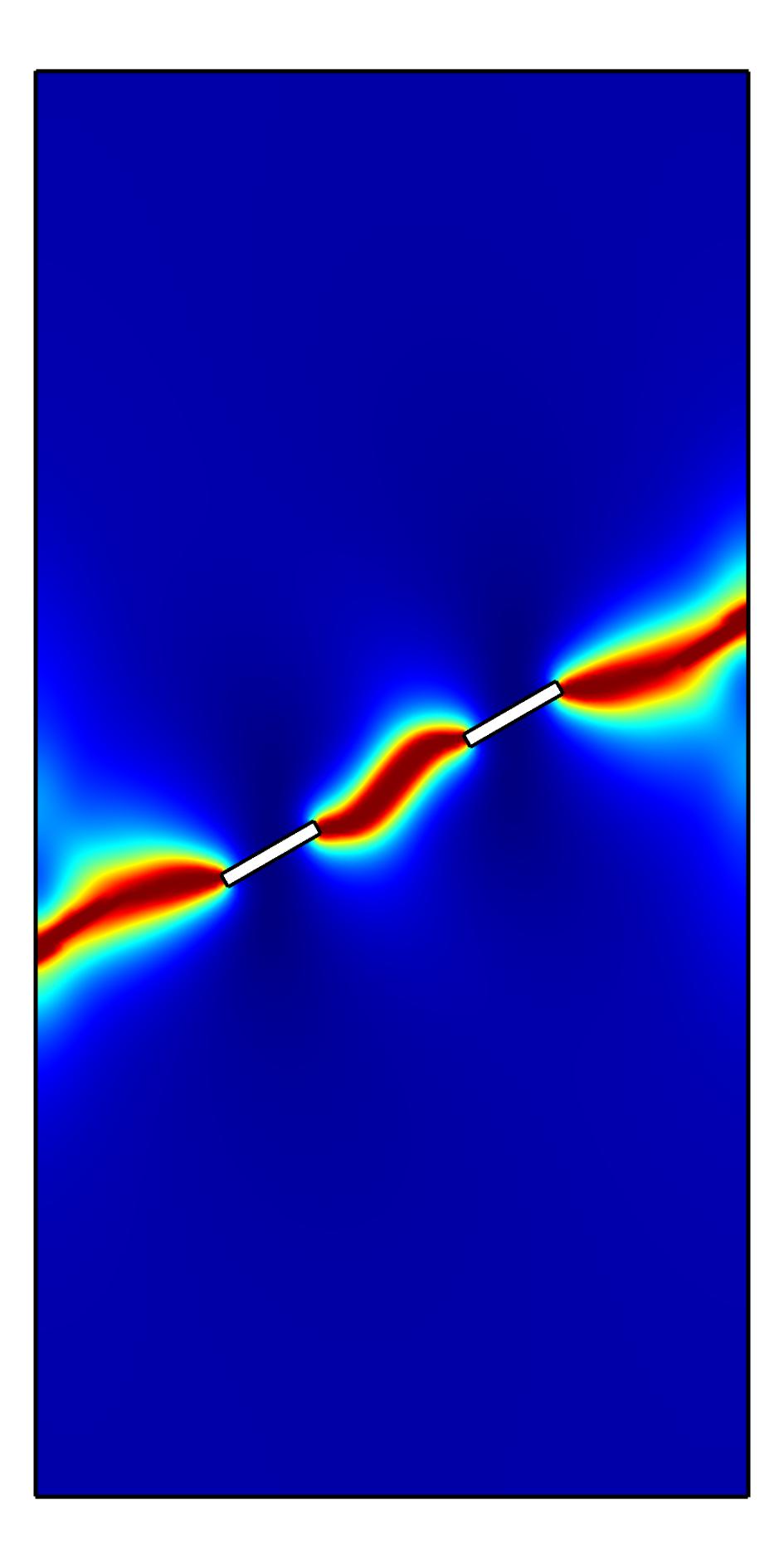}}\\
	\caption{Fracture patterns of a specimen with two parallel inclined flaws (coplanar flaws)}
	\label{Fracture patterns of a specimen with two parallel inclined flaws (coplanar flaws)}
	\end{figure}

\begin{figure}[htbp]
	\centering
	\subfigure[$u = 0.1244$ mm]{\includegraphics[width = 4cm]{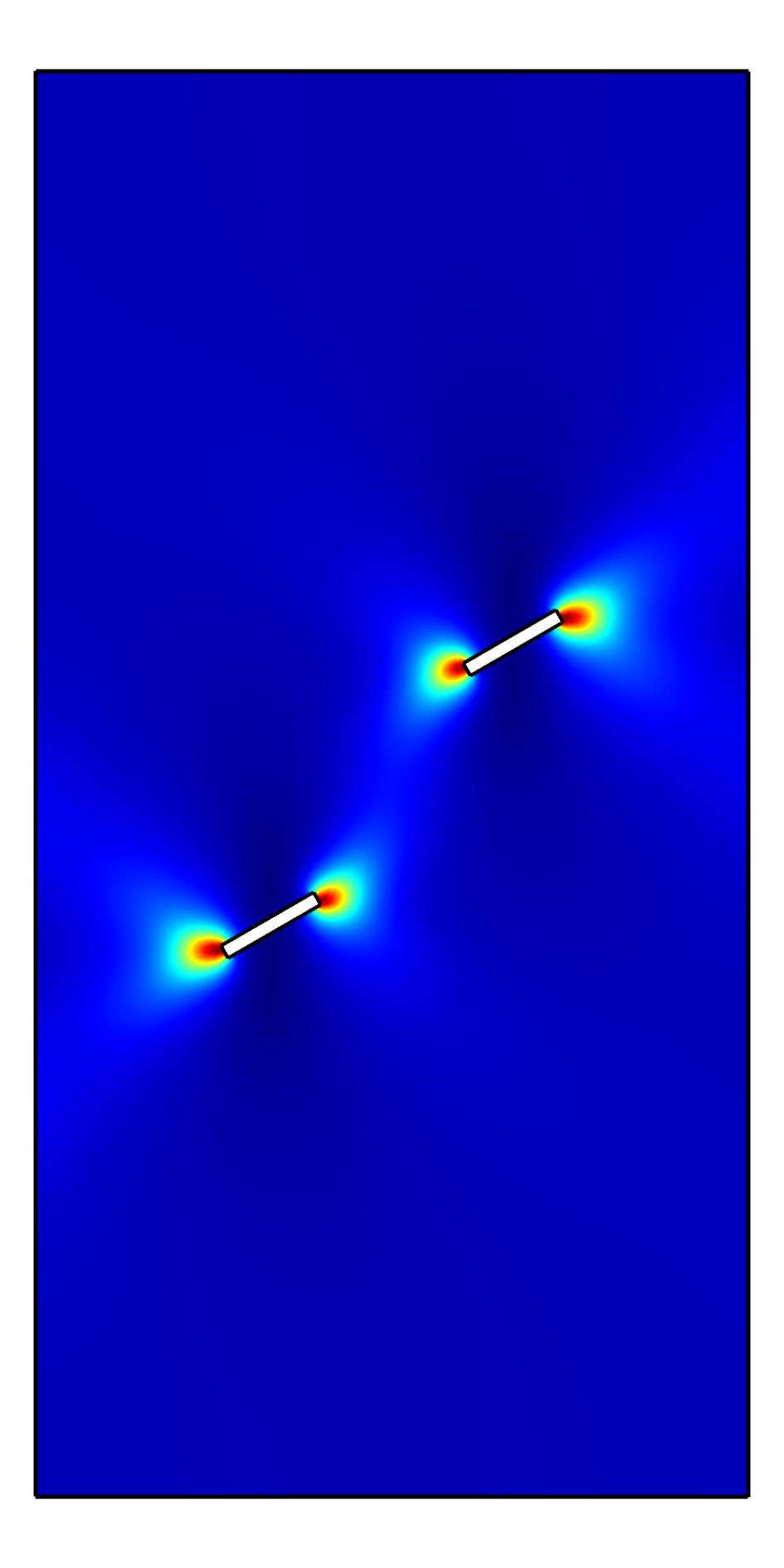}}
	\subfigure[$u = 0.1256$ mm]{\includegraphics[width = 4cm]{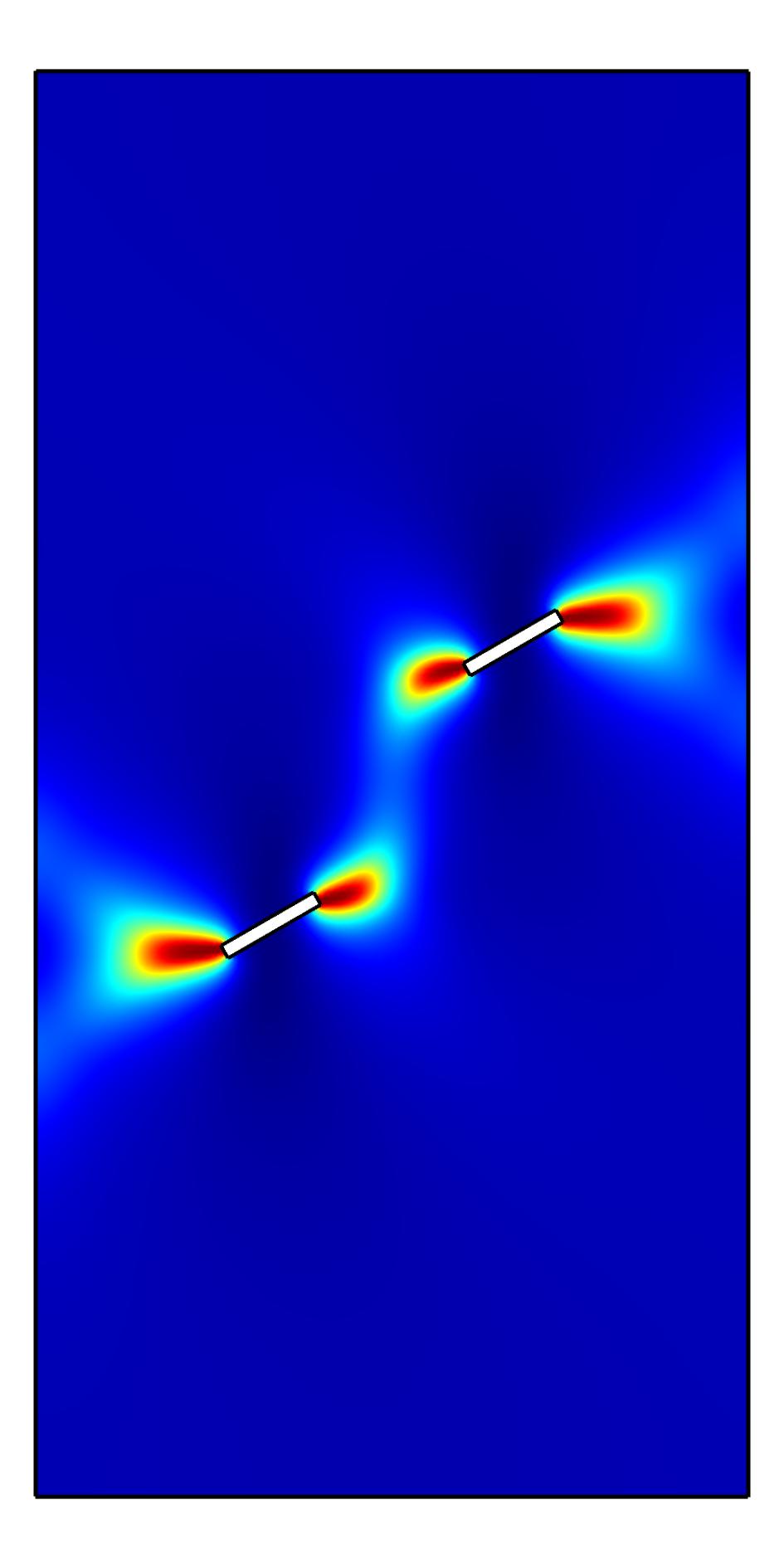}}
	\subfigure[$u = 0.1258$ mm]{\includegraphics[width = 4cm]{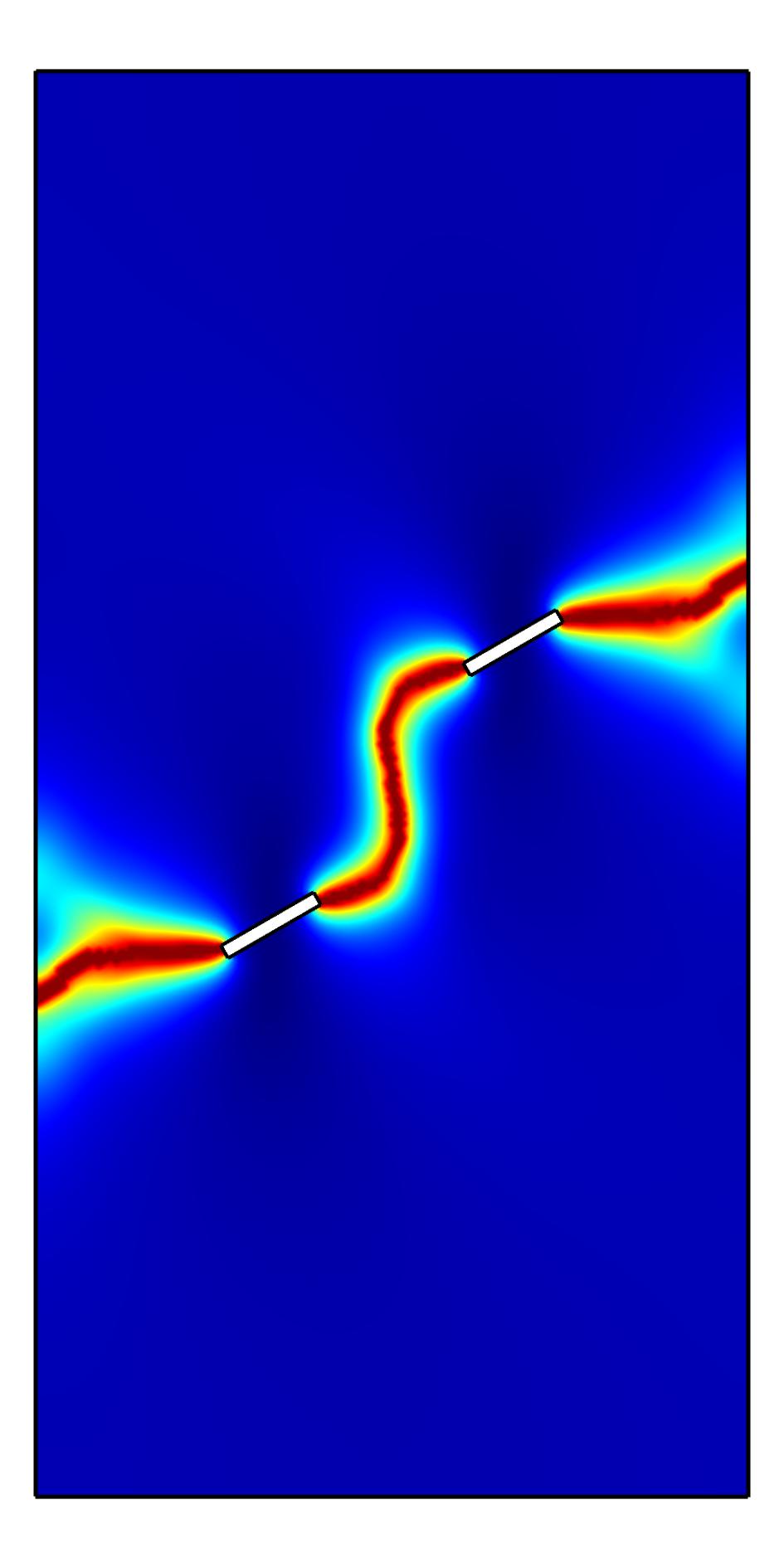}}\\
	\caption{Fracture patterns of a specimen with two parallel inclined flaws (non-coplanar flaws)}
	\label{Fracture patterns of a specimen with two parallel inclined flaws (non-coplanar flaws)}
	\end{figure}

\begin{figure}[htbp]
	\centering
	\includegraphics[width = 10cm]{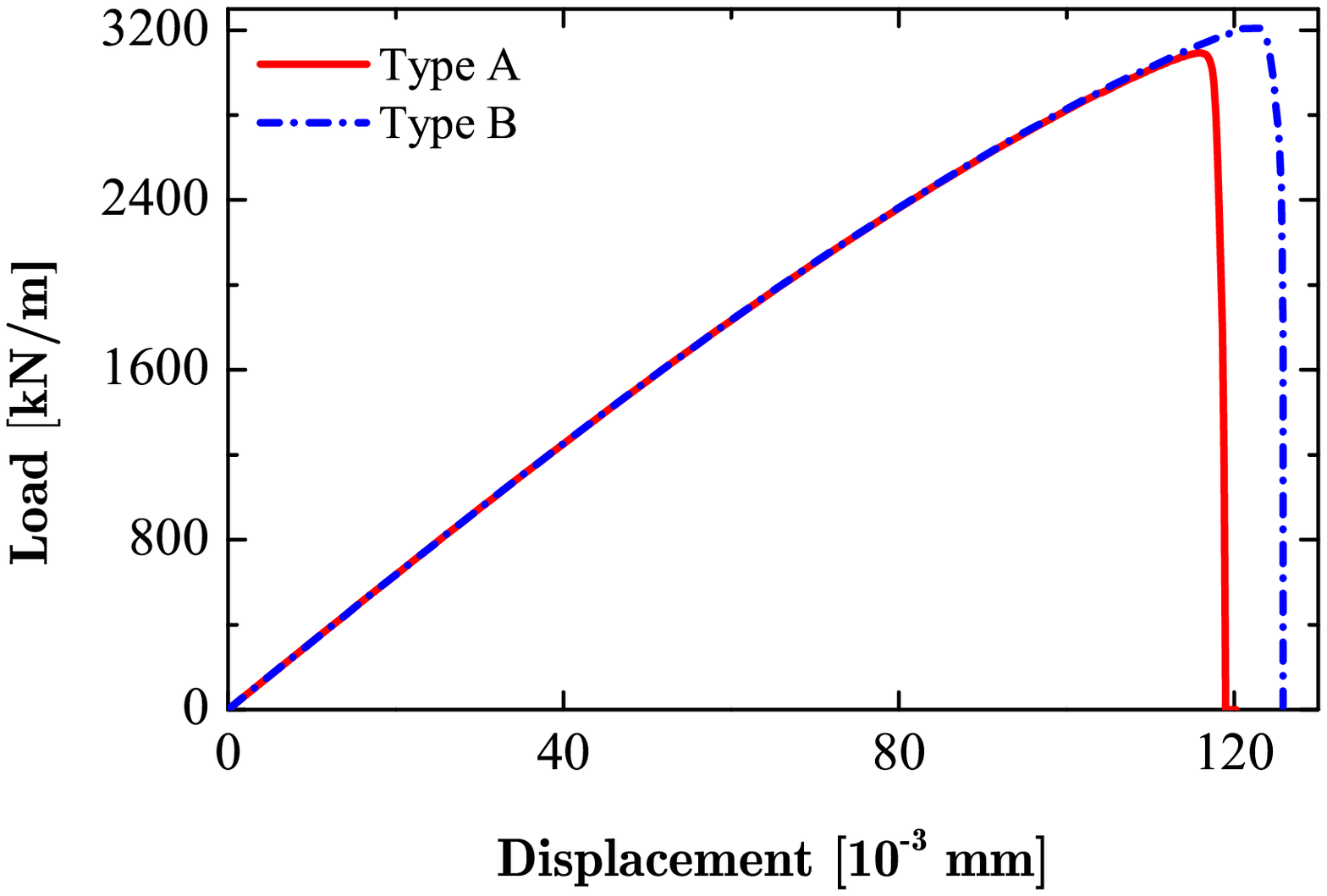}
	\caption{Load-displacement curves of the specimen with two parallel inclined flaws}
	\label{Load-displacement curves of the specimen with two parallel inclined flaws}
	\end{figure}

Figure \ref{Fracture patterns of a specimen with two parallel inclined flaws (coplanar flaws)} shows the compressive-shear fracture propagation in the specimen with two coplanar pre-existing flaws (Type A). With the increase in the compressive loads, shear zones occur around the tips of the inclined flaws and the fractures initiate at $u=0.1178$ mm. When $u=0.1184$ mm, the fractures from the two inclined flaws coalesce. Fractures from the two pre-existing flaws reach the left and right boundaries of the specimen at $u=0.1200$ mm and quasi-coplanar fractures can be observed in Fig. \ref{Fracture patterns of a specimen with two parallel inclined flaws (coplanar flaws)}c. Figure \ref{Fracture patterns of a specimen with two parallel inclined flaws (non-coplanar flaws)} shows the compressive-shear fracture propagation in the specimen with two non-coplanar pre-existing flaws (Type B). The fracture patterns are similar to those in Fig. \ref{Fracture patterns of a specimen with two parallel inclined flaws (coplanar flaws)}. Fractures initiate at $u=0.1244$ mm while the fractures from the two inclined flaws coalesce and reach the left and right boundaries of the specimen at $u=0.1258$ mm. However, the S-shaped coalescence between the two pre-existing flaws is non-coplanar, which is different from the observations for the two coplanar flaws in Fig. \ref{Fracture patterns of a specimen with two parallel inclined flaws (non-coplanar flaws)}. In addition, the load-displacement curves of the specimens with two parallel inclined flaws are shown in Fig. \ref{Load-displacement curves of the specimen with two parallel inclined flaws}. The non-coplanar flaws are shown to have a larger peak load than the coplanar flaws because of the relatively large factual load-bearing area.

\begin{figure}[htbp]
	\centering
	\subfigure[Type I]{\includegraphics[width = 8cm]{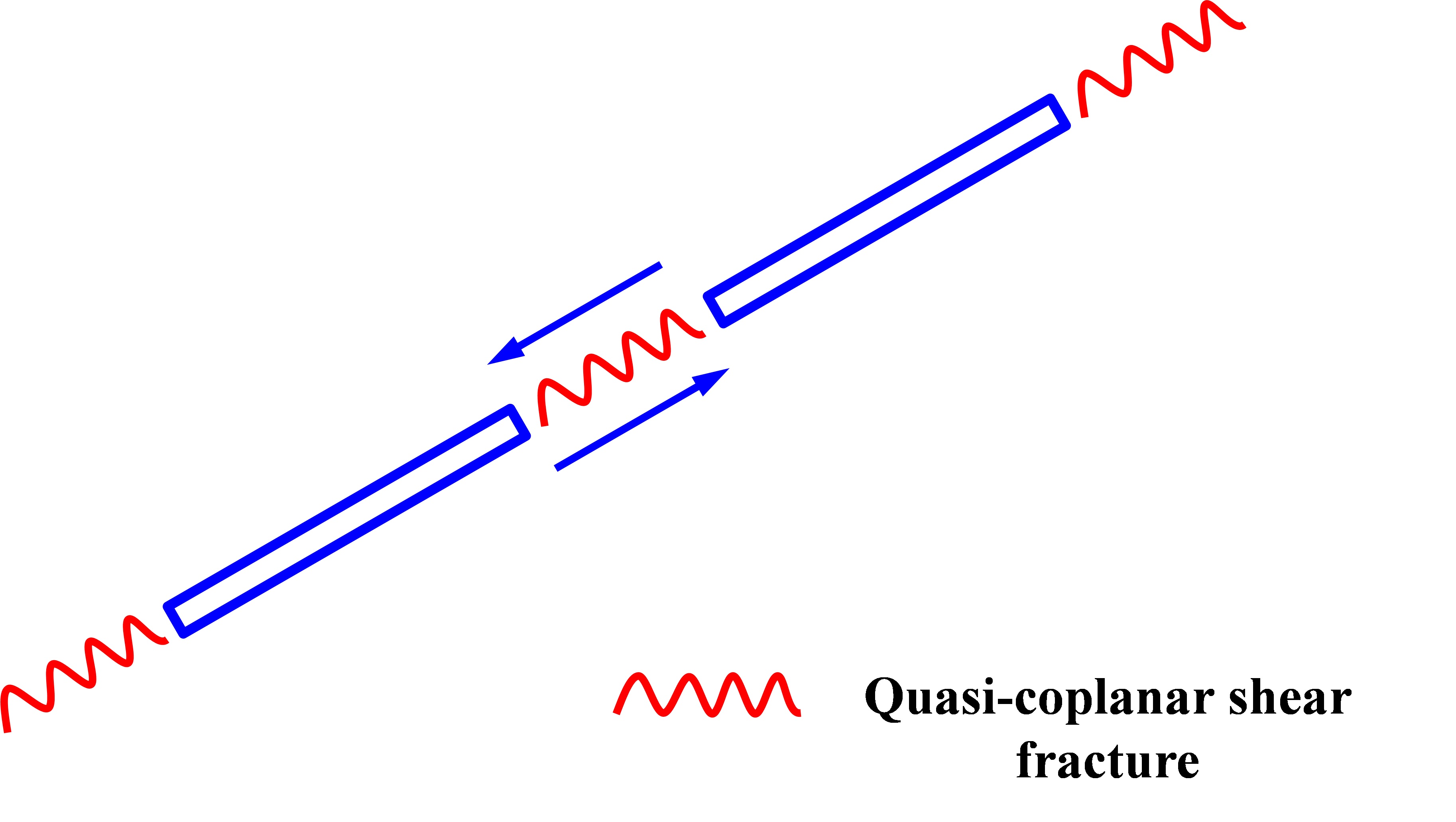}}\\
	\subfigure[Type II]{\includegraphics[width = 8cm]{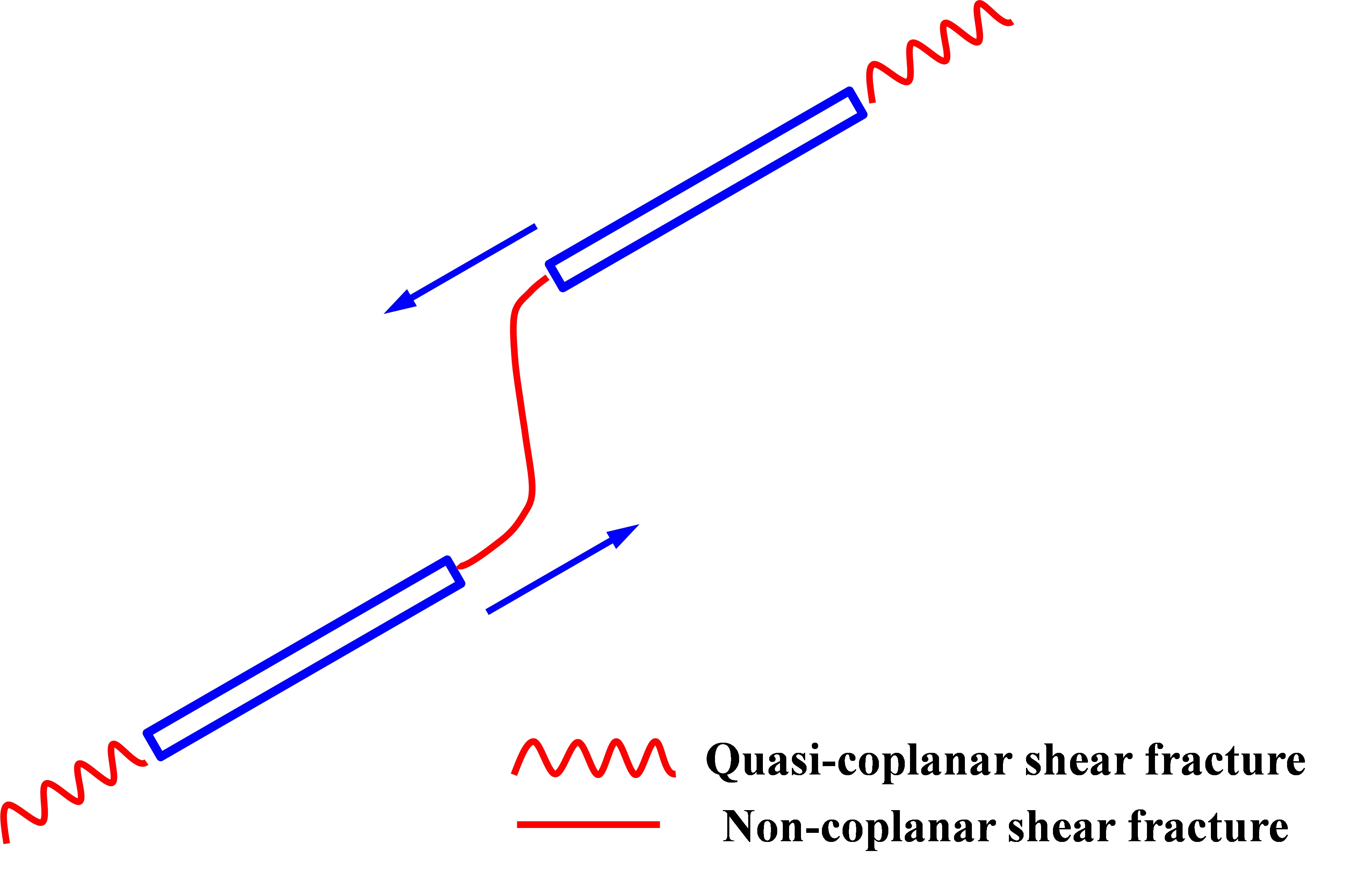}}\\
	\caption{Experimental observations of coalescence of two parallel inclined flaws \citep{sagong2002coalescence}}
	\label{Experimental observations of coalescence of two parallel inclined flaws}
	\end{figure}

\citet{sagong2002coalescence} performed compression tests on rock-like specimens with multiple inclined pre-existing flaws and summarized the patterns of fracture coalescence from the experimental observations. Among the nine patterns, two shear types, namely Types I and II are shown in Fig. \ref{Experimental observations of coalescence of two parallel inclined flaws}. Note that the wing cracks (tensile fractures) are suppressed in Fig. \ref{Experimental observations of coalescence of two parallel inclined flaws} because this study focuses on the compressive-shear fractures and we don't introduce and discuss any tensile fractures. Therefore, the compressive-shear fracture patterns obtained by using the proposed phase field method are in line with the experimental observations, validating the capability of the proposed model in predicting complex compressive-shear fractures in rock-like materials.

\section {Conclusions}\label{Conclusions}

A new phase field model is proposed for simulating brittle compressive-shear fractures in rock-like materials. By using the strain spectral decomposition and with emphasis on the compressive parts of the strain, a new driving force is re-constructed and applied in the evolution equation of phase field. The new driving force can also account for the influence of cohesion and internal friction angle on fracture propagation; this effect cannot be considered in other PFMs. We construct a hybrid formulation for the phase field modeling and implement the proposed model in COMSOL within the framework of conventional finite element method. We then simulate the brittle compressive-shear fractures in rock-like specimens subjected to uniaxial compression. These specimens include intact specimens and specimens with a single inclined flaw or two parallel inclined flaws. The presented numerical results are in good agreement with previous experimental results, validating the feasibility and practicability of the proposed phase field model for simulating brittle compressive-shear fractures.

All the presented numerical examples show that the initiation, propagation, coalescence, and branching of compressive-shear fractures are autonomous without any external fracture criteria or setting a propagation path in advance. This highlights the advantages of the proposed phase field method over other numerical methods in modeling complex compressive-shear fractures in rock-like materials. In addition, because the tensile part of the strain is suppressed in the proposed model, tensile fractures and fractures under tension-shear mode are not included in this new model. In this sense, an improved phase field model that combines the presented new driving force and those previously used will be attractive in the future, especially for the mixed-mode fractures commonly seen in rocks.

\section*{Acknowledgement}
The authors gratefully acknowledge financial support provided by the Natural Science Foundation of China (51474157), and RISE-project BESTOFRAC (734370).

\bibliography{references}

%\onecolumn
%\tableofcontents
\end{document}